\documentclass[aps,superscriptaddress,preprintnumbers,showpacs,showkeys,nofootinbib,floatfix]{revtex4}
\usepackage{graphicx,epsfig,amssymb}
\usepackage{color}
\definecolor{darkgreen}{rgb}{0,0.65,0}
\newcommand{\blue}[1]{{\color{blue} #1}}
\newcommand{\red}[1]{{\color{red} #1}}
\newcommand{\be}{\begin{equation}}
\newcommand{\ee}{\end{equation}}
\newcommand{\ba}{\begin{eqnarray}}
\newcommand{\ea}{\end{eqnarray}}
\newcommand{\la}{\langle}
\newcommand{\ra}{\rangle}
\newcommand{\di}{ {\rm d} }

\begin{document}
\newcommand*{\Jlab}{Thomas Jefferson National Accelerator Facility,
Newport News, VA 23606, U.S.A.}\affiliation{\Jlab}
\newcommand*{\Dubna}{Joint Institute for Nuclear Research, Dubna,
141980 Russia}\affiliation{\Dubna}
\newcommand*{\UConn}{Department of Physics, University of Connecticut,
Storrs, CT 06269, U.S.A.}\affiliation{\UConn}
\newcommand*{\BNL}{RIKEN BNL Research Center, Building 510A, BNL,
Upton, NY 11973, U.S.A.}\affiliation{\BNL}
\newcommand*{\Berkley}{Nuclear Science Division, Lawrence Berkeley National
Laboratory, Berkeley, CA 94720, U.S.A.}\affiliation{\Berkley}

\title{ \boldmath The transverse momentum dependent
 distribution functions in the bag model}
\author{H.~Avakian}\affiliation{\Jlab}
\author{A.~V.~Efremov}\affiliation{\Dubna}
\author{P.~Schweitzer}\affiliation{\UConn}
\author{F.~Yuan}\affiliation{\BNL}\affiliation{\Berkley}

\date{January 2010}
\begin{abstract}
  Leading and subleading twist transverse momentum dependent parton distribution 
  functions (TMDs) are studied in a quark model framework provided by the bag model. 
  A complete set of relations among different TMDs is derived, and the question
  is discussed how model-(in)dependent such relations are.
  A connection of the pretzelosity distribution and quark orbital angular momentum 
  is derived. 
  Numerical results are presented, and applications for phenomenology discussed. 
  In particular, it is shown that in the valence-$x$ region the bag model supports
  a Gaussian Ansatz for the transverse momentum dependence of TMDs.
\end{abstract}
\maketitle
\section{Introduction}
\label{Sec-1:introduction}

TMDs are a generalization 
\cite{Collins:2003fm,Collins:2007ph,Collins:1999dz,Hautmann:2007uw}
of parton distribution functions (PDFs) promising to extend our 
knowledge of the nucleon structure far beyond what we have learned from PDFs
about the longitudinal momentum distributions of partons in the nucleon.
In addition to the latter, TMDs carry also information on transverse parton 
momenta and spin-orbit correlations 
\cite{Collins:1981uk,Ji:2004wu,Collins:2004nx,Cahn:1978se,Konig:1982uk,Chiappetta:1986yg,Collins:1984kg,Sivers:1989cc,Efremov:1992pe,Collins:1992kk,Collins:1993kq,Kotzinian:1994dv,Mulders:1995dh,Boer:1997nt,Boer:1997mf,Boer:1999mm,Bacchetta:1999kz,Brodsky:2002cx,Collins:2002kn,Belitsky:2002sm,Burkardt:2002ks,Pobylitsa:2003ty,Goeke:2005hb,Bacchetta:2006tn,Cherednikov:2007tw,Brodsky:2006hj,Avakian:2007xa,Miller:2007ae,Arnold:2008kf,Brodsky:2010vs,lattice-TMD}.
Here longitudinal and transverse refers to the hard momentum flow in the 
process, for example, in deeply inelastic lepton nucleon scattering (DIS)
the momentum of the virtual photon.

TMDs (and/or transverse momentum dependent fragmentation functions) enter
the description of leading-twist observables in deeply inelastic reactions 
\cite{Collins:1981uk,Ji:2004wu,Collins:2004nx}
on which data are available like: semi-inclusive DIS (SIDIS)
\cite{Arneodo:1986cf,Airapetian:1999tv,Avakian:2003pk,Airapetian:2004tw,Alexakhin:2005iw,Gregor:2005qv,Ageev:2006da,Airapetian:2005jc,Kotzinian:2007uv,Diefenthaler:2005gx,Airapetian:2008sk,Osipenko:2008rv,Giordano:2009hi,Gohn:2009,Airapetian:2009jy},
Drell-Yan process 
\cite{Falciano:1986wk,Conway:1989fs,Zhu:2006gx},
or hadron production in $e^+e^-$ annihilations 
\cite{Abe:2005zx,Ogawa:2006bm,Seidl:2008xc,Vossen:2009xz}.

The interpretation of these data is not straight-forward though.
In SIDIS one deals with convolutions of a priori unknown transverse momentum
distributions in nucleon and fragmentation process, and in practice is forced 
to {\sl assume} models for transverse parton momenta such as the Gaussian 
Ansatz 
\cite{Efremov:2004tp,D'Alesio:2004up,Collins:2005ie,Anselmino:2005nn,Vogelsang:2005cs,Efremov:2006qm,Anselmino:2007fs,Arnold:2008ap,Anselmino:2008sg,Barone:2005kt}.
In the case of subleading twist observables, one moreover faces the problem 
that several twist-3 TMDs and fragmentation functions enter the description 
of one observable 
\cite{DeSanctis:2000fh,Anselmino:2000mb,Efremov:2001cz,Efremov:2001ia,Ma:2002ns,Yuan:2003gu,Gamberg:2003pz,Bacchetta:2004zf,Metz:2004je,Afanasev:2006gw}
(we recall that presently factorization is not proven 
for subleading-twist observables \cite{Gamberg:2006ru}).

In this situation information from models 
\cite{Yuan:2003wk,Gamberg:2007gb,
Yuan:2003gu,Gamberg:2003pz,Metz:2004je,Afanasev:2006gw,
Gamberg:2006ru,Gamberg:2007wm,Jakob:1997wg,Pasquini:2008ax,Efremov:2009ze,Efremov:2009vb,Bacchetta:2008af,She:2009jq,Courtoy:2008dn,Cherednikov:2006zn,Meissner:2007rx,Avakian:2008dz,Boffi:2009sh,Avakian:2009jt,Wakamatsu:2009fn}
is valuable for several reasons.
Models can be used for direct estimates of observables,  
though it is difficult to reliably apply the results, typically 
obtained at low hadronic scales, to experimentally relevant energies 
\cite{Boffi:2009sh}.
Another aspect concerns relations among TMDs observed in models
\cite{Jakob:1997wg,Pasquini:2008ax,Efremov:2009ze,Efremov:2009vb,Avakian:2008dz}.
Such relations, especially when supported by several models, 
could be helpful --- at least for qualitative interpretations of first 
data. Furthermore, model results allow to test assumptions made in literature,
such as the Gaussian Ansatz for transverse momentum distributions
or certain approximations 
\cite{Kundu:2001pk,Goeke:2003az,Kotzinian:2006dw,Avakian:2007mv,Metz:2008ib,Teckentrup:2009tk,Accardi:2009au}.

In addition to such practical applications model studies are of interest 
also because they provide important insights into non-perturbative properties
of TMDs. In this context the probably most interesting recent observation 
in models concerns the pretzelosity distribution function, which in some 
quark models is related to the difference of the helicity and transversity 
distributions \cite{Avakian:2008dz} and, so far, in one model 
to quark orbital momentum \cite{She:2009jq} which is, to best of our knowledge, 
the first 'rigorous' connection of a TMD and quark orbital angular momentum 
in a model.

The purpose of this work is to study TMDs in the framework of the MIT bag 
model. We compute in this model all leading- and subleading-twist, 
time-reversal (T-) even TMDs in Sec.~\ref{Sec-2:TMDs-in-bag}, and address 
then in Sec.~\ref{Sec-3:eq-and-ineq} questions like:
how do relations among TMDs arise in a quark model? How many such relations 
are there in a model? To which extent may one expect such relations to 
be realized in nature?
In Sec.~\ref{Sec-4:pretzelosity-and-OAM} we establish a connection 
of pretzelosity and quark orbital angular momentum in the bag model. 
In Sec.~\ref{Sec-5:results} we present and discuss the numerical results,
using them, among others, for 'testing' the Gaussian Ansatz or 
Wandzura-Wilczek-type approximations
\cite{Avakian:2007mv,Metz:2008ib,Teckentrup:2009tk,Accardi:2009au}.
Finally, in Sec.~\ref{Sec-6:conclusions} we present our conclusions.
Some of the results presented here were shown in the proceeding
\cite{Avakian:2009jt}.

For convenience and in order to make this work self-contained, in the 
remainder of this Introduction we include general definitions 
of TMDs, and introduce relevant notation.

\subsection{General definitions of TMDs}
\label{Sec-1a:TMDs}

Hard processes sensitive to parton transverse momenta like SIDIS
are described in terms of light-front correlators
\be\label{Eq:correlator}
    \phi(x,\vec{p}_T)_{ij} = \int\frac{\di z^-\di^2\vec{z}_T}{(2\pi)^3}\;e^{ipz}\;
    \la N(P,S)|\bar\psi_j(0)\,{\cal W}(0,\,z;\,\mbox{path})\,\psi_i(z)|N(P,S)\ra
    \biggl|_{z^+=0,\,p^+ = xP^+} \;.
    \ee
We use light-cone coordinates $a^\pm=(a^0\pm a^3)/\sqrt{2}$. In SIDIS the
singled-out 3-direction is along the momentum of the hard virtual photon,
and transverse vectors like $\vec{p}_T$ are perpendicular to it.
The path of the symbolically indicated Wilson-link depends on the process
\cite{Collins:2002kn,Belitsky:2002sm,Cherednikov:2007tw}. In the nucleon rest
frame the polarization vector is given by $S=(0,\vec{S}_T,S_L)$ with 
$\vec{S}_T^2+S_L^2=1$.

The information content of the correlator  (\ref{Eq:correlator}) is summarized by
eight leading-twist TMDs \cite{Boer:1997nt}, that can be projected out from the
correlator (\ref{Eq:correlator}) as follows
(color online: \red{red: T-odd}, \blue{blue: T-even})
\ba
    \frac12\;{\rm tr}\biggl[\gamma^+ \;\phi(x,\vec{p}_T)\biggr]
    &=& \hspace{5mm}
    \blue{f_1}-\frac{\varepsilon^{jk}p_T^j S_T^k}{M_N}\,\red{f_{1T}^\perp}
    \label{Eq:TMD-pdfs-I}\\
    \frac12\;{\rm tr}\biggl[\gamma^+\gamma_5 \;\phi(x,\vec{p}_T)\biggr] &=&
    S_L\,\blue{g_1} + \frac{\vec{p}_T\cdot\vec{S}_T}{M_N}\,\blue{g_{1T}^\perp}
    \label{Eq:TMD-pdfs-II}\\
    \frac12\;{\rm tr}\biggl[i\sigma^{j+}\gamma_5 \;\phi(x,\vec{p}_T)\biggr] &=&
    S_T^j\,\blue{h_1}  + S_L\,\frac{p_T^j}{M_N}\,\blue{h_{1L}^\perp} +
    \frac{(p_T^j p_T^k-\frac12\,\vec{p}_T^{\:2}\delta^{jk})S_T^k}{M_N^2}\,
    \blue{h_{1T}^\perp} + \frac{\varepsilon^{jk}p_T^k}{M_N}\,\red{h_1^\perp}\;,
    \label{Eq:TMD-pdfs-III}
\ea
and by the subleading twist TMDs \cite{Bacchetta:2006tn}
\ba
    \frac12\;{\rm tr}\biggl[\,1\;\phi(x,\vec{p}_T)\biggr]         &=&
        \frac{M_N}{P^+}\biggl[
    \hspace{5mm}\blue{e} -\frac{\varepsilon^{jk}p_T^j S_T^k}{M_N}\,\red{e_T^\perp}
    \biggr]
    \label{Eq:sub-TMD-pdfs-I}\\
    \frac12\;{\rm tr}\biggl[i\gamma_5\;\phi(x,\vec{p}_T)\biggr]         &=&
        \frac{M_N}{P^+}\biggl[
    S_L\red{e_L} +\frac{\vec{p}_T\vec{S_T}}{M_N}\,\red{e_T}
    \biggr]
    \label{Eq:sub-TMD-pdfs-II}\\
    \frac12\;{\rm tr}\biggl[\;\,\gamma^\alpha\;\phi(x,\vec{p}_T)\biggr]         &=&
        \frac{M_N}{P^+}\biggl[
    \frac{p_T^j}{M_N}\,\blue{f^\perp}
    +\varepsilon^{jk}S_T^k\red{f_T}
    +S_L\varepsilon^{jk}S_T^k\red{f_L^\perp}
    +\frac{(p_T^j p_T^k-\frac12\,\vec{p}_T^{\:2}\delta^{jk})\varepsilon^{kl}S_T^l}
          {M_N^2}\, \red{f_T^\perp}
    \biggr]
    \label{Eq:sub-TMD-pdfs-III}\\
    \frac12\;{\rm tr}\biggl[\gamma^j\gamma_5 \;\phi(x,\vec{p}_T)\biggr] &=&
    \frac{M_N}{P^+}\biggl[
    S_T^j\,\blue{g_T}  + S_L\,\frac{p_T^j}{M_N}\,\blue{g_L^\perp} +
    \frac{(p_T^j p_T^k-\frac12\,\vec{p}_T^{\:2}\delta^{jk})S_T^k}{M_N^2}\,
    \blue{g_T^\perp} + \frac{\varepsilon^{jk}p_T^k}{M_N}\,\red{g^\perp} \;,
    \biggr]
    \label{Eq:sub-TMD-pdfs-IV}\\
    \frac12\;{\rm tr}\biggl[i\,\sigma^{jk}\gamma_5 \;\phi(x,\vec{p}_T)\biggr] &=&
    \frac{M_N}{P^+}\biggl[
    \frac{S_T^j p_T^k-S_T^k p_T^j}{M_N}\,\blue{h_T^\perp}
    -\varepsilon^{jk}\,\red{h} \biggr]\;,
    \label{Eq:TMD-pdfs-V}
    \\
    \frac12\;{\rm tr}\biggl[\,i\,\sigma^{+-}\gamma_5 \;\phi(x,\vec{p}_T)\biggr] &=&
    \frac{M_N}{P^+}\biggl[
    S_L\,\blue{h_L} + \frac{\vec{p}_T\cdot\vec{S}_T}{M_N}\,\blue{h_T}
    \biggr] \;, 
    \label{Eq:TMD-pdfs-VI}
\ea
where the space-indices $j,k$ refer to the plane transverse with respect to the
light-cone and $\varepsilon^{12} = - \varepsilon^{21} = 1$ and zero else.
Integrating out transverse momenta in the correlator (\ref{Eq:correlator})
leads to the 'usual' parton distributions known from collinear kinematics
$j^a(x) = \int\di^2\vec{p}_T \, j^a(x,\vec{p}_T^{\:2})$ with
$j=f_1,\,g_1,\,h_1,\,e,\,,g_T,\,h_L$ \cite{Ralston:1979ys,Jaffe:1991ra}.
Dirac-structures other than that in
Eqs.~(\ref{Eq:TMD-pdfs-I}--\ref{Eq:TMD-pdfs-VI})
lead to subsubleading-twist terms \cite{Goeke:2005hb}.

For convenience we introduce for a generic TMD $j^q(x,k_\perp)$ the
'(unintegrated) transverse (1)-moments' defined as
\be\label{Eq:def-1mom}
    j^{(1)q}(x,k_\perp) = \frac{k_\perp^2}{2M_N^2}\;j^q(x,k_\perp)\;,\;\;\;
    j^{(1)q}(x) = \int\di^2k_\perp\;\frac{k_\perp^2}{2M_N^2}\;j^q(x,k_\perp)\;.
\ee
Moreover, we shall also make use of the '(1/2)-moments' defined
for a generic TMD as
\be\label{Eq:def-f(1/2)}
        f_1^{(1/2)q}(x) = 
        \int\di^2k_\perp\;\frac{k_\perp}{2M_N}\;f_1^q(x,k_\perp)\;.
\ee

\newpage
\section{TMDs in the bag model}
\label{Sec-2:TMDs-in-bag}

In the MIT bag model, the quark field has the following general form
\cite{Chodos:1974je,Jaffe:1974nj,Celenza:1982uk},
\begin{equation}\label{bw}
    \Psi_\alpha(\vec{x},t)=\sum\limits_{n>0,\kappa=\pm 1,m=\pm 1/2} N(n\kappa)
    \{ b_\alpha(n\kappa m)\psi_{n\kappa jm}(\vec{x},t)+
    d_\alpha^\dagger(n\kappa m)\psi_{-n-\kappa jm}(\vec{x},t)\} \ ,
    \end{equation}
where $b_\alpha^\dagger$ and $d_\alpha^\dagger$ create quark and anti-quark
excitations in the bag with the wave functions
\begin{equation}\label{Eq:wave-func-0}
    \psi_{n,-1,\frac{1}{2}m}(\vec{x},t)=\frac{1}{\sqrt{4\pi}}
    \left ( \begin{array}{r}
    i j_0(\frac{\omega_{n,-1}|\vec{x}|}{R_0})\chi_m\\
    -\vec{\sigma}\cdot\hat{{x}} \; j_1(\frac{\omega_{n,-1}|\vec{x}|}{R_0})\chi_m
    \end{array} \right )
    e^{-i\omega_{n,-1} t/R_0} \ .
    \end{equation}
For the lowest mode, we have $n=1$, $\kappa=-1$, and $\omega_{1,-1}\approx 2.04$
denoted as $\omega\equiv\omega_{1,-1}$ in the following. In the above equation,
$\vec{\sigma}$ is the $2\times 2$ Pauli matrix, $\chi_m$ the Pauli spinor, $R_0$
the bag radius, $\hat{{x}}=\vec{x}/|\vec{x}|$, and $j_i$, are spherical Bessel
functions. Taking the Fourier transformation, we have the momentum space
wave function for the lowest mode,
\begin{equation}
    \varphi_{m}(\vec{k})=i\sqrt{4\pi}N R_0^3
    \left (\begin{array}{r} t_0(k)\chi_m\\
    \vec{\sigma}\cdot\hat{k} \;t_1(k)\chi_m
    \end{array} \right ) \ ,
    \label{wp}
    \end{equation}
where $\hat{k} = \vec{k}/k$ with $k=|\vec{k}|$ and the normalization factor $N$ is,
\begin{equation}
        N=\left(\frac{\omega^3}{2R_0^3(\omega-1)\sin^2\omega}\right)^{1/2} \ .
    \end{equation}
The two functions $t_i$, $i=0,1$ are defined as
\begin{equation}
    \label{Eq:t0-t1}
    t_i(k)=\int\limits_0^1 u^2 du j_i(ukR_0)j_i(u\omega) \ .
\end{equation}
From the above equations, we see that the bag model wave function Eq.~(\ref{wp})
contains both $S$ and $P$ wave components. Especially, $t_0$ represents the $S$-wave
component, whereas $t_1$ represents the $P$-wave component of the proton wave
functions.

With the above wave functions, we can calculate all quark TMDs.
For convenience we define the constant $A$, which will be common to all TMDs,
and the momenta $k_z$ and $k$ as
\be\label{Eq:notation}
        A=\frac{16\omega^4}{\pi^2(\omega -1)j_0^2(\omega)\,M_N^2}\,,\;\;\;
    k=\sqrt{k_z^2+k_\perp^2}\;,\;\;\;
    k_z=xM_N-\omega/R_0\;,\;\;\;\widehat{k}_z=\frac{k_z}{k}\;,\;\;\;
    \widehat{M}_N=\frac{M_N}{k}\;,
\ee
where $M_N$ is the proton mass, and the bag radius is fixed such that 
$R_0 M_N=4\omega$. Moreover, we assume $SU(6)$ spin-flavor symmetry of the
proton wave function, such that spin-independent TMDs of definite flavor
are given in terms of respective 'flavor-less' expressions multiplied
by a 'flavor factor' $N_q$, and spin-dependent TMDs of definite flavor
follow from multiplying the respective 'flavor-less' expressions
by a 'spin-flavor factor' $P_q$ with
\begin{equation}\label{Eq:wafe-function-SU(6)-pol}
    N_u = 2\;,\;\;\; N_d=1\;,\;\;\;
    P_u =  \frac{4}{3}\;,\;\;\;
    P_d = -\frac{1}{3}\;.
\end{equation}
We recall that in the quark model formulated for a general (odd) number of colors 
$N_c$, these flavor factors are given by $N_u = (N_c+1)/2$ and $N_d=(N_c-1)/2$ 
while $P_u = (N_c+5)/6$ and $P_d=(-N_c+1)/6$ \cite{Karl:1984cz}.

We mention that the MIT bag model gives rise also to antiquark distributions, 
but to unphysical ones, since $f_1^{\bar q}(x)<0$, which violates positivity. 
The TMDs receive non-vanishing support also from the regions $|x|\ge 1$.
Though non-physical these contributions must be included when evaluating 
sum rules like $\int\di x\,f_1^q(x) = N_q$ or the momentum sum rule 
$\sum_q\int\di x\,xf_1^q(x) = 1$, i.e.\  sum rules are satisfied only 
when integrating over the whole $x$-axis.

In literature it was discussed how to deal with these caveats,
see for example \cite{Schreiber:1991tc}.
In this work, we limit ourselves to the discussion of quark TMDs at 
$0\le x\le 1$, which should not be confused with 'valence distributions', 
for example $f_{1\,\rm val}^q(x)=f_1^q(x)-f_1^{\bar q}(x)$.
When discussing sum rules, however, integration over the whole 
$x$-axis is implied.

Since there are no explicit gluon degrees of freedom,
T-odd TMDs vanish in this model \cite{Yuan:2003wk}. In principle, 
one can simulate the effect of the gauge link, which is crucial in QCD
for T-odd effects \cite{Brodsky:2002cx,Collins:2002kn,Belitsky:2002sm},
for example by introducing 'one-gluon-exchange' 
\cite{Yuan:2003wk,Courtoy:2008dn} or invoking instanton effects 
\cite{Cherednikov:2006zn}.
In this work we shall not consider such extensions of the bag model,
and restrict ourselves to the description of T-even distributions.

\newpage
\subsection{Results for TMDs in the bag model}
\label{Sec-2a:results}

In the notation introduced above, the results for the T-even leading twist 
TMDs are given by
\ba
   f_1^q(x,k_\perp) &=& N_q  A\biggl[t_0^2+2\widehat{k}_z\,t_0t_1
            +t_1^2\biggr]               \label{Eq:f1}\\
   g_1^q(x,k_\perp) &=& P_q\,A\biggl[t_0^2+2\widehat{k}_z\,t_0t_1
                    +(2\widehat{k}_z^2-1)\,t_1^2\biggr]  \label{Eq:g1}\\
   h_1^q(x,k_\perp) &=& P_q\,A\biggl[t_0^2+2\widehat{k}_z\,t_0t_1
            +\widehat{k}_z^2\,t_1^2\biggr]      \label{Eq:h1}\\
   g_{1T}^{\perp q}(x,k_\perp) &=& P_q\,A\biggl[\phantom{-}2\widehat{M}_N
    (t_0t_1+\widehat{k}_z\,t_1^2)\biggr] \label{Eq:g1Tperp}\\
   h_{1L}^{\perp q}(x,k_\perp) &=& P_q\,A\biggl[-2\widehat{M}_N
    (t_0t_1+\widehat{k}_z\,t_1^2)\biggr] \label{Eq:h1Lperp}\\
   h_{1T}^{\perp q}(x,k_\perp) &=& P_q\,A\biggl[-2\widehat{M}_N^{\,2} \,t_1^2\biggr]
    \label{Eq:h1Tperp}
\ea
and for the subleading twist TMDs we obtain
\ba
e^q          (x,k_\perp)&=&N_q  A\biggl[t_0^2-t_1^2 \biggr] \label{Eq:e}\\
f^{\perp q}  (x,k_\perp)&=&N_q  A\biggl[2\widehat{M}_N\,t_0t_1\biggr]\label{Eq:fperp}\\
g_T^q        (x,k_\perp)&=&P_q\,A\biggl[t_0^2-\widehat{k}_z^2\,t_1^2\biggr]\label{Eq:gT}\\
g_L^{\perp q}(x,k_\perp)&=&P_q\,A\biggl[2\widehat{M}_N\,\widehat{k}_z\,t_1^2\biggr]
                                    \label{Eq:gLperp}\\
g_T^{\perp q}(x,k_\perp)&=&P_q\,A\biggl[2\widehat{M}_N^2\,t_1^2\biggr]\label{Eq:gTperp}\\
h_L^q(x,k_\perp)        &=&P_q\,A\biggl[t_0^2+(1-2\widehat{k}_z^2)t_1^2\,
                                 \biggr] \hspace{16mm}\label{Eq:hL}\\
h_T^{\perp q}(x,k_\perp)&=&P_q\,A\biggl[ 2\widehat{M}_N\;t_0t_1\biggr]\label{Eq:hTperp}\\
h_T^{      q}(x,k_\perp)&=&P_q\,A\biggl[-2\widehat{M}_N\widehat{k}_z\,t_1^2\biggr]
\label{Eq:hT}
\ea 

In the following Sections we shall discuss these results in detail.

\newpage
\section{Equalities and inequalities among TMDs}
\label{Sec-3:eq-and-ineq}

In QCD all TMDs are independent functions. But in quark models,
due to absence of gauge field degrees of freedom, certain relations 
among different TMDs appear which must be satisfied in any 
consistent relativistic quark~model. We discuss these ``model-independent''
quark-model relations in Sec.~\ref{Sec-3a:LIRs}.
Of course, depending on a quark model further relations may appear,
and the bag model results (\ref{Eq:f1}-\ref{Eq:hT}) provide a nice 
illustration why this happens which is demonstrated in 
Secs.~\ref{Sec-3b:linear-relation-in-bag-model}
and \ref{Sec-3c:non-linear-relation-in-bag-model} in detail.
In Sec.~\ref{Sec-3d:validity-of-relations} we compare to results
from other models. This comparison helps to establish to which extent 
which relations might be expected to be useful in nature.
Sec.~\ref{Sec-3e:particular-relation} is devoted to the discussion
of one particular relation.

\subsection{Relations valid in all quark models}
\label{Sec-3a:LIRs}

Certain relations among TMDs must be valid in any quark model of the 
nucleon lacking gluon degrees of freedom \cite{Teckentrup:2009tk}. 
In such ``no-gluon models'' the absence of the Wilson-link 
implies that in the general Lorentz-decomposition of the 
unintegrated quark-correlator certain amplitudes do not appear,
namely the $B_i$-amplitudes ($i=1,\,2,\,\dots\,20$)
in the notation of \cite{Goeke:2005hb}. 
This gives rise to the following relations
\cite{Mulders:1995dh,Teckentrup:2009tk} 
\ba
\label{eq:LIR1} g_T(x) \; &\stackrel{\rm LIR}{=}& \; 
	        g_1(x) + \frac{\di }{\di  x} g^{\perp(1)}_{1T}(x)\, ,\\ 
\label{eq:LIR2} h_L(x) \; &\stackrel{\rm LIR}{=}& \; 
	      	h_1(x) - \frac{\di }{\di  x} h^{\perp(1)}_{1L}(x) \, , \\
\label{eq:LIR4} h_T(x) \; &\stackrel{\rm LIR}{=}& \; 
		- \frac{\di }{\di  x} h^{\perp(1)}_{1T}(x) \, , \\
\label{eq:LIR3} g_L^\perp(x) + \frac{\di }{\di  x} g_T^{\perp(1)}(x) \; 
		&\stackrel{\rm LIR}{=}& \; 0 \, ,\\ 
\label{eq:LIR5} h_T(x,p_T)-h_T^\perp(x,p_T) \; &\stackrel{\rm LIR}{=}& 
		\; h^{\perp}_{1L}(x,p_T) \, ,
\ea
which {\sl must} hold in any consistent relativistic quark model.
These so-called ``Lorentz-invariance relations''  (LIRs) are not valid 
in models with gauge field degrees of freedom \cite{Kundu:2001pk} and in 
QCD \cite{Goeke:2003az}. The applications of LIRs in phenomenology were 
discussed in \cite{Metz:2008ib,Teckentrup:2009tk}. There it was 
also shown, by exploring QCD equations of motion, that some LIRs hold in 
an approximation consisting of the neglect of quark-gluon-quark-correlator 
and current quark mass terms. Whether such an approximation is is justified
in nature is, of course, a different question. For discussions of specific
cases see \cite{Metz:2008ib,Teckentrup:2009tk,Avakian:2007mv,Accardi:2009au,Wandzura:1977qf,Zheng:2004ce,Balla:1997hf,Gockeler:2000ja,Anikin:2001ge}.
For quark model calculations, the practical application of the relations
(\ref{eq:LIR1}--\ref{eq:LIR5}) is immediate: they provide a valuable
cross check for the numerical results. 

In App.~\ref{App-A:prove-LIRs} we provide analytical proofs that the
LIRs (\ref{eq:LIR1}--\ref{eq:LIR5}) are satisfied in the bag model.
We also checked that the numerical results satisfy the LIRs,
which provides a welcome cross-check for the numerical calculation.

\newpage
\subsection{Linear relations in bag model}
\label{Sec-3b:linear-relation-in-bag-model}

In the bag model, there are 9 linear relations among the 14 
(twist-2 and 3) T-even TMDs, which can be written as follows
\ba
&&  {\cal D}^q\,f_1^q(x,k_\perp) + g_1^q(x,k_\perp) = \;2 h_1^q(x,k_\perp)
    \label{Eq:rel-I} \\
&&  {\cal D}^q\,e^q(x,k_\perp) + h_L^q(x,k_\perp) = 2g_T^q(x,k_\perp)
    \label{Eq:rel-II}\\
&&  {\cal D}^q\,f^{\perp q}(x,k_\perp) = h_T^{\perp q}(x,k_\perp)
    \label{Eq:rel-III}\\
&&  \nonumber\\
&&  \hspace{5mm}g_{1T}^{\perp q}(x,k_\perp) = -\, h_{1L}^{\perp q}(x,k_\perp)\;\label{Eq:rel-IV}\\
&&  \hspace{5mm}g_T   ^{\perp q}(x,k_\perp)    = -\, h_{1T}^{\perp q}(x,k_\perp)\label{Eq:rel-V}\\
&&  \hspace{5mm}g_L   ^{\perp q}(x,k_\perp)    = -\, h_T   ^{      q}(x,k_\perp)\label{Eq:rel-VI}\\
&&  \nonumber\\
&&  \hspace{5mm}g_1^q(x,k_\perp) - h_1^q(x,k_\perp) = h_{1T}^{\perp(1)q}(x,k_\perp)
    \label{Eq:measure-of-relativity}\\
&&  \hspace{5mm}g_T^{      q}(x,k_\perp)-h_L^{      q}(x,k_\perp) =  h_{1T}^{\perp(1)q}(x,k_\perp)\label{Eq:rel-VIII}\\
&&  \hspace{5mm}h_T^{      q}(x,k_\perp)-h_T^{\perp q}(x,k_\perp) =  h_{1L}^{\perp q}(x,k_\perp)\label{Eq:rel-IX}
\ea
where the 'dilution factor' is defined as
\be\label{Eq:dilution-factor}
    {\cal D}^q=\frac{P_q}{N_q}\;.
\ee

In the relations (\ref{Eq:measure-of-relativity},~\ref{Eq:rel-VIII})
some TMDs need to be multiplied  by the model-independent factor
$k_\perp^2/(2M_N^2)$, which is a `legitimate linear operation' in our context
(the meaning of that will be explained shortly). 
The '(unintegrated) transverse moments' are defined in Eq.~(\ref{Eq:def-1mom}).

Why are there 9 linear relations? In fact,
naively, one could have expected even more relations, since all TMDs
are expressed in terms of only two functions, $t_0$ and $t_1$ representing
the contributions from the S and P-wave components of the proton wave function,
Eqs.~(\ref{wp},~\ref{Eq:t0-t1}). However, having {\sl linear relations} in mind,
the combinations $t_0^2$, $t_0t_1$, $t_1^2$ are to be considered as independent
structures. But there are more {\sl independent} structures than that.
By inspecting Eqs.~(\ref{Eq:f1}-\ref{Eq:hT}) we see
that the actual number of linearly independent structures in the TMDs is 5,
namely
\be\label{Eq:indepdendent-structures}
{{\displaystyle
\left.\begin{array}{l}
    \displaystyle {\rm I}.\;\;t_0^2         \cr \phantom{a} \cr
    \displaystyle {\rm II}.\;\;\widehat{k}_z\,t_0t_1 \;\;\cr \phantom{a} \cr
    \displaystyle {\rm III}.\;\;\;t_1^2,\,\widehat{k}_z^2t_1^2
    \end{array}
    \right\}
    \leftrightarrow
    \left\{\begin{array}{l}
    f_1^q,\,g_1^q,h_1^q,h_{1T}^{\perp q}\;\,  (\mbox{twist 2})\cr
    e^q,g_T^q,h_L^q,g_T^{\perp q}       \;\,  (\mbox{twist 3})
    \end{array}\right. \;\;\Rightarrow\;\;\; 
    \mbox{relations (\ref{Eq:rel-I}, \ref{Eq:rel-II}, \ref{Eq:rel-V}, \ref{Eq:measure-of-relativity}, \ref{Eq:rel-VIII})} }
\atop
{\phantom{\frac11}}}\atop
{\left.\begin{array}{l}
    \displaystyle {\rm IV}.\;\;t_0t_1            \cr \phantom{a} \cr
    \displaystyle {\rm V}.\;\;\widehat{k}_z\,t_1^2\;\;
    \end{array}\right\}
    \leftrightarrow
    \left\{
    \begin{array}{l}
    g_{1T}^{\perp q},h_{1L}^{\perp q} \hspace{1.5cm} (\mbox{twist 2})\cr
    f^{\perp q},g_{ L}^{\perp q},h_{ T}^{\perp q}, h_{ T}^{ q}\;\,(\mbox{twist 3})
    \end{array}\right. \;\;\Rightarrow\;\;\; 
    \mbox{relations (\ref{Eq:rel-III}, \ref{Eq:rel-IV}, \ref{Eq:rel-VI}, \ref{Eq:rel-IX})} }
    \;\;\;\;\;
\ee
where we show respectively to which TMDs the different structures contribute.
We observe that in some sense there are two 'disconnected subspaces': one is 
due to the structures I, II, III, and the other due to the structures IV, V.

The structures II and IV, $\widehat{k}_zt_0t_1$ and $t_0t_1$, are linearly 
independent, as there is no way of relating one with the other in a
model-independent way. Indeed, in order to do this, one should multiply 
a TMD by a factor including~$k_z$ which explicitly depends on parameters 
of the bag model, as is evident from Eq.~(\ref{Eq:notation}),
and we discard such a manipulation as a model-dependent operation. 
For the same reason the structures in III and V are linearly independent.

But $\widehat{k}_z^2t_1^2$ and $t_1^2$ in point~III are linearly dependent: 
if we multiply $t_1^2$ 
(actually in relevant expressions $\widehat{M}_N^2t_1^2$ appears) 
by the {\sl model-independent} factor $k_\perp^2/M_N^2$
and add $\widehat{k}_z^2\,t_1^2$ we obtain just $t_1^2$ which happens in
Eqs.~(\ref{Eq:measure-of-relativity},~\ref{Eq:rel-VIII}).
Clearly, the multiplication of TMDs by $k_\perp^2/(2M_N^2)$ is a
{\sl model-independent} manipulation leading to transverse moments
in (\ref{Eq:def-1mom}).

To summarize, there are 5 linearly independent structures in the bag model
and 14 T-even TMDs. This implies 9 linear relations, 
and Eqs.~(\ref{Eq:rel-I}--\ref{Eq:rel-IX}) represent one way of
writing these relations. 
These findings mean that one can choose, in the bag model, a basis of
5 linearly independent TMDs, and construct the other TMDs from this
basis. 

\subsection{Non-linear relations in the bag model}
\label{Sec-3c:non-linear-relation-in-bag-model}

In Eq.~(\ref{Eq:indepdendent-structures}) we found that the TMDs in the bag
model form two independent 'subspaces.' Let us summarize:
\ba\label{Eq:indepdendent-structures-2}
    \mbox{subspace A (I, II, III):} &
    \underbrace{f_1^q\;g_1^q\;h_1^q\;h_{1T}^{\perp q}}_{\mbox{\footnotesize twist 2}}\;\;
    \underbrace{e^q  \;g_T^q\;h_L^q\;g_T^{\perp q}   }_{\mbox{\footnotesize twist 3}} & 
    \;\;\Rightarrow\;\;\; 
    \mbox{relations (\ref{Eq:rel-I}, \ref{Eq:rel-II}, \ref{Eq:rel-V}, \ref{Eq:measure-of-relativity}, \ref{Eq:rel-VIII}),} 
    \\
    \mbox{subspace B (IV, V):}  \;\;\;\; &
    \underbrace{g_{1T}^{\perp q}\;h_{1L}^{\perp q}}_{\mbox{\footnotesize twist 2}}\;\;
    \underbrace{f^{\perp q}\;g_{ L}^{\perp q}\;h_{ T}^{\perp q}\;h_{ T}^{ q}}_{\mbox{\footnotesize twist 3}} &
    \;\; \Rightarrow\;\;\; 
    \mbox{relations (\ref{Eq:rel-III}, \ref{Eq:rel-IV}, \ref{Eq:rel-VI}, \ref{Eq:rel-IX}).} \nonumber
\ea
In other words, there is no linear relation which would transform 
TMDs from subspace A into TMDs in subspace B. However, there are 
{\sl non-linear relations} which can do that, for example, 
\ba\label{Eq:non-lin-2}
      h_1^q(x,k_\perp)\,h_{1T}^{\perp q}(x,k_\perp) &=& -
      \frac{1}{2}\,\biggl[h_{1L}^{\perp q}(x,k_\perp)\biggr]^2,\\ 
   \label{Eq:non-lin-1}
      g_T^q(x,k_\perp)\,g_T^{\perp q}(x,k_\perp) &=& \phantom{-}
      \frac{1}{2}\,\biggl[g_{1T}^{\perp q}(x,k_\perp)\biggr]^2 -
      g_{1T}^{\perp q}(x,k_\perp)\,g_L^{\perp q}(x,k_\perp)\,.
\ea
The results are presented such that on the left-(right-)hand-sides
(L(R)HS) only TMDs from subspace A (B) appear. 

The Eqs.~(\ref{Eq:non-lin-2},~\ref{Eq:non-lin-1}) are independent
in the sense that it is impossible to convert one into the other 
upon use of the linear relations (\ref{Eq:rel-I}--\ref{Eq:rel-IX}).
In order to convince oneself of that, notice that on RHS of 
(\ref{Eq:non-lin-2}) $h_{1L}^{\perp q} \in$ subspace B appears from
which alone one cannot construct the TMDs $\in$ subspace B 
on RHS of (\ref{Eq:non-lin-1}): from $h_{1L}^{\perp q}$
one obtains $g_{1T}^{\perp q}$ via (\ref{Eq:rel-IV}) but not 
$g_{L}^{\perp q}$. However, for example, $g_{1T}^{\perp q}$ and
 $g_L^{\perp q}$ span a basis which allows to construct all
TMDs $\in$ subspace B. Similarly one finds that from the 
over-complete set of TMDs on LHS of 
(\ref{Eq:non-lin-2},~\ref{Eq:non-lin-1}) all TMDs $\in$ subspace A
follow. To summarize, the non-linear relations 
(\ref{Eq:non-lin-2},~\ref{Eq:non-lin-1}) are independent, 
and these are the only independent non-linear relations.

Of course, upon the use of the linear relations 
(\ref{Eq:rel-I}--\ref{Eq:rel-IX}) one could generate 
further non-linear relations. One advantage of the presentation 
(\ref{Eq:non-lin-2},~\ref{Eq:non-lin-1}) is that they connect
only chirally even, or only chirally odd TMDs.

With the 9 linear relations (\ref{Eq:rel-I}--\ref{Eq:rel-IX}) 
and the 2 non-linear relations (\ref{Eq:non-lin-2},~\ref{Eq:non-lin-1}) 
we find altogether 11 relations among 14 TMDs in the bag model.
This may reflect that eventually all TMDs can be traced back to 
the free structures proportional to $t_0^2$, $t_0t_1$, $t_1^2$.

It should be noticed that in 
Secs.~\ref{Sec-3b:linear-relation-in-bag-model}
and \ref{Sec-3c:non-linear-relation-in-bag-model}
we permitted only manipulations of the kind: adding TMDs, multiplying 
them, and forming (1)-moments. If one includes differentiation of TMDs 
one obtains additional relations, see Sec.~\ref{Sec-3a:LIRs} and
App.~\ref{App-A:prove-LIRs}.

\newpage
\subsection{How general are quark model relations among TMDs?}
\label{Sec-3d:validity-of-relations}

The deeper reason, why  in the bag model relations among TMDs appear, 
is ultimately related to Melosh rotations which connect longitudinal 
and transverse nucleon and quark polarization states in a 
Lorentz-invariant way \cite{Efremov:2002qh}.

An important issue, when observing relations among TMDs in a model,
concerns their presumed validity beyond that particular model framework.
For that it is instructive to compare first to other models.  
In fact, some of the relations (\ref{Eq:rel-I}--\ref{Eq:rel-IX}) 
were discussed previously in literature in various models.
Let us review briefly.
\begin{itemize}
\item 
	Eq.~(\ref{Eq:rel-I}):
	its  $k_\perp$-integrated version was discussed in 
	bag model in \cite{Jaffe:1991ra} and 
	\cite{Signal:1997ct,Barone:2001sp} and in light-cone 
	constituent models in \cite{Pasquini:2005dk}.
	The unintegrated version was discussed in
	bag and light-cone constituent models
	\cite{Pasquini:2008ax,Avakian:2008dz}.
\item 
	Eq.~(\ref{Eq:rel-II}):
	its integrated version was observed in the
	bag model previously in \cite{Signal:1997ct}.
\item 
	Eq.~(\ref{Eq:rel-IV}): 
	was first observed in the spectator model of \cite{Jakob:1997wg} 
	and later also in light-cone constituent models \cite{Pasquini:2008ax} 
	and the covariant parton model of Ref.~\cite{Efremov:2009ze}.
\item 
	Eq.~(\ref{Eq:rel-VI}): 
	was found in the spectator model of Ref.~\cite{Jakob:1997wg}.
\item 
	Eq.~(\ref{Eq:measure-of-relativity}): 
	was first observed in the bag \cite{Avakian:2008dz}.
	It is valid also in the spectator \cite{Jakob:1997wg},
	light-cone constituent \cite{Pasquini:2008ax},
	and covariant parton \cite{Efremov:2009ze} models.
\item
	Eqs.~(\ref{Eq:rel-III},~\ref{Eq:rel-V},~\ref{Eq:rel-VIII},~\ref{Eq:rel-IX}):
	are new in the sense of not having been mentioned previously in literature. 
	But the latter 3 are satisfied by the spectator model results from 
	\cite{Jakob:1997wg}.
\item 
        The non-linear relation (\ref{Eq:non-lin-2}), which connects all 
        T-even, chirally-odd leading-twist TMDs was observed in the covariant parton 
        model approach \cite{Efremov:2009ze}. Eq.~(\ref{Eq:non-lin-1})
        was not discussed so far in literature.
\end{itemize}
The detailed comparison, in which models these relations hold
and in which they are violated, gives some insight into the
question to which extent these relations are model-dependent.

Let us discuss first Eqs.~(\ref{Eq:rel-I}--\ref{Eq:rel-III}),
which include the 'dilution factor' (\ref{Eq:dilution-factor}) and
connect polarized and unpolarized TMDs. For these relations 
SU(6)-spin-flavor symmetry is necessary, but not sufficient.
In fact, this type of relations holds only in `simplest models' such 
as the bag model version used here or light-cone constituent models 
\cite{Avakian:2008dz,Pasquini:2008ax}. What these models have in common
is that the nucleon wave-function is constructed from 'flavor-blind'
quark wave-functions multiplied by appropriate spin-flavor factors 
in Eq.~(\ref{Eq:wafe-function-SU(6)-pol}).
The SU(6) symmetry, however, does not need to be realized in a model
that simply. For example, the spectator model of \cite{Jakob:1997wg} is 
SU(6) symmetric. But it does not support (\ref{Eq:rel-I}--\ref{Eq:rel-III})
which are spoiled by the different masses of the (scalar and axial-vector) 
spectator diquark systems.
Interestingly, it is possible to recover these relations in \cite{Jakob:1997wg}
in the limiting case of the scalar and axial-vector diquark 
masses becoming equal (justified in large-$N_c$ limit).
We mention that (\ref{Eq:rel-I},~\ref{Eq:rel-II}) also are not supported
in the covariant parton model approach of \cite{Efremov:2009ze}.
However, also in that approach it is possible to 'restore' these relations
by introducing additional, restrictive assumptions, see \cite{Efremov:2009ze}
for a detailed discussion. 
We conclude that the relations (\ref{Eq:rel-I}--\ref{Eq:rel-III}) require
strong model assumptions. It is difficult to estimate to which extent
such relations could be useful approximations in nature, though they
could hold in the valence-$x$ region with an accuracy of (20--30)$\,\%$,
see the comparison of similar SU(6) predictions and data for the 
polarized neutron and proton structure functions in \cite{Boffi:2009sh}.

From the point of view of model dependence, it is 'safer' \cite{Avakian:2008dz}
to compare relations which include only polarized or only unpolarized TMDs.
We know no example for the latter, however, the relations
(\ref{Eq:rel-IV}--\ref{Eq:rel-IX}) are of the former type.
It is gratifying to observe that these relations are satisfied not only
in the bag model, this work and \cite{Avakian:2008dz}, but also in the
spectator model version of Ref.~\cite{Jakob:1997wg}.
The relations among the leading twist TMDs,
Eqs.~(\ref{Eq:rel-IV},~\ref{Eq:measure-of-relativity}),
hold also in light-cone constituent \cite{Pasquini:2008ax},
and covariant parton \cite{Efremov:2009ze} models.
They are also valid in the non-relativistic model \cite{Efremov:2009ze}.
Though they do not prove anything, these observations indicate that 
such relations could be valid in a wider class of quark models.

Of course,  quark model relations among TMDs have limitations, 
even in quark models. In \cite{Bacchetta:2008af} various versions 
of spectator models were used, and in some versions the relations 
were not supported (\ref{Eq:rel-IV},~\ref{Eq:measure-of-relativity}).
It is instructive to observe that also the quark-target model 
\cite{Meissner:2007rx} not supported the relations 
(\ref{Eq:rel-IV},~\ref{Eq:measure-of-relativity}).
In fact, the inclusion of gauge fields brings us a step closer 
to QCD as compared to quark models, at least in what concerns 
the involved degrees of freedom. Finally, in QCD none of such
relations is valid, and all TMDs are independent structures.

It would be interesting to `test' such quark model relations
in other models, lattice QCD, and in experiment. The latter,
in fact, provide a test for the usefulness of quark models 
themselves --- or, more precisely, their application to TMDs.

\newpage
\subsection{A special linear relation among collinear functions }
\label{Sec-3e:particular-relation}

By taking linear combinations of (\ref{Eq:rel-I}--\ref{Eq:rel-IX}) one
can obtain many more linear relations. It is worth to discuss in some more detail 
one particularly interesting relation, which can be obtained in this way.
By eliminating the transverse moment of the pretzelosity
distribution from Eqs.~(\ref{Eq:measure-of-relativity},~\ref{Eq:rel-VIII}),
and integrating over transverse momenta, we obtain
\be\label{Eq:new-interesting}
	g_1^q(x)-h_1^q(x) = 
	g_T^q(x)-h_L^q(x) \,.
\ee
This relation holds also in its unintegrated form.
There are several reasons, why this relation is interesting. 

First, it involves only collinear parton distribution functions,
which is the only relation of such type in bag model. 
(Actually (\ref{Eq:rel-I},~\ref{Eq:rel-II}) are also of such type, 
but they include the 'dilution factor' and are supported only
in models with simplest spin-flavor structures, see 
Sec.~\ref{Sec-3d:validity-of-relations}.)
The QCD evolution equation for all these functions are known,
and they are different, which shows the limitation of this relation: 
even if for some reason (\ref{Eq:new-interesting}) was valid in QCD 
at a certain renormalization scale $\mu_0$, 
it would break down at any other scale  $\mu\neq\mu_0$.
This is by no means surprising, and we expect such 
limitations for all model relations. 

Second, for the first Mellin moment this relation 
is valid  model-independently. Hereby we strictly 
speaking presume the validity of the Burkardt-Cottingham sum rule,
which is equivalent to the statement $\int\di x\,g_T^q(x)=\int\di x\,g_1^q(x)$, 
and an analog sum rule for $h_L^q(x)$ and $h_1^q(x)$.
In QCD there are doubts especially concerning the validity of  
the Burkardt-Cottingham sum rule. However, it is valid in many
models such as bag \cite{Jaffe:1991ra} or chiral
quark soliton model~\cite{Wakamatsu:2000ex}.

Third, though it certainly is not exact in QCD, it would be
interesting to learn whether (\ref{Eq:new-interesting})
is satisfied in nature approximately. Also this relation 
can be tested on the lattice, especially for low Mellin moments 
and in the flavour non-singlet case.
Lattice QCD calculations for Mellin moments of $g_T^q(x)$
were reported in \cite{Gockeler:2000ja}.

Forth, the relation (\ref{Eq:new-interesting}) can be tested in 
models where collinear parton distribution functions were studied.
Some results can be found in literature.
For example, calculations of parton distribution functions 
in bag models \cite{Jaffe:1991ra,Signal:1997ct} support this 
relation (the bag model version of \cite{Jaffe:1991ra} coincides
with the one used here). 
Moreover, the spectator model \cite{Jakob:1997wg} supports 
this relation: it is equivalent to $g_2^q(x)=\frac12 h_2^q(x)$ in
the notation of \cite{Jakob:1997wg}, and also the unintegrated
version of (\ref{Eq:new-interesting}) is valid there.
One counter-example is known though: the chiral quark-soliton model
does not support this relation \cite{Schweitzer:2001sr,Wakamatsu:2000ex}. 
This observation could provide a hint in which models 
(\ref{Eq:new-interesting}) is valid.
The models where (\ref{Eq:new-interesting}) holds include only 
the components in the nucleon wave-function with the quark orbital angular
momenta up to $L=0,\,1,\, 2$ at most. The chiral quark soliton model,
which does not support (\ref{Eq:new-interesting}), contains {\sl all} 
quark angular momenta $L=0,\,1,\,2,\,3,\,4,\,\dots$ but 
this point deserves further investigation.

Fifth, an important aspect of model relations is 
that they inspire interpretations. The relation (\ref{Eq:new-interesting})
means that the difference between $g_T^q$ and $h_L^q$ is to the same
extent a 'measure of relativistic effects in the nucleon' as the 
difference between  helicity and transversity \cite{Jaffe:1991ra}.
Both these differences are related to the transverse moment of
pretzelosity, see Eqs.~(\ref{Eq:measure-of-relativity},~\ref{Eq:rel-VIII})
and \cite{Avakian:2008dz}.

\subsection{Inequalities}
\label{Sec-3f:ineq}

Finally, we discuss inequalities among leading twist TMDs, which are 
valid in QCD and all models \cite{Bacchetta:1999kz}
\ba
    f_1^q(x,k_\perp)  \ge 0 \;,\;\;\;
    |g_1^q(x,k_\perp)|\le f_1^q(x,k_\perp)  \;,\;\;\; 
    |h_1^q(x,k_\perp)|\le f_1^q(x,k_\perp)  \;,\;\;\;\label{Ineq:f1-g1-h1}&&
    \phantom{\biggl|}\\
    |h_1^q(x,k_\perp)|\le\frac12\biggl(f_1^q(x,k_\perp)+g_1^q(x,k_\perp)\biggr)\;,
    \;\;\; \label{Ineq:Soffer}&&\\
    |h_{1T}^{\perp q}(x,k_\perp)|
    \le\frac12\biggl(f_1^q(x,k_\perp)-g_1^q(x,k_\perp)\biggr)\;,
    \;\;\; \label{Ineq:pretzel}&&\\
    g_{1T}^{\perp(1)q}(x,k_\perp)^2+f_{1T}^{\perp(1)q}(x,k_\perp)^2 \le
    \frac{k_\perp^2}{4M_N^2}\,\biggl(f_1^q(x,k_\perp)^2-g_1^q(x,k_\perp)^2\biggr)
    \;,\;\; \label{Ineq:g1Tperp}&&\\
    h_{1L}^{\perp(1)q}(x,k_\perp)^2+h_1^{\perp(1)q}(x,k_\perp)^2 \le
    \frac{k_\perp^2}{4M_N^2}\,\biggl(f_1^q(x,k_\perp)^2-g_1^q(x,k_\perp)^2\biggr)
    \;,\;\; \label{Ineq:h1Lperp}&&
\ea
where we have to keep in mind that in the present quark model framework the 
inequalities simplify, due the absence of the T-odd TMDs $f_{1T}^{\perp q}$ 
and $h_1^{\perp q}$. In App.~\ref{App-B:prove-ineq} we demonstrate explicitly 
that the bag model expressions for the quark TMDs satisfy 
(\ref{Ineq:f1-g1-h1}--\ref{Ineq:h1Lperp}). It is interesting 
to remark that for the nucleon, except for $f_1^q(x,k_\perp)\ge 0$, 
all the other inequalities in (\ref{Ineq:f1-g1-h1}--\ref{Ineq:h1Lperp})
are 'true' (i.e.\ never saturated) inequalities,
see App.~\ref{App-B:prove-ineq}  for a detailed discussion.

\newpage
\section{Pretzelosity and quark orbital angular momentum}
\label{Sec-4:pretzelosity-and-OAM}

In quark models, in contrast to gauge theories, one may unambiguously
define the quark orbital angular momentum operator as
$\hat{L}_q^i = \bar\psi_q\varepsilon^{ikl}\hat{r}^k\hat{p}^l\psi_q$
where for clarity the 'hat' indicates a quantum operator.
This definition follows (in the absence of gauge fields) uniquely, 
for instance, from identifying that part of the generator of rotations 
not associated with the intrinsic quark spin.
For the following it will be convenient to introduce a 'non-local version'
of this operator, by defining (keep in mind that we work in non-gauge theory)
\be
        \hat{L}_q^i(0,z) = \bar\psi_q(0)\varepsilon^{ikl}\hat{r}^k\hat{p}^l\psi_q(z)\,.
\ee
In the bag model it is convenient to work in momentum space where 
$\hat{r}^k=i\,\frac{\partial\;}{\partial p^k}$ and $\hat{p}^l=p^l$.
Next let us define, in analogy to Eq.~(\ref{Eq:correlator}) the following quantity
\be\label{Eq:OAM}
    L_q^j(x,p_T) = \int\frac{\di z^-\di^2\vec{z}_T}{(2\pi)^3}\;e^{ipz}\;
    \la N(P,S^3)|\hat{L}_q^i(0,z)|N(P,S^3)\ra
    \biggl|_{z^+=0,\,p^+ = xP^+} \;.
    \ee
In order to find a connection to TMDs we must consider a longitudinally polarized
nucleon, choosing the polarization vector as $S=(0,0,1)$ for definiteness, and we 
must focus on the $j=3$ component in (\ref{Eq:OAM}), i.e.\ on the component
of the angular momentum operator along the light-cone space-direction.
This is because the transverse momenta $\vec{p}_T$ of the quarks generate 
orbital angular momentum which is oriented perpendicular to $\vec{p}_T$ 
(and to the transverse position of quarks inside the nucleon, which can be 
quantified rigorously in the impact parameter space in terms of generalized 
parton distribution functions, but we do not need this notion here).

In a quark model, where the ambiguities of gauge field theories are absent,
the partonic interpretation of (\ref{Eq:OAM}) is the following.
For example, in a longitudinally polarized nucleon
$L_q^3(x,p_T)\di^2\vec{p}_T\di x$ tells how much the orbital angular momentum
of a quark of flavour $q$, which carries the longitudinal momentum fraction $x$ 
and the transverse momentum $p_T=|\vec{p}_T|$, contributes to the nucleon spin.
(In QCD such an interpretation for the light-cone plus-component $L_q^+$
would also be possible, in an appropriately fixed gauge.)

Evaluating the expression (\ref{Eq:OAM}) in the bag model we
obtain
\be\label{Eq:OAM-vs-pretzelosity}
    L_q^3(x,p_T) = (-\,1)\,h_{1T}^{\perp(1)q}(x,p_T)\;.
\ee
In order to demonstrate the consistency of this result we compute the 
contribution to the total angular momentum of the nucleon $J^3_q$ due to 
flavour $q$. $J_q^3$ is composed of contributions from intrinsic quark spin,
$S_q^3=\frac12\int\di x g_1^q(x)$, and quark orbital angular momentum 
$L^3_q=\int\di x\int\di^2\vec{p}_T L_q^3(x,p_T)$. We obtain
\ba\label{Eq:spin-decomposition}
        2J_q^3 
    &=& \int\di x\int\di^2k_\perp\biggl[\,g_1^q(x,k_\perp) 
        -2\,h_{1T}^{\perp(1)q}(x,k_\perp)\,\biggr] \nonumber\\
    &=& P_q\,\frac{A}{M_N}\int\di^3k  \biggl[t_0^2+2\widehat{k}_z\,t_0t_1
                    +(2\widehat{k}_z^2-1+2\,\frac{k_\perp^2}{k^2})\,t_1^2\biggr]\nonumber\\
    &=& P_q\,\frac{A}{M_N}\int\di^3k  \bigl[t_0^2+t_1^2\bigr]\nonumber\\
    &=& P_q \phantom{\int}
\ea
where we first substituted $x\to k_z\equiv xM_N-\omega/R$ 
(recalling that $x$-integration is carried over entire $x$-axis, 
Sec.~\ref{Sec-2:TMDs-in-bag}), then used that under the integral 
over $\di^3k$ for symmetry reasons $\widehat{k}_z=k_z/k$ drops out 
while $k_z^2\to\frac13\,k^2$ and $k_\perp^2\to\frac23\,k^2$, and 
finally observed the same integral which appears in the normalization 
of the unpolarized distribution.

Eq.~(\ref{Eq:spin-decomposition}) is the correct SU(6) quark model result
for the contributions of various flavours to the total nucleon spin. 
Notice that $J_u^3+J_d^3=\frac12$.
This confirms that the connection of pretzelosity and the quark orbital
angular momentum content of the nucleon is consistent. 
Thus, our results, supported by the light-cone SU(6) quark-diquark 
model \cite{She:2009jq}, suggest that
\be\label{Eq:OAM-vs-pretzelosity-II}
    L_q^3 = (-\,1)\,\int\di x\;h_{1T}^{\perp(1)q}(x)\;.        
\ee
It is important to observe that the relation of  pretzelosity and 
orbital angular momentum, Eqs.~(\ref{Eq:OAM-vs-pretzelosity}) and 
(\ref{Eq:OAM-vs-pretzelosity-II}), is at the level of matrix elements 
of operators, and there is no a priori operator identity which would 
make such a connection.

\newpage

\section{Numerical results}
\label{Sec-5:results}

In this Section we discuss the numerical results for the TMDs.
In Sec.~\ref{Sec-5a:integrated-TMDs} we present the results for
the integrated TMDs as functions of $x$. 
In Sec.~\ref{Sec-5b:pT-dependence} we focus on their $k_\perp$-behaviour.
Finally, in Sec.~\ref{Sec-5c:WW-type} we investigate the question whether the 
bag model results support the so-called Wandzura-Wilczek(-type) approximations.

\subsection{Results for the integrated TMDs}
\label{Sec-5a:integrated-TMDs}

As the flavour dependence governed by the spin-flavour SU(6) symmetry is
trivial, we will show only results
for the $u$-flavor. For unpolarized TMDs the $d$-quark distributions 
are factor 2 smaller than the $u$-quark distributions. In the case of the
polarized TMDs, the $d$-quark distributions are factor 4 smaller 
and have opposite sign compared to the $u$-quark distributions, according 
to Eq.~(\ref{Eq:wafe-function-SU(6)-pol}).
All results discussed below refer to the low scale of the bag model.

\begin{figure}[b!]

\vspace{-5mm}

             \includegraphics[width=6cm]{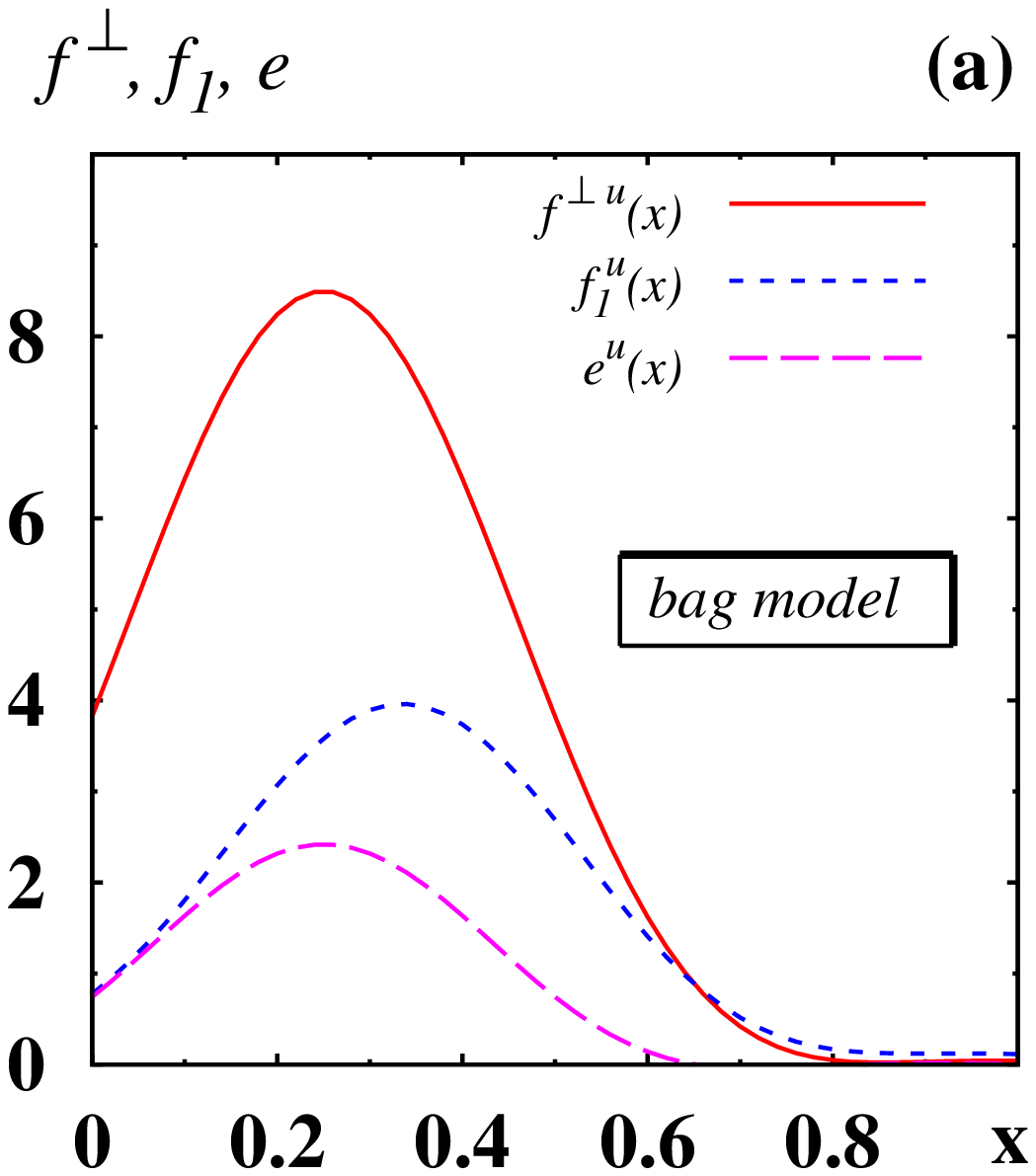}
\hspace{-5mm}\includegraphics[width=6cm]{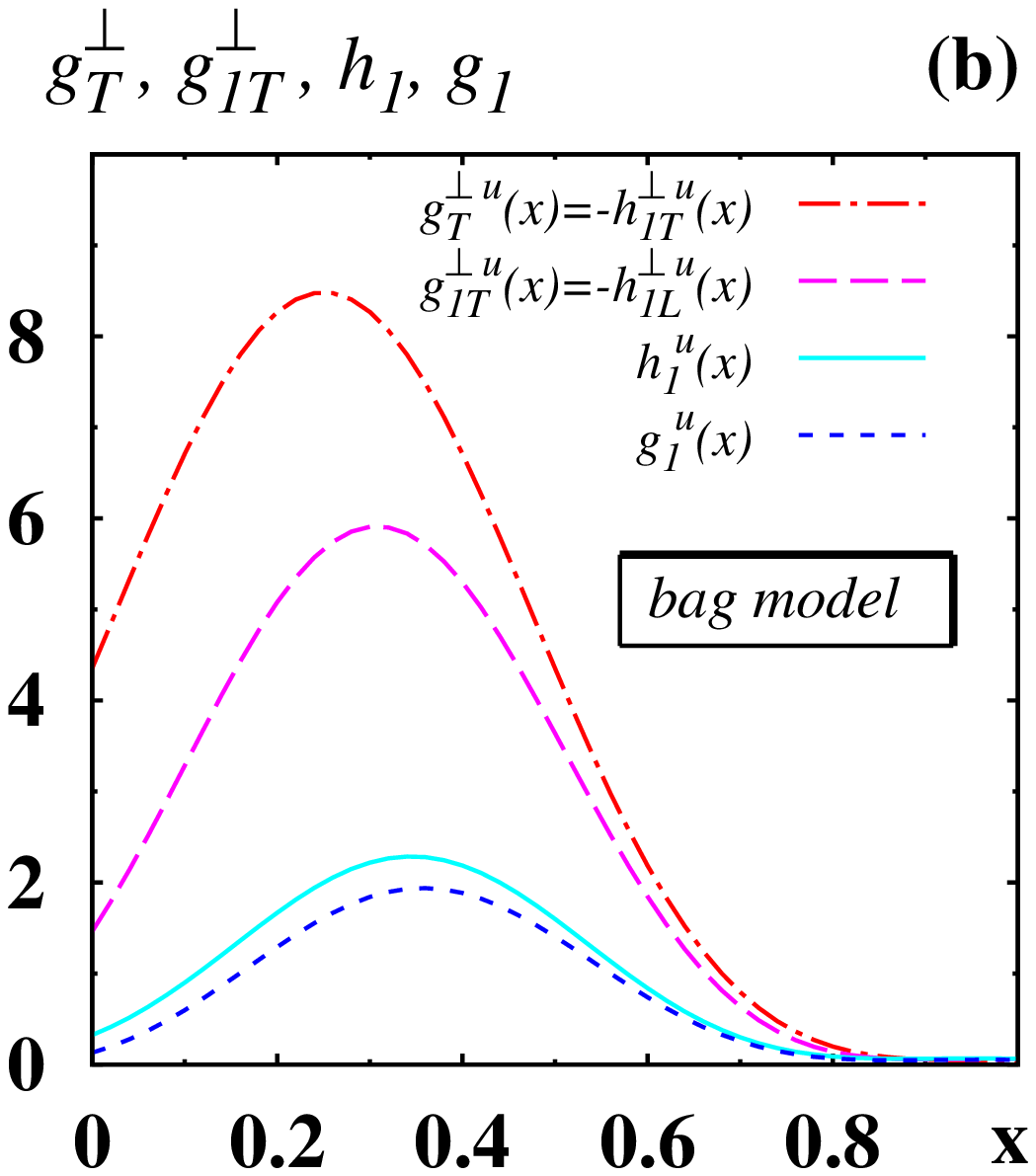}
\hspace{-5mm}\includegraphics[width=6cm]{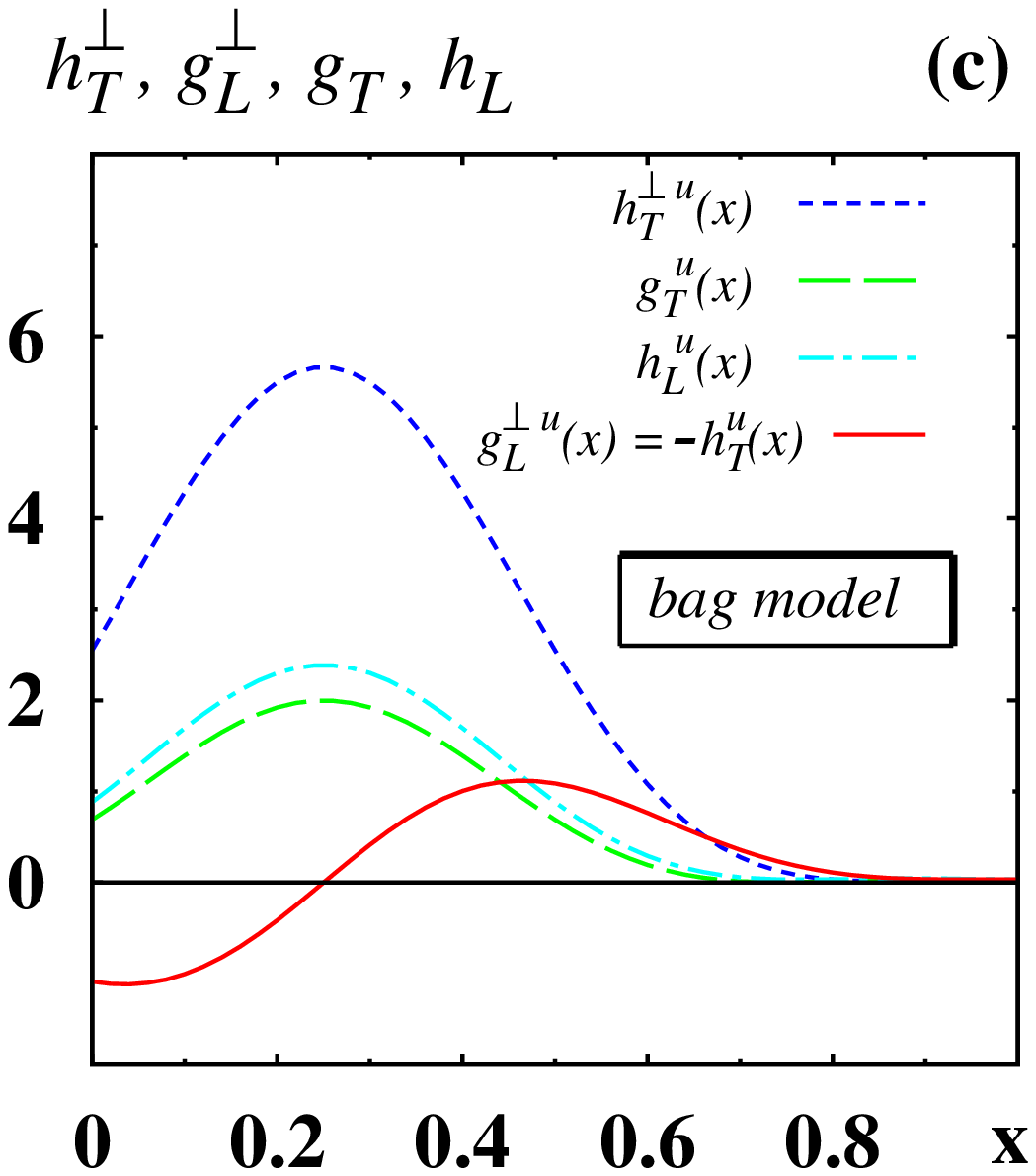}
\caption{\label{Fig01:fperp-f1-e}
    (a)
    The unpolarized functions $f^{\perp u}(x)$, $f_1^u(x)$, $e^u(x)$ vs.\ $x$
    from the bag model at the low scale. The $d$-quark distributions
    are factor two smaller compared to the unpolarized $u$-quark distributions
    according to the SU(6)-flavour factors in
    Eqs.~(\ref{Eq:wafe-function-SU(6)-pol},~\ref{Eq:fperp}).
    (b)
    The polarized functions $g_T^{\perp u}(x)=-h_{1T}^{\perp u}(x)$, 
    $g_{1T}^{\perp u}(x)=-h_{1L}^{\perp u}(x)$, $h_1^u(x)$, $g_1^u(x)$ 
    vs.\ $x$.
    The $d$-quark distributions are factor four smaller and have opposite sign 
    compared to the $u$-quark distributions according to the SU(6)-flavour 
    factors in Eqs.~(\ref{Eq:wafe-function-SU(6)-pol},~\ref{Eq:fperp}).
    (c)
    The polarized functions $h_T^{\perp u}(x)$, 
    $g_L^{\perp u}(x)=-h_T^u(x)$, $g_T^u(x)$, $h_L^u(x)$ vs.\ $x$.
    The $d$-quark functions are as in (b).}
\end{figure}

Let us start the discussion of the numerical results with unpolarized TMDs.
Fig.~\ref{Fig01:fperp-f1-e}a shows the twist-2 unpolarized distribution
function $f_1^u(x)$, and the subleading twist functions $f^{\perp u}(x)$,
$e^u(x)$. Only $f^{\perp q}(x)=\int\di^2k_\perp f^{\perp q}(x,k_\perp)$ 
is new in this figure. 
The remarkable observation is that $f^{\perp q}(x)$ is rather large, even 
larger than $f_1^q(x)$ for $x\lesssim 0.7$. However, one has to keep
in mind that there are no positivity bounds for twist-3 
TMDs. Moreover, it is $\frac{k_\perp}{M}f^{\perp q}(x,k_\perp)$ which 
enters in cross sections, and typically $\la k_\perp\ra\ll M_N$, 
which eventually guarantees positivity of cross sections.
We remark that QCD equations of motion \cite{Efremov:2002qh} 
imply a $\delta(x)$-contribution in $e^q(x)$, which is found
in some \cite{Schweitzer:2003uy,Burkardt:2001iy} but not all
effective approaches \cite{Jakob:1997wg,Mukherjee:2009uy},
including the bag model, see \cite{Jaffe:1991ra,Signal:1997ct} 
and Fig.~\ref{Fig01:fperp-f1-e}a.

%
%

Fig.~\ref{Fig01:fperp-f1-e}b shows the polarized functions 
$g_T^{\perp u}(x)$, $g_{1T}^{\perp u}(x)$, $h_1^u(x)$, $g_1^u(x)$.
The TMDs $h_{1T}^{\perp a}(x)$ and $h_{1L}^{\perp a}(x)$
are simply related to the shown TMDs according to 
$h_{1T}^{\perp a}(x)= -g_T^{\perp a}(x)$ and 
$h_{1L}^{\perp a}(x)=-g_{1T}^{\perp a}(x)$, such that the
results for these TMDs do not need to be shown.
Also the results for $d$-quark distributions are not shown,
as explained above.
The results for the TMDs
$g_T^{\perp a}(x)$,
$g_{1T}^{\perp a}(x)$, $h_{1L}^{\perp a}(x)$,
$h_{1T}^{\perp a}(x)$ are new, and it is interesting to observe 
that they are rather sizable, but again there are no positivity 
constraints on these objects.  

Fig.~\ref{Fig01:fperp-f1-e}c shows the polarized functions 
$g_L^{\perp u}(x)$, $g_T^{\perp u}(x)$, $h_L^u(x)$, $g_T^u(x)$.
The TMD $h_T^u(x)$ is not shown, being related to 
$g_L^{\perp u}(x)$ as $h_T^u(x)=-g_L^{\perp u}(x)$.
We see that $h_T^{\perp u}$ is rather sizable, it is even
bigger than $f_1^u(x)$ (the same scale is used in 
Figs.~\ref{Fig01:fperp-f1-e}a--c). Again there is no 
positivity constraint for this TMD, which would object this.

The large size of the integrated twist-2 TMDs 
$g_{1T}^{\perp a}(x)$, $h_{1L}^{\perp a}(x)$, $h_{1T}^{\perp a}(x)$ 
can be understood qualitatively in the non-relativistic limit which was 
formulated for an arbitrary number of colours $N_c$ in \cite{Efremov:2009ze}, 
and can eventually be traced back the convention of using the nucleon mass 
$M_N$ in order to compensate the dimension of the $k_\perp$ factor(s) in the 
decomposition of the correlators in Eq.~(\ref{Eq:TMD-pdfs-I}--\ref{Eq:TMD-pdfs-VI}).

It is interesting to notice that $g_L^{\perp q}(x)=-h_T^q(x)$ are the 
only TMDs in the bag model which have a zero in the valence-$x$ region.
This observation is actually not surprising but a consequence of the fact
that the LIRs (\ref{eq:LIR4},~\ref{eq:LIR3}) hold, and 
$g_T^{\perp(1)q}(x)=-h_{1T}^{\perp(1)q}(x)$ have extrema in the valence-$x$ 
region.

\subsection{Transverse momentum dependence}
\label{Sec-5b:pT-dependence}

In the context of TMDs the most interesting aspect is, of course, 
their transverse momentum dependence. In principle, all information
is contained in the two-dimensional functions $j(x,k_\perp)$ for a
generic TMD, but here we shall content ourselves to discuss
'one- or zero-dimensional' projections of that information.

The first point we address is: what are the typical transverse momenta 
of unpolarized quarks in the bag TMDs? For that we define for a generic TMD 
$j^q(x,k_\perp)$ the following quantities
\be\label{Eq:def-avpT-avpT2}
	\la p_T\ra =  
	\frac{\int\di x\int\di^2k_\perp\;k_\perp\,j^q(x,k_\perp)}
	     {\int\di x\int\di^2k_\perp\;j(x,k_\perp)}\;,\;\;\;
	\la p_T^2\ra =  
	\frac{\int\di x\int\di^2k_\perp\;k_\perp^2\,j^q(x,k_\perp)}
	     {\int\di x\int\di^2k_\perp\;j(x,k_\perp)}\;.
\ee
Due to the simple spin flavor structure of the MIT bag model
the $\la p_T\ra$ and $\la p_T^2\ra$  are flavor-independent for all TMDs.

The first observation is that depending on the TMD $\la p_T\ra$ and 
$\la p_T^2\ra$  in Eq.~(\ref{Eq:def-avpT-avpT2}) may not exist in the 
bag model, because the momentum-space wave-function components $t_i(k)$, 
Eq.~(\ref{Eq:t0-t1}), do not vanish sufficiently fast at large $k$.
This is the case especially for $f_1^q(x,k_\perp)$.

\begin{figure}[b!]
             \includegraphics[width=6cm]{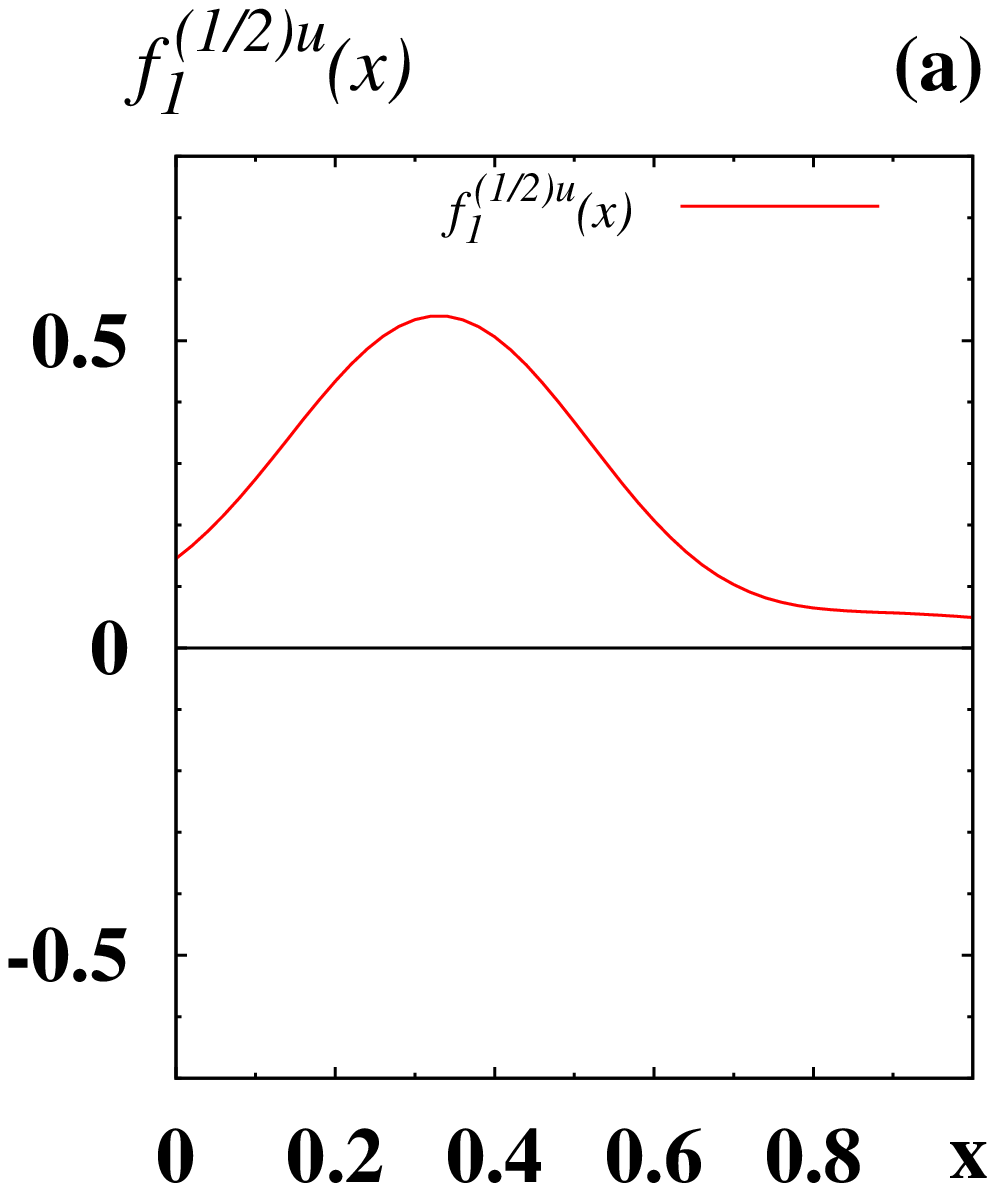}
\hspace{-5mm}\includegraphics[width=6cm]{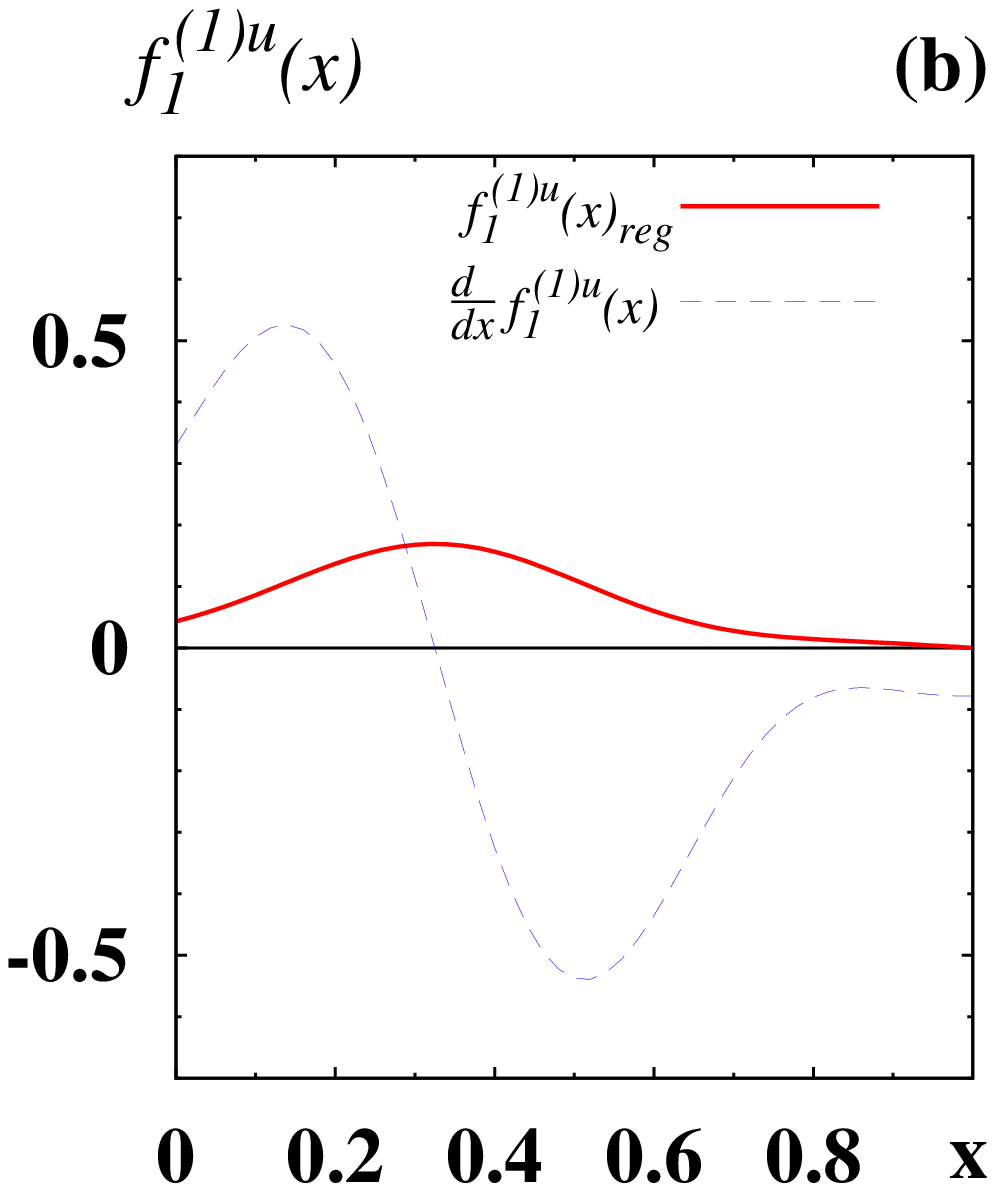}
\hspace{-5mm}\includegraphics[width=6cm]{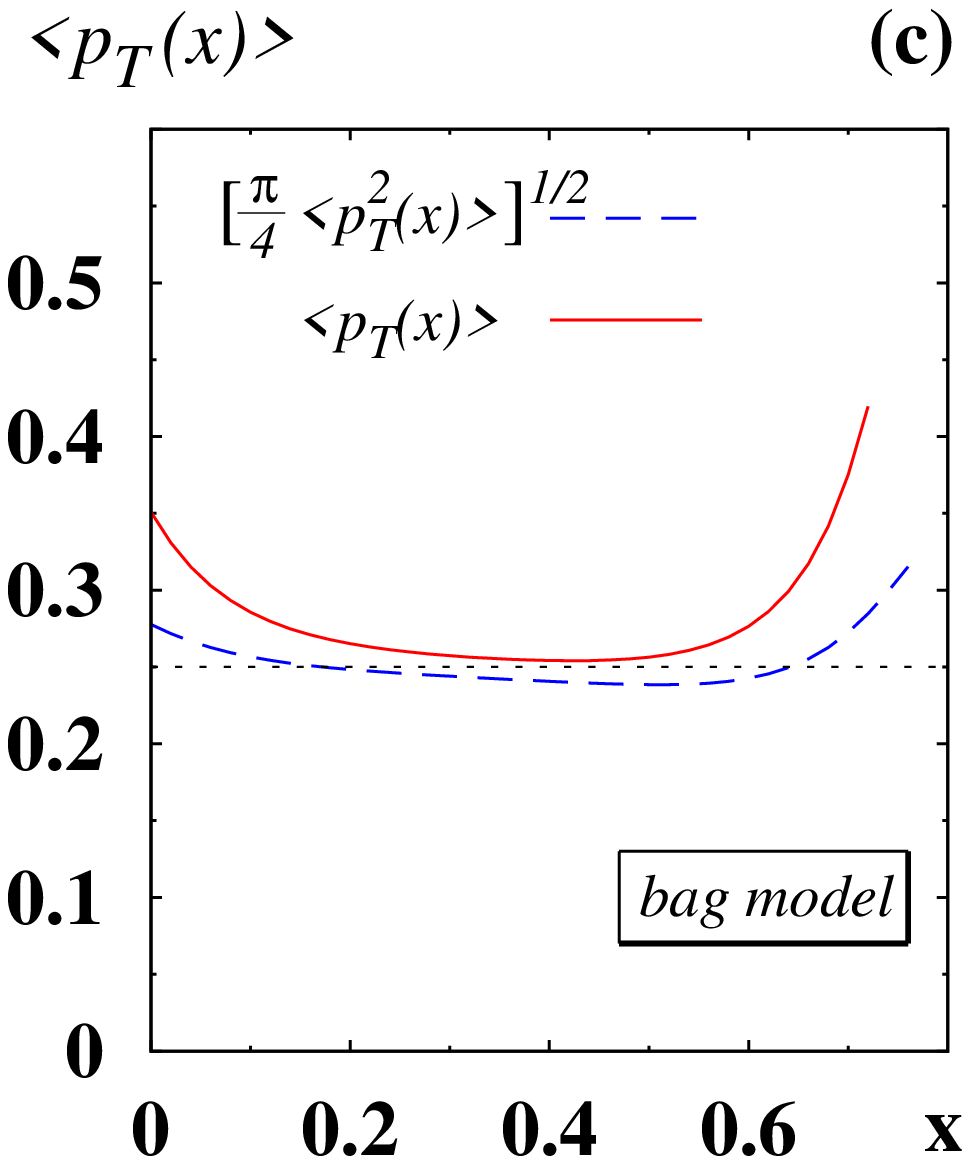}
\caption{\label{Fig02:pT-f1}
    For the unpolarized TMD $f_1^q(x,k_\perp)$ 
    (a) the (1/2)-moment defined in Eq.~(\ref{Eq:def-avpT(x)}),
    (b) the derivative of the (1)-moment and the {\sl regularized}
    (1)-moment as discussed in the text, and
    (c) $\la p_T(x)\ra$ in comparison to $(\pi\la p_T^2(x)\ra/4)^{1/2}$.
    In the Gauss-model the two quantities would be equal. (The dotted
    marks the value $\la p_T(x)\ra=0.25\,{\rm GeV}$.)} 
\end{figure}

For the same reason also the (1)-moment  $f_1^{(1)q}(x)$ does not exist.
However, the (1/2)-moment  $f_1^{(1/2)q}(x)$ defined according to
(\ref{Eq:def-f(1/2)}) exists, and can be used to introduce
an $x$-dependent average transverse momentum $\la p_T(x)\ra$ as
\be\label{Eq:def-avpT(x)}
	\la p_T(x)\ra = 2M_N\;\frac{f_1^{(1/2)q}(x)}{f_1^q(x)}
        \;.
\ee
Fig.~\ref{Fig02:pT-f1}a shows the result for $f_1^{(1/2)q}(x)$.
(The divergence of $\la p_T\ra$ from (\ref{Eq:def-avpT-avpT2}) 
emerges when one tries to integrate $f_1^{(1/2)q}(x)$ over $x$, 
recalling that this integration extends to the entire $x$-axis, 
see Sec.~\ref{Sec-2:TMDs-in-bag}.)

Now the (1)-moment  $f_1^{(1)q}(x)$ is divergent, but its derivative
with respect to $x$ exists, see the dotted line in Fig.~\ref{Fig02:pT-f1}b. 
Hereby it is understood that the (1)-moment is computed with a finite
cutoff $\Lambda_{\rm cut}\gg M_N$, then the derivative is taken, and
only then the limit $\Lambda_{\rm cut}\to\infty$ is performed.

%
%

By integrating the well-defined $\frac{\di}{\di x}f_1^{(1)q}(x)$ we 
can compute a {\sl regularized} (1)-moment $f_1^{(1)q}(x)_{reg}$. 
The result depends on some arbitrary integration constant, which we fix 
such that the (1)-moment vanishes at $x=1$. This choice is reasonable 
but not unique,
if we recall that in the MIT bag model TMDs in general have a non-zero 
(though small) support for $|x|\ge1$, see Sec.~\ref{Sec-2:TMDs-in-bag}.
Our main conclusions in this respect, to be presented below in this Section, 
depend weakly on the chosen value of the integration constant, provided 
reasonable choices are made 
(such as, for example, $f_1^{(1)q}(x)_{reg}=f_1^{(1/2)q}(x)$ at $x=1$).
The result for $f_1^{(1)q}(1)_{reg}$ defined in this way is shown
as solid line in Fig.~\ref{Fig02:pT-f1}b. 

With $f_1^{(1)q}(x)_{reg}$ we are in the position to define an 
$x$-dependent average transverse momentum square $\la p_T^2(x)\ra$ as 
\be\label{Eq:def-avpT2(x)}
	\la p_T^2(x)\ra = 2M_N^2\;\frac{f_1^{(1)q}(x)_{reg}}{f_1^q(x)}\;.
\ee

Fig.~\ref{Fig02:pT-f1}c shows  $\la p_T(x)\ra$ as solid line.
We observe that in the valence-$x$ region at the low hadronic scale 
$\la p_T(x)\ra$ very weakly depends on $x$. Numerically we find 
\be\label{Eq:pT-unpol-bag}
     \la p_T(x)\ra \approx 0.25\,{\rm GeV} \;\;\;
     \mbox{for $0.2\lesssim x\lesssim 0.5$}.
\ee
(The $\la p_T(x)\ra=0.25\,{\rm GeV}$ is marked as dotted line
in Fig.~\ref{Fig02:pT-f1}c.)
This is similar to results from the light-cone constituent model 
\cite{Pasquini:2008ax} which also refer to a very low hadronic scale.
In fact, keeping in mind the $p_T$-broadening effects due to gluon radiation 
with increasing normalization scale \cite{Collins:1984kg}, this is a 
reasonable result at a low scale.
(We remark that in parton model approaches one finds comparably low values
for $\la p_T(x)\ra$ (albeit there the results refer to high scales) 
\cite{Jackson:1989ph,Zavada:1996kp} models.)

In phenomenology at high scales, however, larger values are required
\cite{Collins:2005ie,Anselmino:2005nn,D'Alesio:2004up}. For example,
the interpretations of SIDIS data from EMC \cite{Arneodo:1986cf}
or HERMES \cite{Airapetian:1999tv} require 
\be\label{Eq:pT-unpol-phenomenology}
	\la p_T(x)\ra_{\rm Gauss}^{\phantom X} = \cases{
	0.64\,{\rm GeV} & from EMC data in \cite{Anselmino:2005nn},\cr
	0.56\,{\rm GeV} & from HERMES data in \cite{Collins:2005ie},}
\ee
where the index ``Gauss'' indicates that the Gaussian model has been assumed
in these studies.  The Gaussian model means that
$f_1^q(x,p_T)=f_1^q(x)\,
\exp(-p_T^2/\la p_T^2(x)\ra_{\rm Gauss})/(\pi \la p_T^2(x)\ra_{\rm Gauss})$.
The width $\la p_T^2(x)\ra_{\rm Gauss}$ could be a function of $x$, but in
practice it is often assumed to be a constant. Such an Ansatz works with 
sufficient precision for many practical applications in phenomenological 
studies \cite{Collins:2005ie,Anselmino:2005nn,D'Alesio:2004up}.
In the Gaussian model the relation holds
\be\label{Eq:Gauss}
    \la p_T(x)\ra_{\rm Gauss} =
    \left[\frac{\pi}{4}\la p_T^2(x)\ra_{\rm Gauss}\right]^{1/2}\;.
\ee
Of course, in no model considered so far such a factorized 
$x$- and transverse parton momentum dependence was ever observed,
and in the bag model we do not observe it either.
However, it is interesting to ask, for example, to which
extent the relation (\ref{Eq:Gauss}) is supported in 
a model. With $\la p_T^2(x)\ra$ defined in (\ref{Eq:def-avpT2(x)})
we obtain for the expression on the RHS of (\ref{Eq:Gauss}) the
result plotted as dashed line in Fig.~\ref{Fig02:pT-f1}c.
The remarkable observation is that (\ref{Eq:Gauss}) is 
supported within an accuracy of ${\cal O}(10\,\%)$
in the valence-$x$ region. We remark that this conclusion is 
insensitive to the way the integration constant in 
$f_1^{(1)q}(x)_{reg}$ is fixed, provided this is done in a reasonable way
(see above).

%
\begin{figure}[b!]
             \includegraphics[width=6cm]{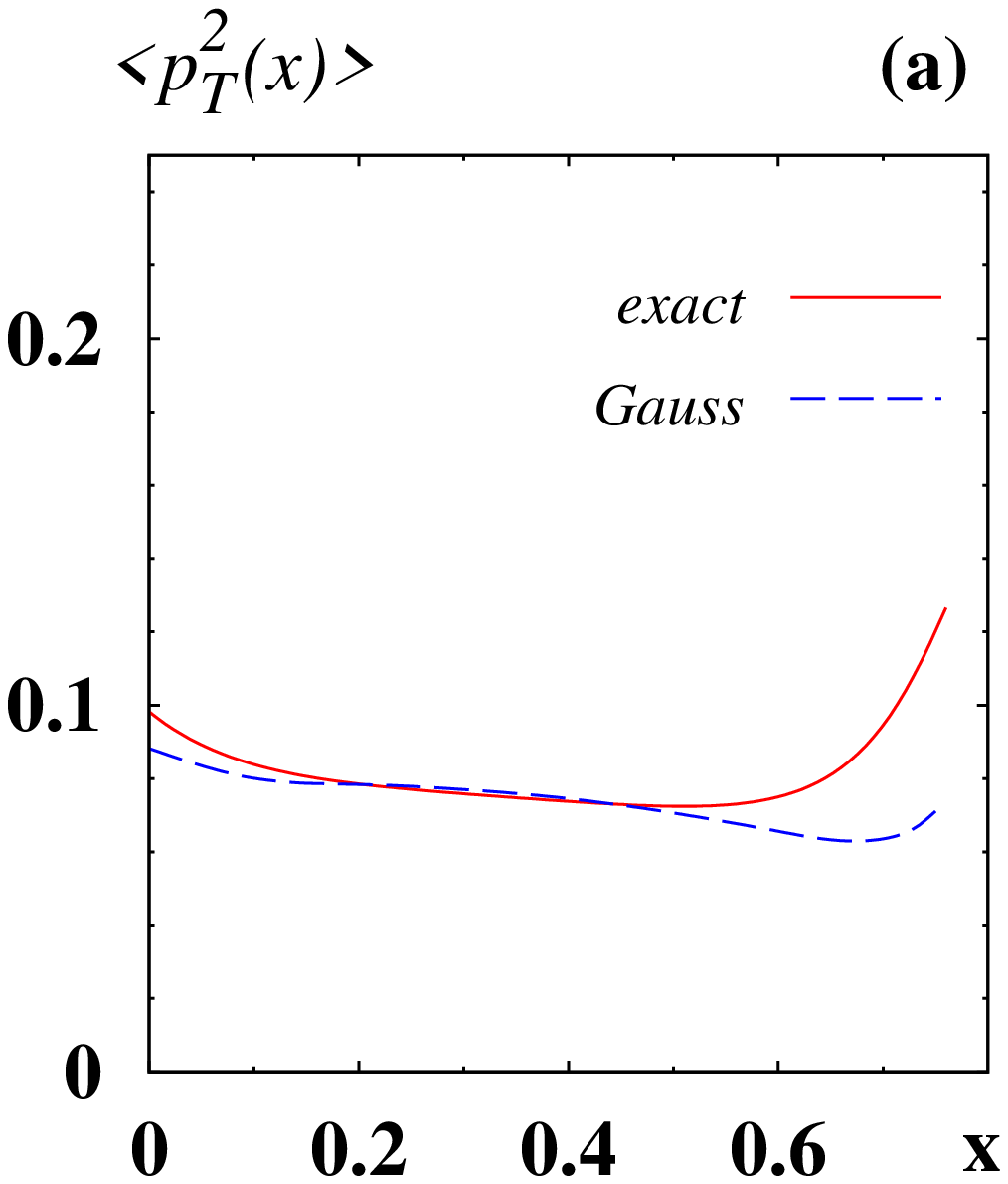}
\hspace{-5mm}\includegraphics[width=6cm]{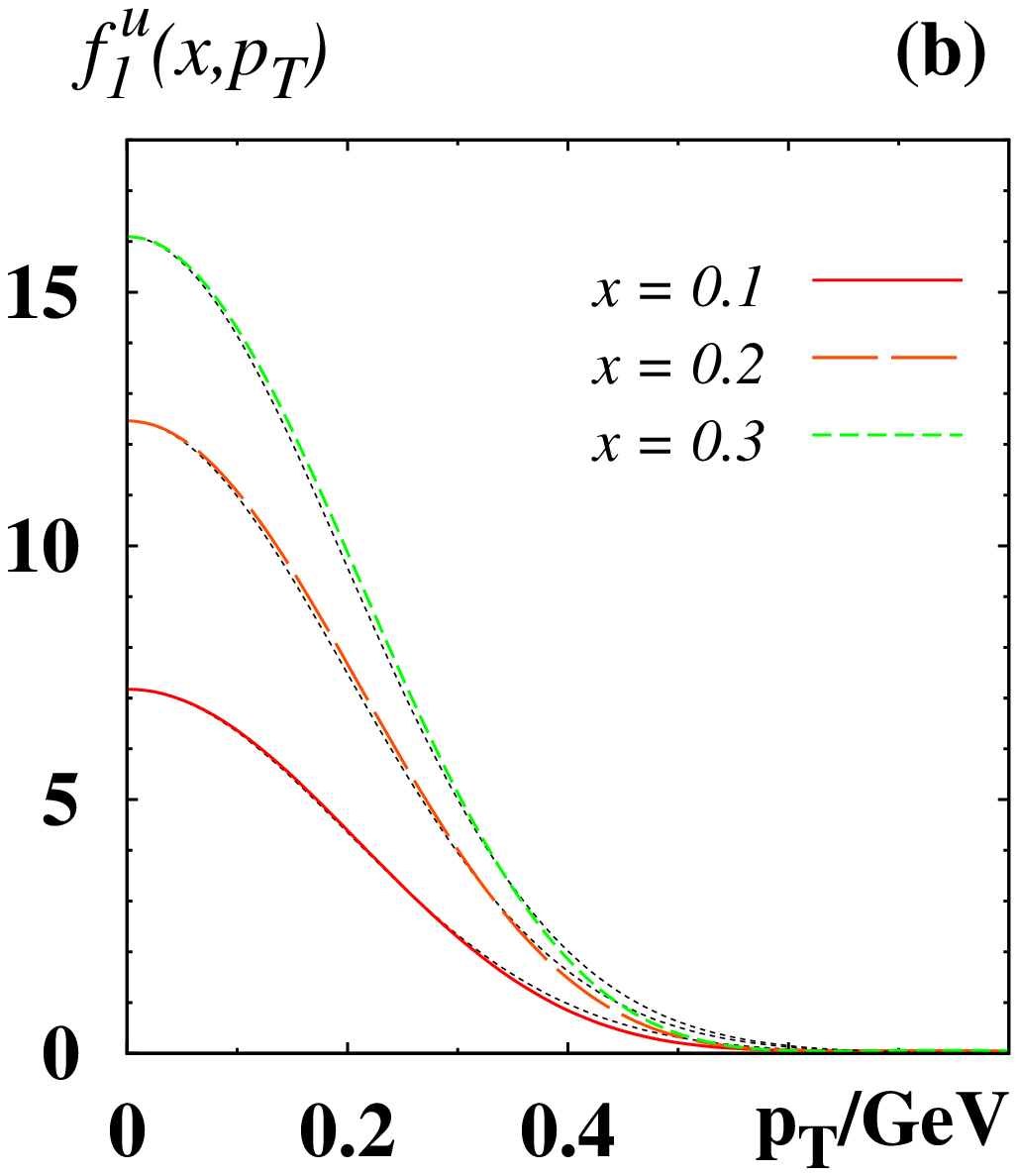}
\hspace{-5mm}\includegraphics[width=6cm]{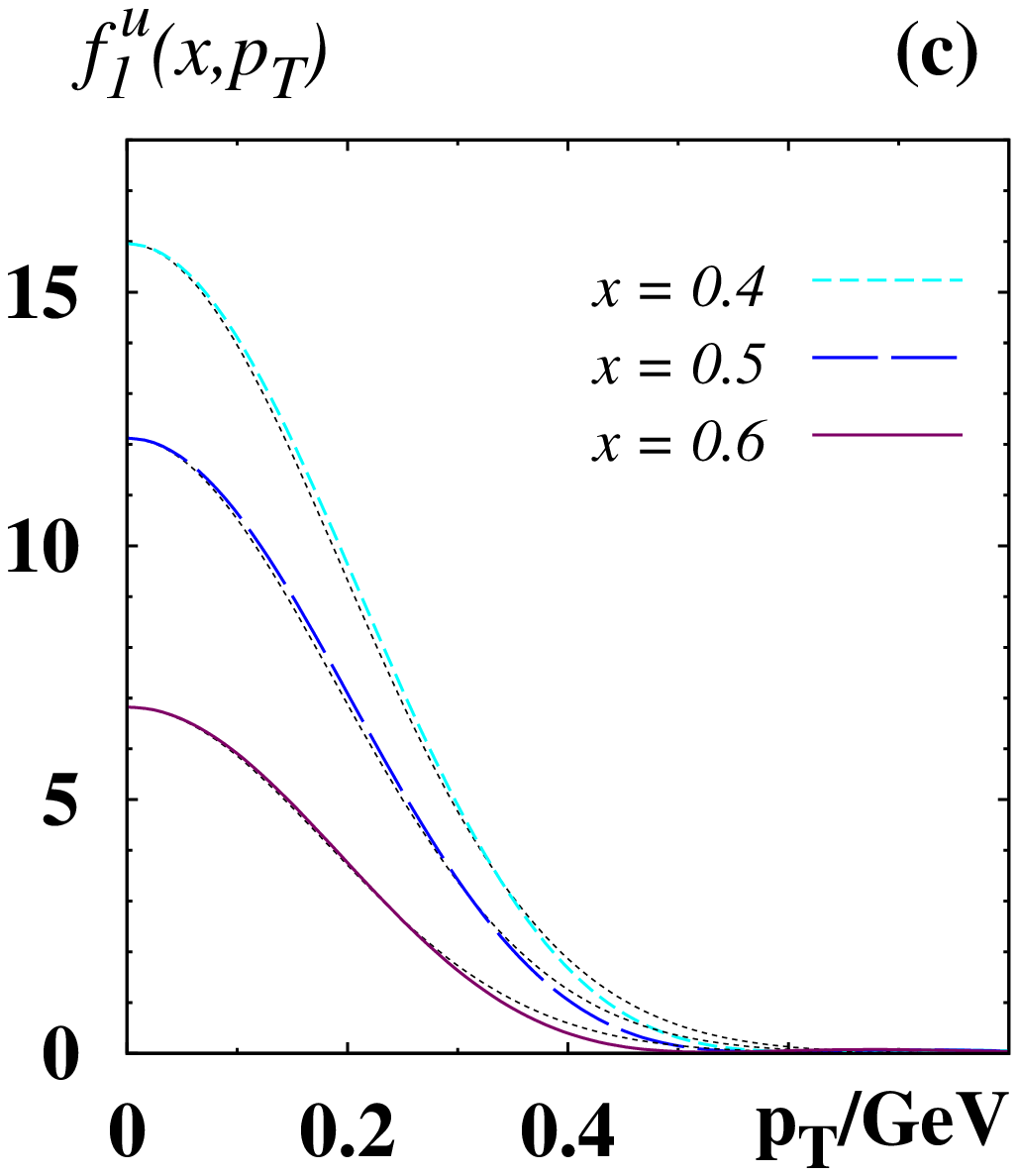}
\caption{\label{Fig03:f1-vs-Gauss}
    (a) 
    $\la p_T^2\ra$ of $f_1^q$ as function of $x$.
    Solid line: computed using the exact definition in Eq.~(\ref{Eq:def-avpT2(x)}). 
    Dashed line: using the Gauss model relation, Eq.~(\ref{Eq:Gauss-width}).  
    (b) and (c)
    $f_1^q(x,p_T)$ vs.\ $p_T$ for selected values of $x$.
    The thin dotted lines are the respective Gauss model approximations
    obtained from the Gaussian widths from Fig.~\ref{Fig03:f1-vs-Gauss}a.}
\end{figure}
%

However, the bag model supports the Gaussian model much more
than that, in the following sense. In the Gauss model we have
$f_1^q(x,p_T)=f_1^q(x,0)\,\exp(-p_T^2/\la p_T^2(x)\ra_{\rm Gauss})$ where,
by definition,  
$f_1^q(x,0)=f_1^q(x)/(\pi \la p_T^2(x)\ra_{\rm Gauss})$.
When dealing with a model with non-Gauss-like transverse momentum
dependence, this can be used to 'fit' the Gaussian width 
\be\label{Eq:Gauss-width}
     \la p_T^2(x)\ra_{\rm Gauss} = \pi \;\frac{f_1^q(x,0)}{f_1^q(x)}
\ee
such that the Gaussian model is exact at $p_T=0$. By continuity
arguments the Gaussian model can be expected to be a good
approximation to the exact model results also for $p_T>0$ in 
{\sl some} vicinity close to $p_T=0$.
The question is: how large is the $p_T$-region in which the 
Gaussian model with the width (\ref{Eq:Gauss-width}) will provide 
a useful approximation to the $p_T$-dependence of the unpolarized
TMD?

Let us first compare the results for the mean transverse momentum
square $\la p_T^2(x)\ra$ defined in Eq.~(\ref{Eq:def-avpT2(x)})
and the expression from the Gauss model $\la p_T^2(x)\ra_{\rm Gauss}$, 
Eq.~(\ref{Eq:Gauss-width}). It has to be noticed that these are
a priori completely different quantities. It is therefore remarkable
that the results agree so well, especially for valence $x$,
see Fig.~\ref{Fig03:f1-vs-Gauss}a.

In order to see to which extent the bag model is compatible with a 
Gauss-like shape of the transverse momentum distributions, we 
plot the $p_T$-dependence of $f_1^u(x,p_T)$ for $0\le p_T < M_N$
for selected values of
$x=0.1$, $0.2$, $0.3$ in Fig.~\ref{Fig03:f1-vs-Gauss}b, and
$x=0.4$, $0.5$, $0.6$ in Fig.~\ref{Fig03:f1-vs-Gauss}c.
In Figs.~\ref{Fig03:f1-vs-Gauss}b and c, we also plot the 
respective Gaussian approximations (as thin-dotted lines).
The result is astonishing: In the valence-$x$ region the 
exact curves and their Gaussian approximations agree 
excellently!

Not visible in Figs.~\ref{Fig03:f1-vs-Gauss}b and c is that
the first worthwhile mentioning deviations from the Gaussian 
behaviour start to become apparent only at larger $p_T>0.5\,{\rm GeV}$. 
The crucial difference is obviously in the large-$p_T$ asymptotics: 
$f_1^q(x,p_T)\sim \alpha(x,p_T)/p_T^4$ with $|\alpha(x,p_T)| < {\rm const}$. 
(The function $\alpha(x,p_T)$ oscillates, with a period defined by the 
periods of the spherical Bessel functions in  (\ref{Eq:wave-func-0}), 
around some value which depends on the TMD but not on $x$.
For all TMDs the respective functions $\alpha(x,p_T)$ are bound from
above and below.)

Of course, non-perturbative models aiming at an effective description 
of the nucleon properties at hadronic scale are not able to address 
the large-$p_T$ region, where one may apply perturbative QCD.
Effective models can, however, provide valuable insights for transverse
momenta up to the order of magnitude of the hadronic scale, i.e.\ 
for $p_T<M_N$. In this $p_T$-region the bag model supports the concept 
of a Gaussian distribution of the transverse parton momenta in the 
case of $f_1^q(x,p_T)$.

Let us now turn our attention to other TMDs, keeping the discussion 
shorter after the detailed investigation of~$f_1^q$.
The TMDs $g_1^q$, $h_1^q$, $e^q$, $g_T$, $h_L^q$, which exist also as 
collinear parton distribution functions, have the same large-$p_T$
behavior like $f_1^q$, and consequently also have divergent 
(1)-moments (which can be regularized as in the case of $f_1^q$).
In contrast to this $g_{1T}^{\perp q}$, $h_{1L}^{\perp q}$, 
 $f^{\perp q}$, $h_T^{\perp q}$ behave like $\alpha(x,p_T)/p_T^5$ at
large $p_T$ and have well-defined convergent (1)-moments.
Finally $h_{1T}^{\perp q}$, $g_L^{\perp q}$, $g_T^{\perp q}$, $h_T^q$ behave 
like $\alpha(x,p_T)/p_T^6$ and at large $p_T$have well-defined convergent 
(1)-moments, too.

%
\begin{figure}[b!]
             \includegraphics[width=6cm]{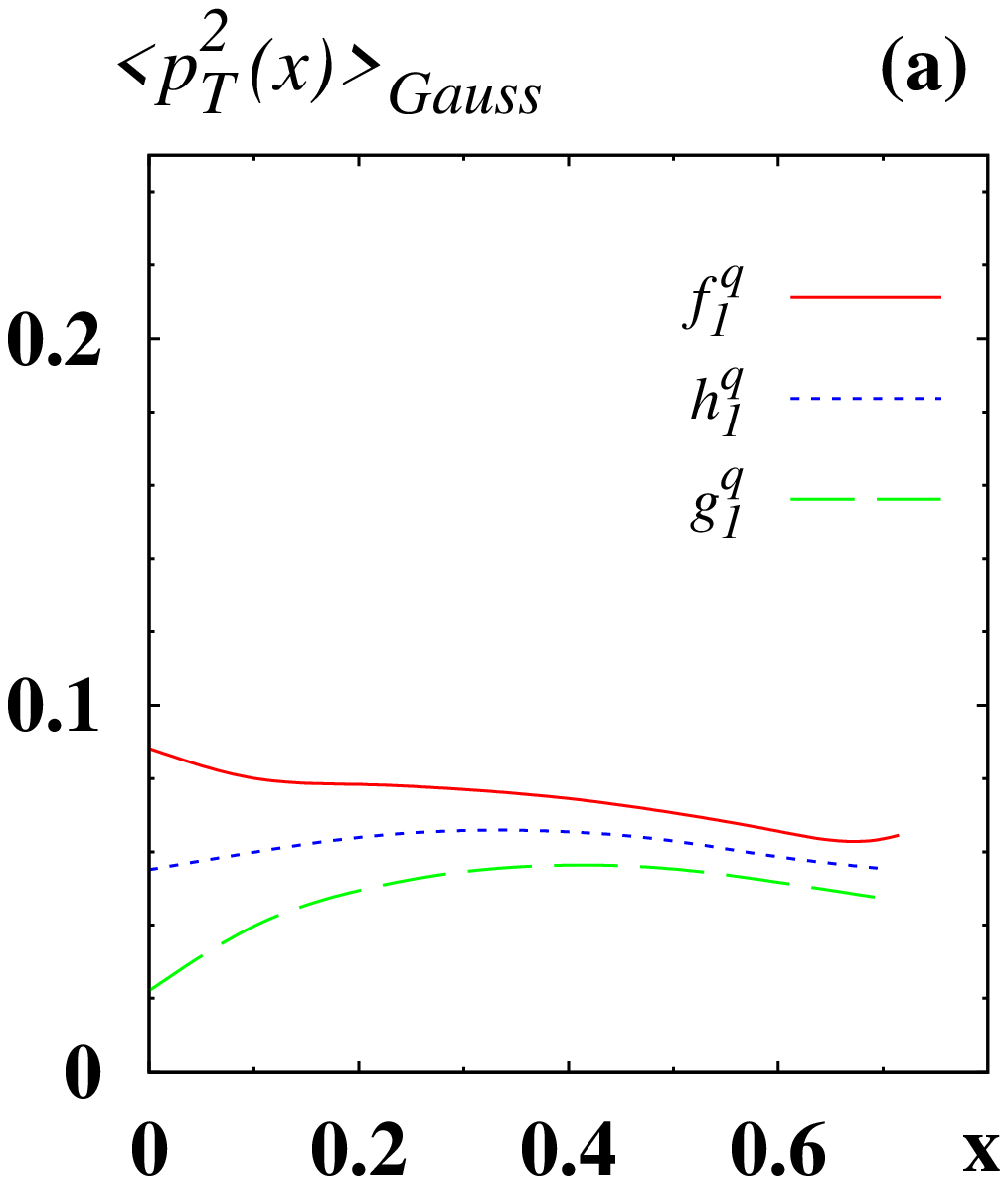}
\hspace{-5mm}\includegraphics[width=6cm]{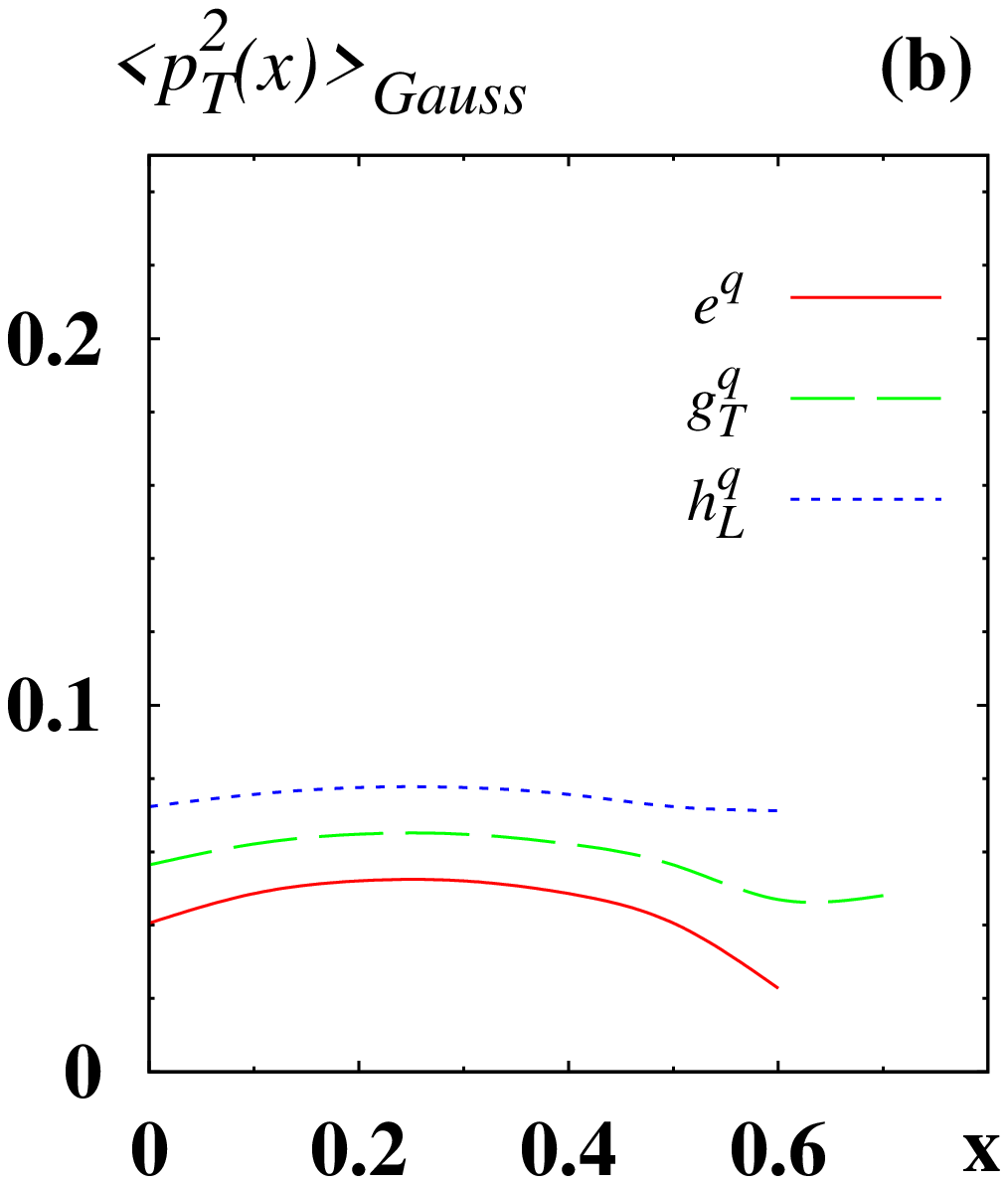}
\hspace{-5mm}\includegraphics[width=6cm]{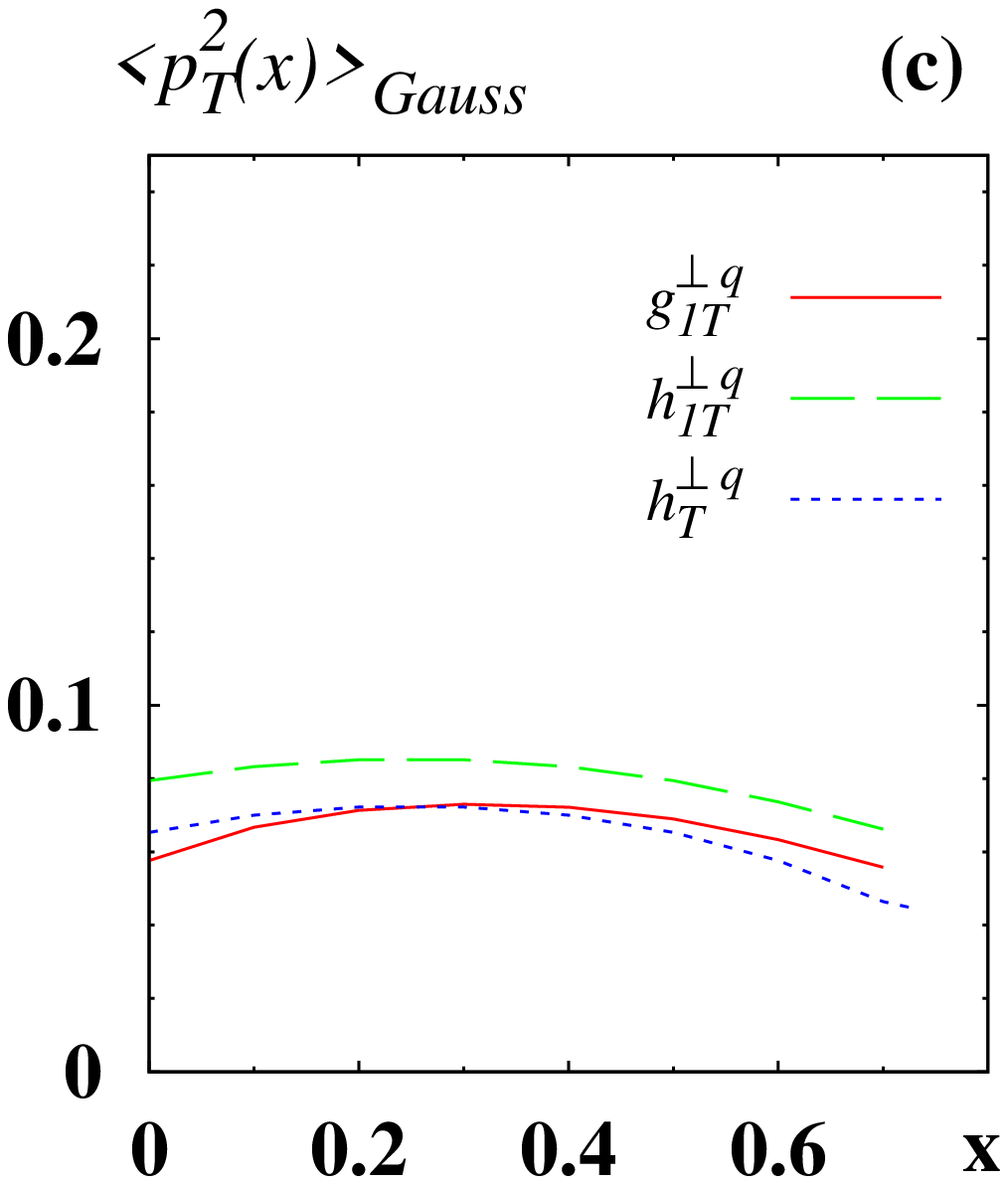}
\caption{\label{Fig04:avpT2-vs-Gauss}
    The Gaussian widths as defined in Eq.~(\ref{Eq:Gauss-width}) 
    vs.\ $p_T$ for various TMDs.}
\end{figure}
%

Is it possible to approximate the $p_T$-dependence also of the other TMDs
in a Gaussian model? The answer is yes.
Fig.~\ref{Fig04:avpT2-vs-Gauss} 
shows the Gaussian widths $\la p_T^2(x)\ra_{\rm Gauss}$ as defined in 
Eq.~(\ref{Eq:Gauss-width}) for the all TMDs. (Widths of 
$h_{1L}^{\perp q}$, $f^{\perp q}$, $g_T^{\perp q}$ are not shown,
because they correspond to those of respectively
$g_{1T}^{\perp q}$, $h_T^{\perp q}$, $h_{1T}^{\perp q}$ thanks to the relations 
(\ref{Eq:rel-III}--\ref{Eq:rel-V}). 
Also the widths of $g_{L}^{\perp q}=-h_T^q$ are not shown, because of the 
{\sl discarded} relation $g_{L}^{\perp q}=-k_z/M_N\,h_{1T}^{\perp q}$ implies 
the same $p_T$-behaviour of $h_{1T}^{\perp q}$, $g_{L}^{\perp q}$, $h_T^q$.) 

We observe a modest $x$-dependence of the various 
$\la p_T^2(x)\ra_{\rm Gauss}$, see Figs.~\ref{Fig04:avpT2-vs-Gauss}a--c.
Important for the widths of $g_1^q$, $h_1^q$ is that they are
not larger than that of $f_1^q$ in order
to comply with positivity, which is of course the case, see
Fig.~\ref{Fig04:avpT2-vs-Gauss}a. However, the widths of
other TMDs are not bound in this way by $\la p_T^2(x)\ra_{\rm Gauss}$ 
of $f_1^q$. In fact, the $\la p_T^2(x)\ra_{\rm Gauss}$ of the pretzelosity
distribution $h_{1T}^{\perp q}$ exceeds the width of $f_1^q$, 
see Fig.~\ref{Fig04:avpT2-vs-Gauss}c. 
Taken literally this would imply a violation of positivity, but 
we have to keep in mind that the $\la p_T^2(x)\ra_{\rm Gauss}$ 
only approximate the true $p_T$-behavior of TMDS in the model, 
and the exact model results always satisfy positivity,
see Sec.~\ref{Sec-3f:ineq} and App.~\ref{App-B:prove-ineq}.

\newpage

\begin{figure}[b!]
             \includegraphics[width=6cm]{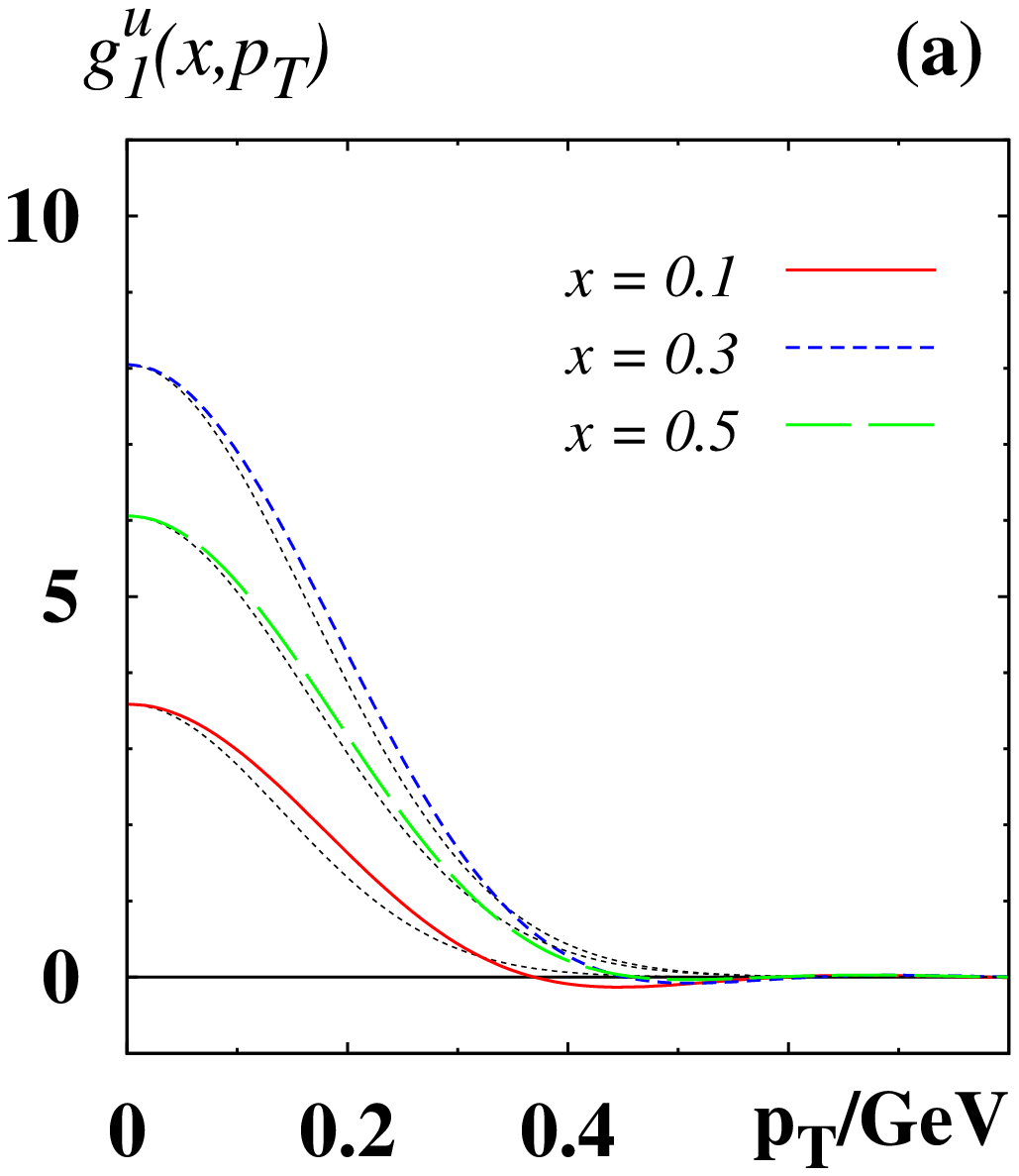}
\hspace{-5mm}\includegraphics[width=6cm]{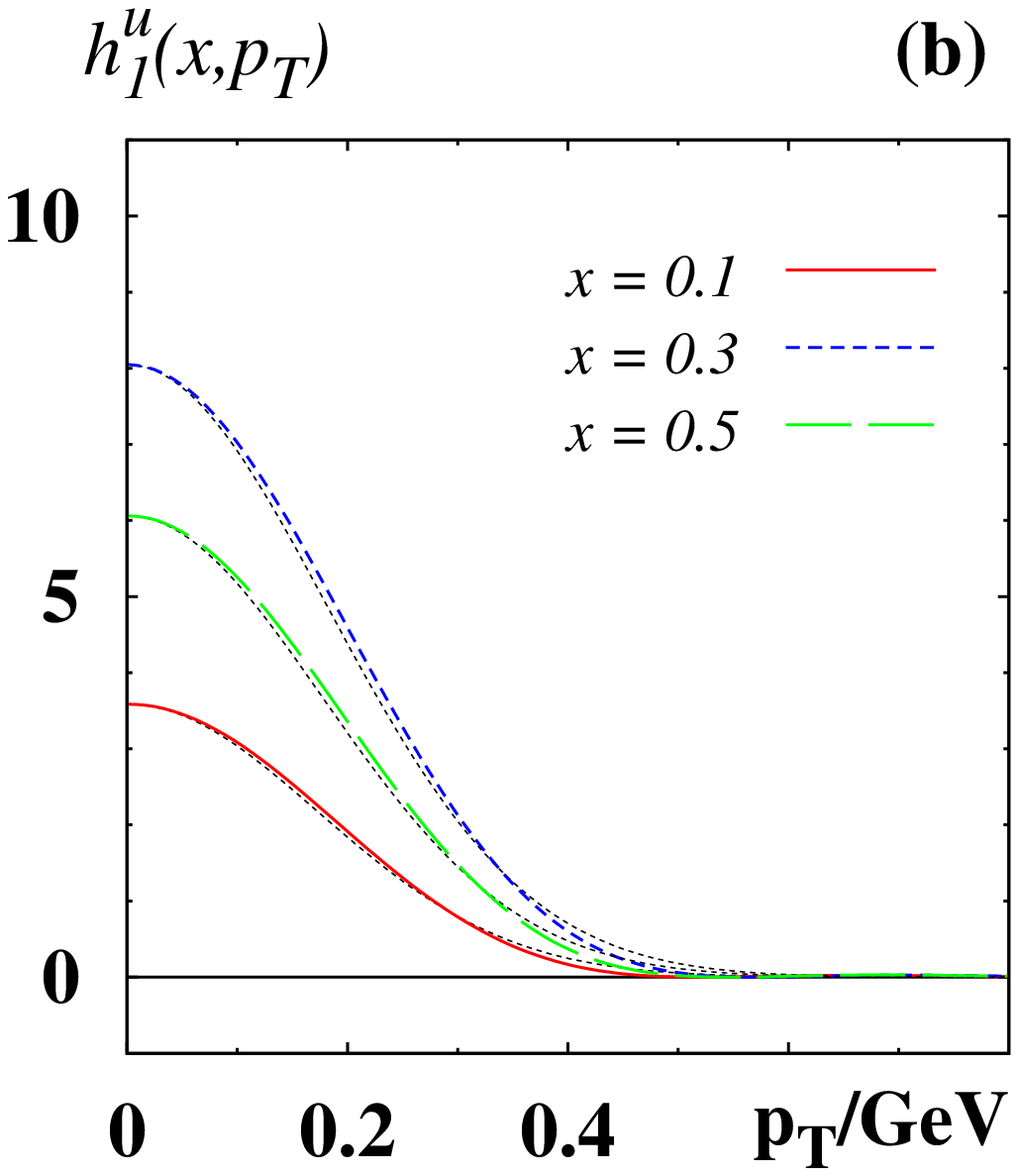}
\hspace{-5mm}\includegraphics[width=6cm]{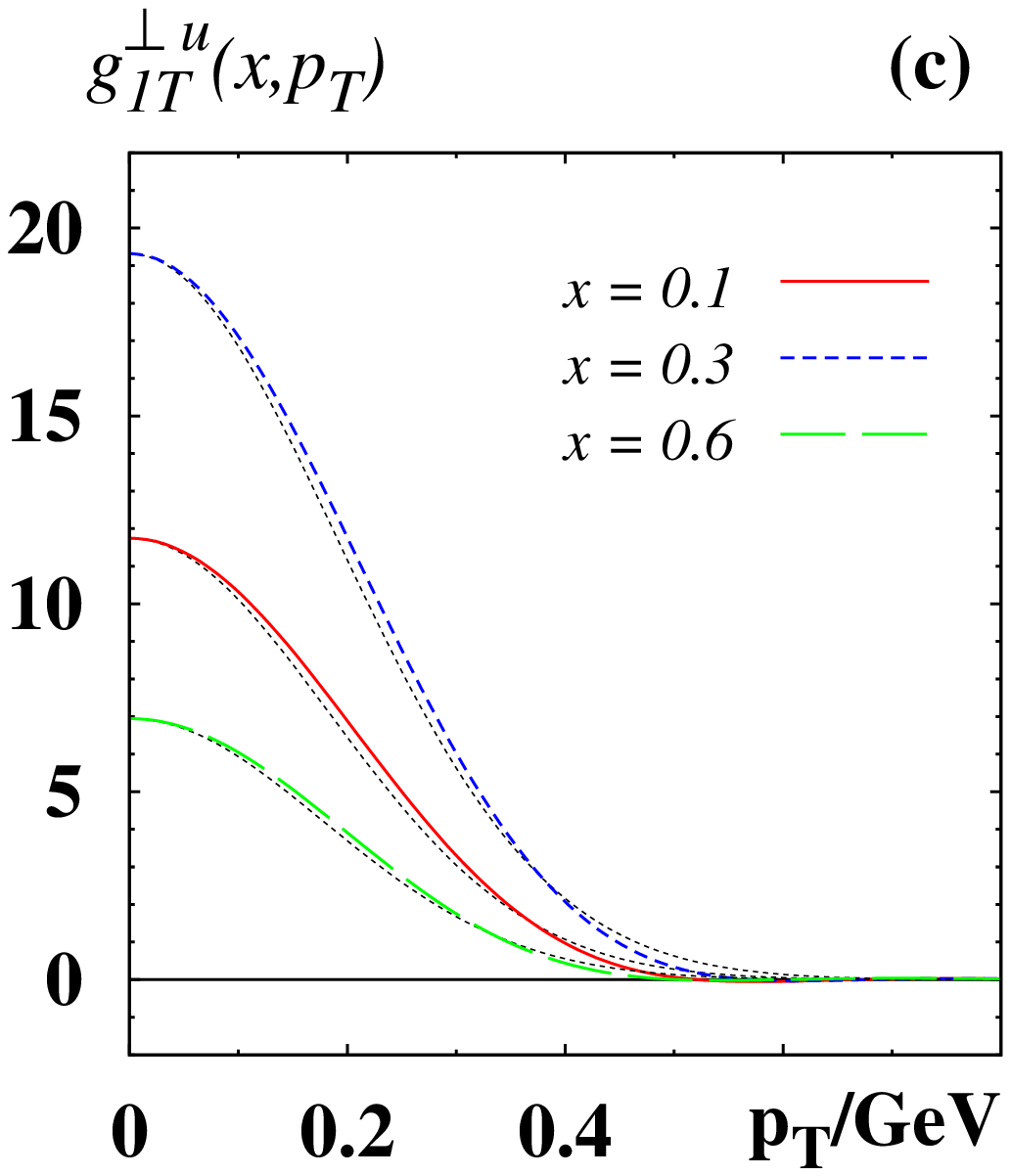}

             \includegraphics[width=6cm]{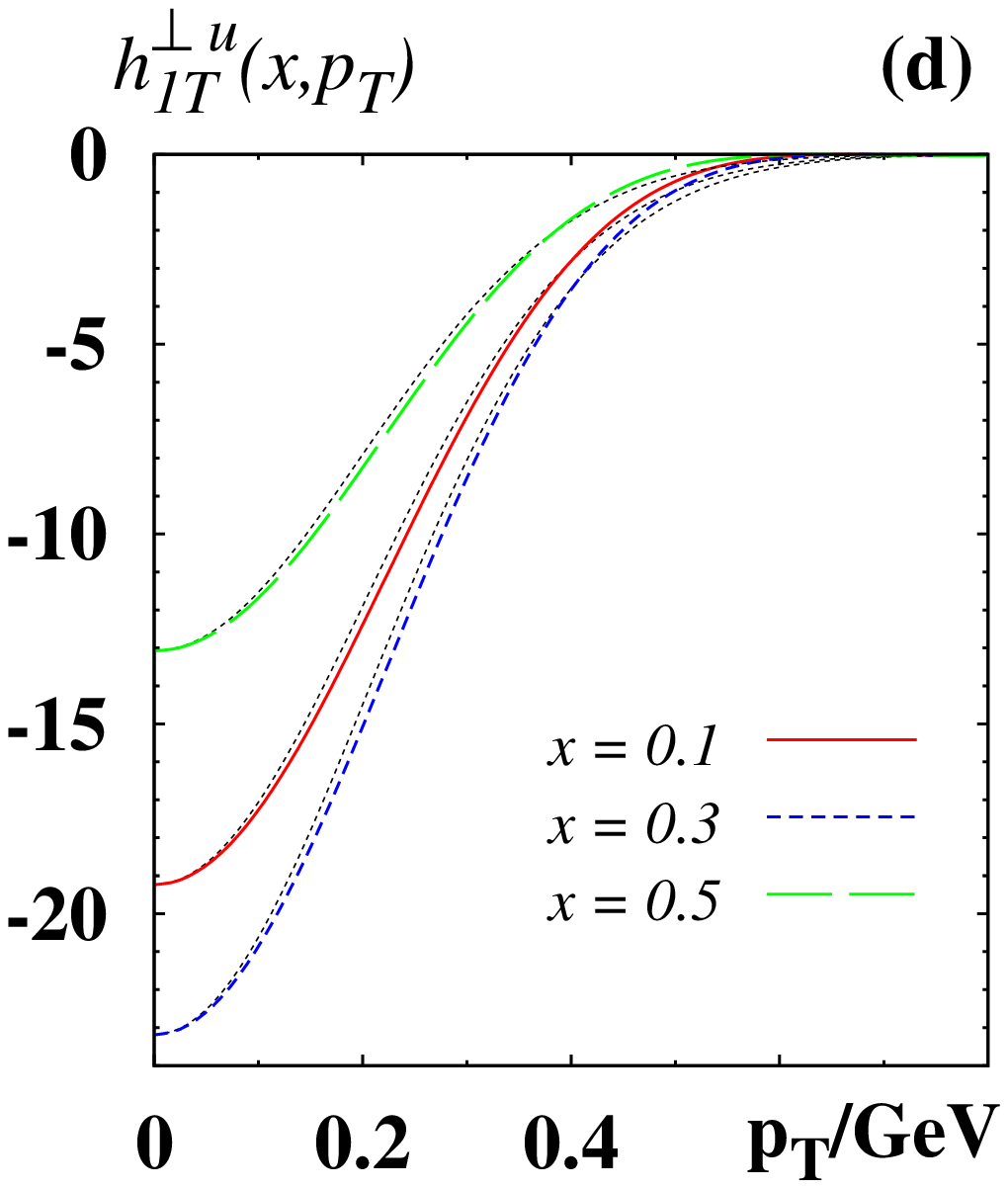}
\hspace{-5mm}\includegraphics[width=6cm]{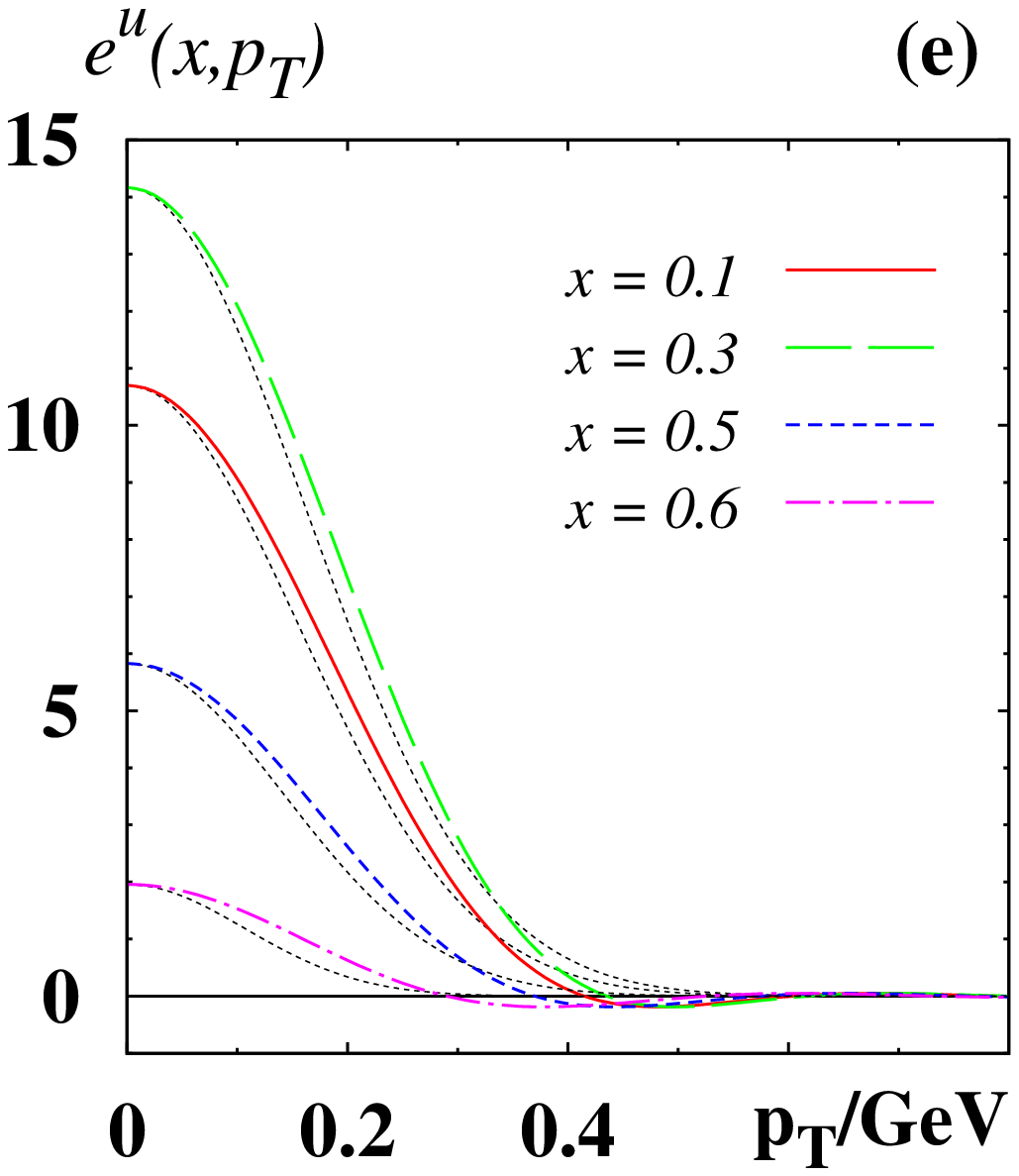}
\hspace{-5mm}\includegraphics[width=6cm]{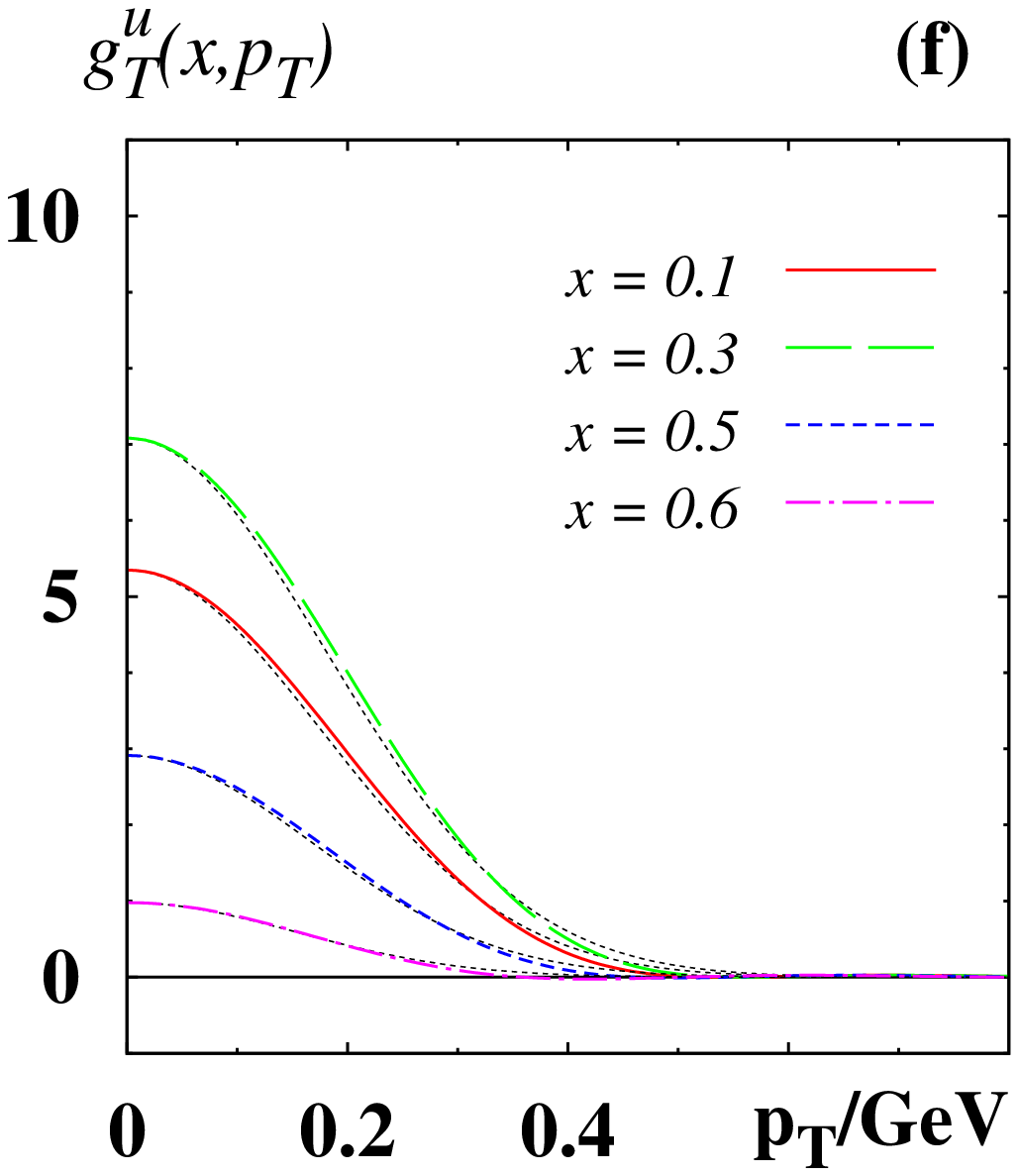}

             \includegraphics[width=6cm]{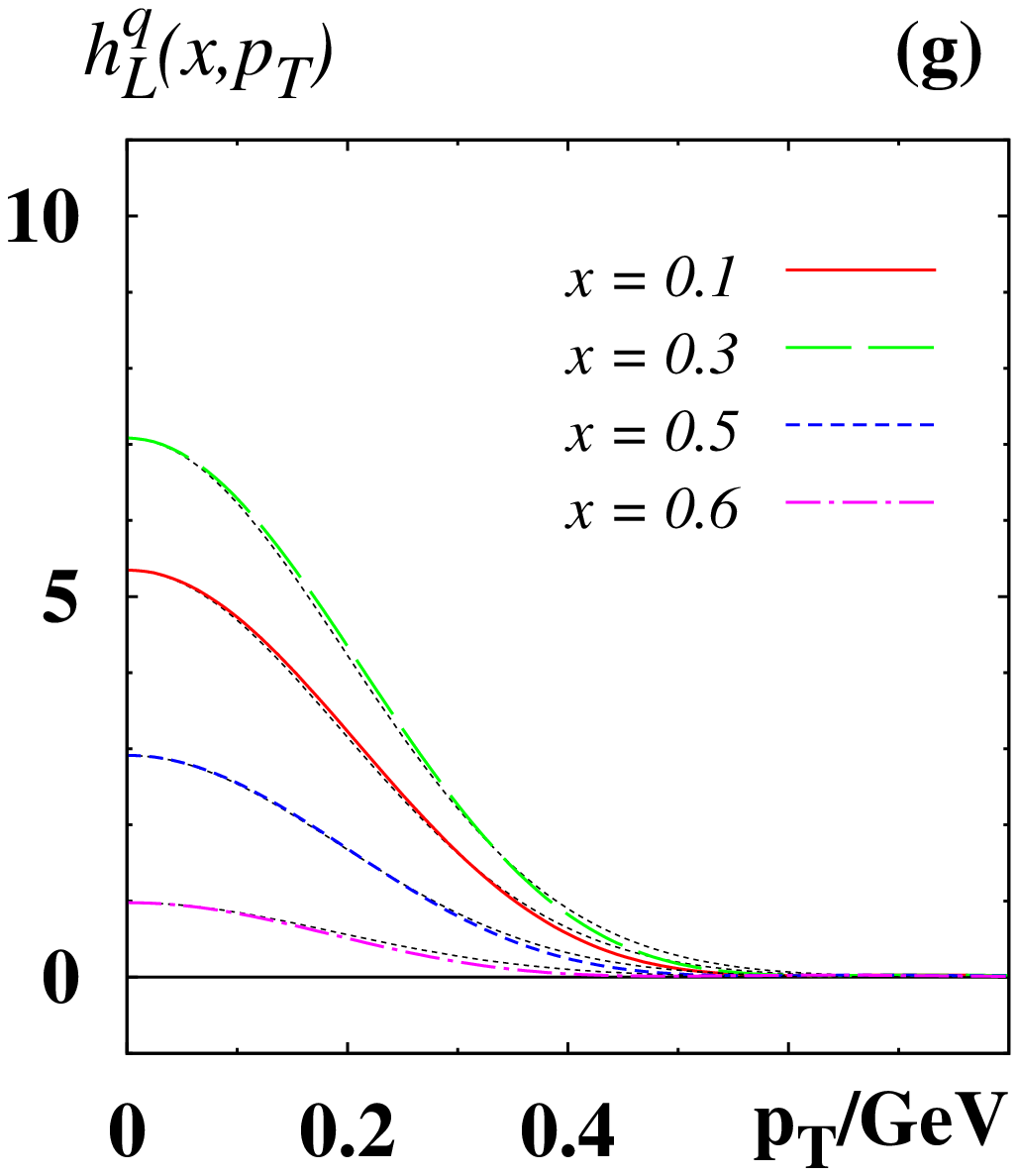}
\hspace{-5mm}\includegraphics[width=6cm]{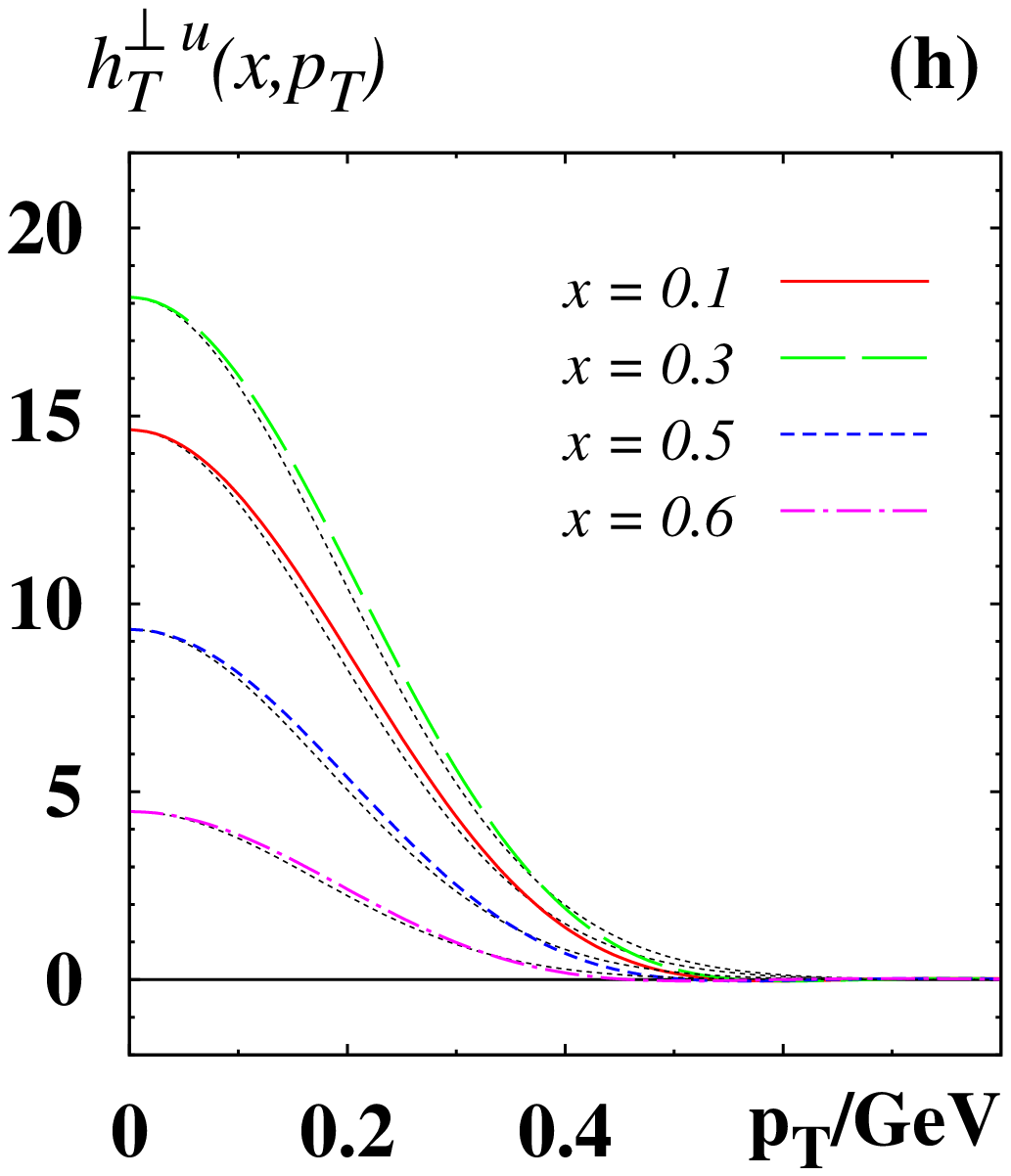}
\hspace{-5mm}\includegraphics[width=6cm]{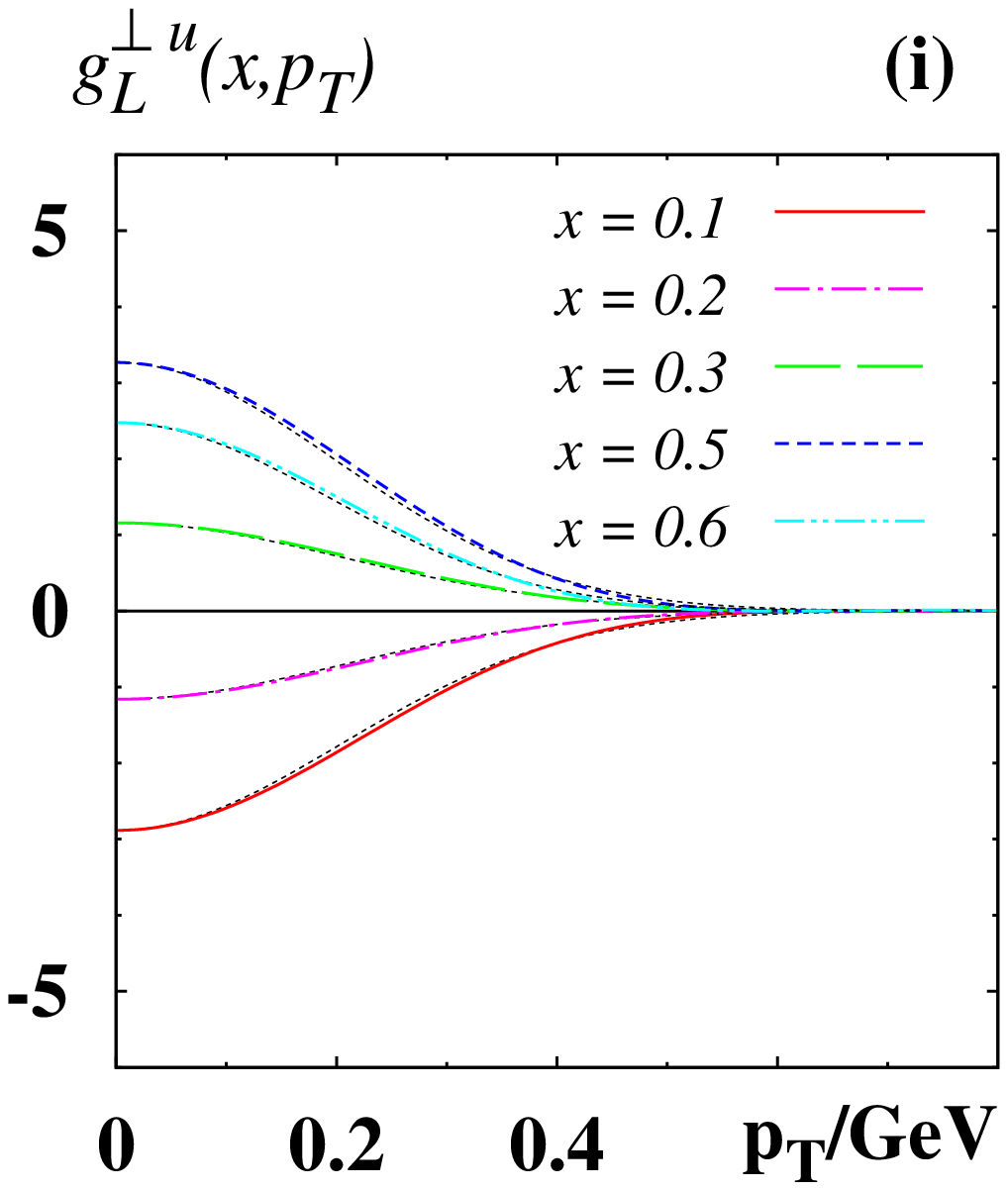}

\caption{\label{Fig05:TMDS-vs-Gauss}
    The $p_T$-dependence of various TMDs for selected values of $x$.
    The thin dotted lines are the respective Gauss model approximations
    obtained from the Gaussian widths from Fig.~\ref{Fig04:avpT2-vs-Gauss}.}
\end{figure}

Fig.~\ref{Fig05:TMDS-vs-Gauss} shows the $p_T$-dependence of the TMDs 
for selected values of $x$, chosen to optimize the clarity of the plots. 
The thin dotted lines are the respective Gauss model approximations
obtained from the Gaussian widths from Fig.~\ref{Fig04:avpT2-vs-Gauss}.
We observe in general a good agreement, including even $g_L^{\perp u}$
which has a zero in $x$, see Fig.~\ref{Fig04:avpT2-vs-Gauss}i.

%
%

Some TMDs, most notably for $g_1^q$ and $e^q$ in 
Fig.~\ref{Fig04:avpT2-vs-Gauss}a and \ref{Fig04:avpT2-vs-Gauss}e,
have for fixed values of $x$ a zero in $p_T$.
One is tempted to suspect a model-artifact which can be
traced back to the oscillatory behaviour of the Bessel functions
in (\ref{Eq:wave-func-0}).
However, in a covariant parton model calculation it was observed
that $g_1^u(x,p_T)$ becomes negative for some values of $x$, $p_T$
\cite{Efremov:2009vb}. This detail may deserve further attention.
But the effect is in any case small, and in phenomenological
applications the total result, for example, in SIDIS after convolution
with a fragmentation function, is strongly dominated by contributions 
from the valence-$x$ region and $p_T\lesssim 0.4\,{\rm GeV}$, where
the TMDs are sizable. 

Thus, to draw an intermediate conclusion, in the valence-$x$ region 
the bag model strongly supports the Gaussian model with a weakly 
$x$-independent Gaussian widths $\la p_T^2(x)\ra_{\rm Gauss}$.

Thus, the bag model results also encourage to use the so convenient
Gaussian model for TMDs also in future (until the data will teach us 
the opposite).
But how to use in practice these model predictions?
Indeed, one cannot use the bag model predictions literally, since, 
for example, the Gaussian width of $f_1^q$ from bag model underestimates
what is needed in phenomenology, see  Eqs.~(\ref{Eq:pT-unpol-bag})
and (\ref{Eq:pT-unpol-phenomenology}).

However, the $p_T$-broadening mechanism \cite{Collins:1984kg} 
is in lowest order approximation polarization independent. 
Therefore, model predictions for ratios of widths can be 
expected to be more reliable and useful for practical applications.
In view of the weak $x$-dependence of the $\la p_T^2(x)\ra_{\rm Gauss}$
we chose $x_v=0.3$ as a typical valence-$x$ value and summarize the
results for the $\la p_T^2(x)\ra_{\rm Gauss}$, in units of 
the width of $f_1^q$, in Table~\ref{Table-avpT2}.

It is worth to compare to the results from the light-cone 
constituent quark model \cite{Pasquini:2008ax,Boffi:2009sh}. 
In that model the wave-functions fall off with $p_T$ sufficiently fast,
such that $\la p_T^2\ra$ defined in Eq.~(\ref{Eq:def-avpT-avpT2}) 
exists. For the $f_1^q$ in that model $\la p_T^2\ra=0.080\,{\rm GeV}^2$ 
is close to our $\la p_T^2(x_v)\ra_{\rm Gauss}=0.077\,{\rm GeV}^2$.
For $g_1^q$, $h_1^q$ the results are comparably similar. However, 
in the case of $g_{1T}^{\perp q}$, $h_{1L}^{\perp q}$, $h_{1T}^{\perp q}$
the light-cone constituent quark model yields smaller widths compared
to the bag model, see Table~\ref{Table-avpT2}.
Twist-3 TMDs were not studied in \cite{Pasquini:2008ax,Boffi:2009sh}.

\ 

\

\begin{table}[h!]
\begin{tabular}{|ll|c|c|}
  \hline
  $\;\;\;$&&\ \hspace{20mm} & \ \hspace{20mm}  \\
  &TMD $j$ \ & $\;\;\displaystyle\la p_T^2\ra\;\;$ & $\;\;\displaystyle\la p_T^2(x_v)\ra_{\rm Gauss}\;\;$ \\
  & & Ref.~\cite{Boffi:2009sh} & (bag, here) \\
  \hline
 &&&\\
  &$f_1^q$           &  1$\phantom{.00}$ & 1$\phantom{.00}$ \\  &&&\\
  &$g_1^q$           &  0.74 & 0.71 \\   &&&\\
  &$h_1^q$           &  0.79 & 0.85 \\   &&&\\   &$g_{1T}^{\perp q}$, 
   $h_{1L}^{\perp q}$  &  0.74 & 0.95 \\   &&&\\
  &$h_{1T}^{\perp q}$  &  0.63 & 1.11 \\   &&&\\
  &$e^q$             &  -    &  0.68 \\  &&&\\
  &$g_T^q$           &  -    &  0.84 \\  &&&\\
  &$h_L^q$           &  -    &  1.01 \\  &&&\\   &$g_T^{\perp q}$, $g_L^{\perp q}$, 
   $h_T^q$           &  -    &  1.11 \\  &&&\\   &$f^{\perp q}$, 
   $h_T^{\perp q}$    &  -    &  0.94 \\
 &&&\\
\hline
\end{tabular}
\caption{\label{Table-avpT2}
   Average transverse momentum squares in T-even TMDs from 
   light-cone constituent quark model \cite{Boffi:2009sh},
   and the bag model (results obtained here). 
   The $\la p_T^2\ra$ from \cite{Boffi:2009sh} are defined 
   according to (\ref{Eq:def-avpT-avpT2}). The bag model
   results for the Gaussian widths are defined according 
   to (\ref{Eq:Gauss-width}) and taken at the valence-$x$ 
   point $x_v=0.3$.
   All results are in units of the respective value for $f_1^q$, 
   which is $\la p_T^2\ra=0.080\,{\rm GeV}^2$ in the case of
   \cite{Boffi:2009sh}, and
   $\la p_T^2(x_v)\ra_{\rm Gauss}^{(f_1)}=0.077\,{\rm GeV}^2$
   in the case of the bag model.}
\end{table}

\newpage
\subsection{WW-type approximations}
\label{Sec-5c:WW-type}

By exploring the equations of motion, twist-3 parton distributions 
typically can be decomposed into pieces related to leading-twist TMDs, 
current quark mass terms, and quark-gluon-quark correlators.
The latter are often referred to as 'pure twist-3' or 'interaction dependent'
terms, and are marked by a tilde. For T-even TMDs one obtains
\cite{Mulders:1995dh} (we suppress the arguments $x$ and $k_\perp$)
\ba
xe^q           &=& x\tilde{e}^q + \frac{m_q}{M}\,f_1^q\;,               \label{eq:WW-e} \\
xf^{\perp q}   &=& x\tilde{f}^{\perp q}+ f_1^q\;,\phantom{\frac{1}{1}}  \label{eq:WW-fperp}\\
xg_L^{\perp q} &=& x\tilde{g}_L^{\perp q} + g_1^q+\frac{m_q}{M}\,h_{1L}^{\perp q},\\
xg_T^{\perp q} &=& x\tilde{g}_T^{\perp q}+g_{1T}^{\perp q} +\frac{m_q}{M}\,h_{1T}^{\perp q}\:,\\
xg_T^q         &=& x\tilde{g}_T^q +\frac{\vec{p}_T^{\:2}}{2M^2}\,g_{1T}^{\perp q}+\frac{m_q}{M}\,h_1^q\:,\\
xh_L^q         &=& x\tilde{h}_L^q -\frac{\vec{p}_T^{\:2}}{M^2}\,h_{1L}^{\perp q}+\frac{m_q}{M}\,g_1^q\;,\\
xh_T^q         &=& x\tilde{h}_T^q - h_1^q -\frac{\vec{p}_T^{\:2}}{2 M^2}\, h_{1T}^{\perp q} 
                   +\frac{m_q}{M}\,g_{1T}^{\perp q}\;,\\
xh_T^{\perp q} &=& x\tilde{h}_T^{\perp}+h_1^q-\frac{\vec{p}_T^{\:2}}{2 M^2}\, h_{1T}^{\perp q}\;,\\
xg_T^{\prime q}&=& x\tilde{g}_T^{\prime q} + \frac{m_q}{M}h_1^q - \frac{m_q}{M}\,
\frac{\vec{p}_T^{\:2}}{2 M^2}\, h_{1T}^{\perp q} \:.
\ea 
where in the last equation the notation is used 
$g_T^{\prime q}\equiv g_T^q -\frac{\vec{p}_T^{\:2}}{2 M^2}\, g_{T}^{\perp q}$ 
and analog for $\tilde{g}_T^{\prime q}$ \cite{Bacchetta:2006tn}. 
If we systematically assume that pure twist-3 and quark mass terms are 
small, which we indicate symbolically and generically by ${\cal O}(\varepsilon)$,
and integrate over transverse momenta, 
then we obtain the following Wandzura-Wilczek-type approximations
\ba
xe^q(x)           &=& {\cal O}(\varepsilon), \label{Eq:WW-type-1}\\
xf^{\perp q}(x)   &=& f_{1}^q(x) + {\cal O}(\varepsilon), \phantom{\frac11}\\
xg_L^{\perp q}(x) &=& g_{1}^q(x)  + {\cal O}(\varepsilon), \\
xg_T^{\perp q}(x) &=& g_{1T}^{\perp q}(x) + {\cal O}(\varepsilon),  \phantom{\frac11}\\
xg_T^q(x)         &=& g_{1T}^{\perp(1)q}(x)+{\cal O}(\varepsilon),\label{Eq:WW-type-5}\\
xh_L^q(x)         &=& -2 \,h_{1L}^{\perp(1)q}(x) + {\cal O}(\varepsilon),\phantom{\frac11}\label{Eq:WW-type-6}\\
xh_T^q(x)         &=& - h_1^q(x) - h_{1T}^{\perp(1)}(x)+ {\cal O}(\varepsilon),\\
xh_T^{\perp q}(x) &=& \phantom{-}h_1^q(x) - h_{1T}^{\perp(1)}(x)+ {\cal O}(\varepsilon),\phantom{\frac11} \\
xg_T^{\prime q}(x)&=& {\cal O}(\varepsilon).\label{Eq:WW-type-9}
\ea 
What these approximations have in common with the classic Wandzura-Wilczek (WW) 
approximation is that pure twist-3 and current quark mass terms are neglected. 
However, the neglected operators are different in all cases.

%
\begin{figure}[b!]
    \includegraphics[width=5cm]{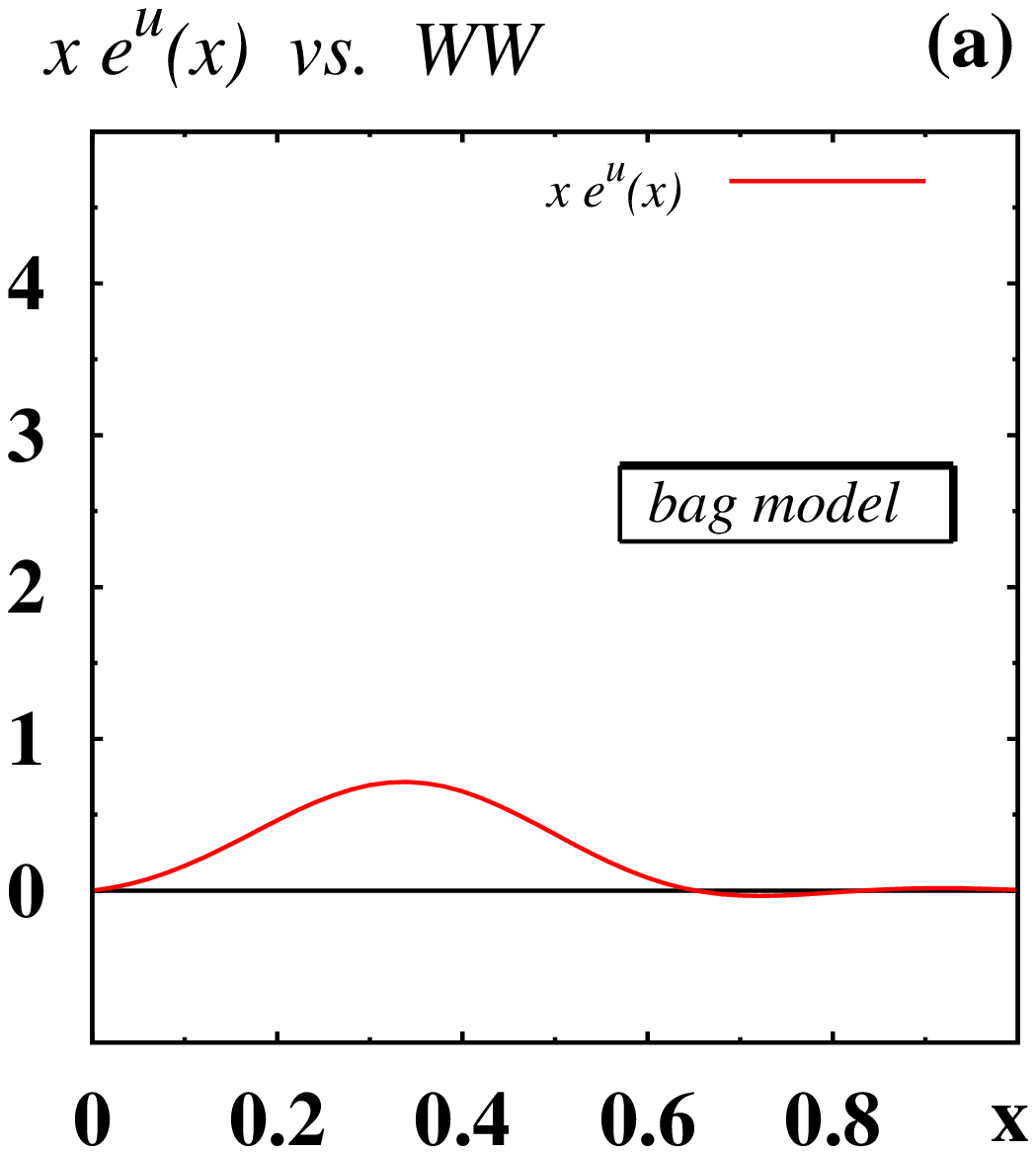}    
    \includegraphics[width=5cm]{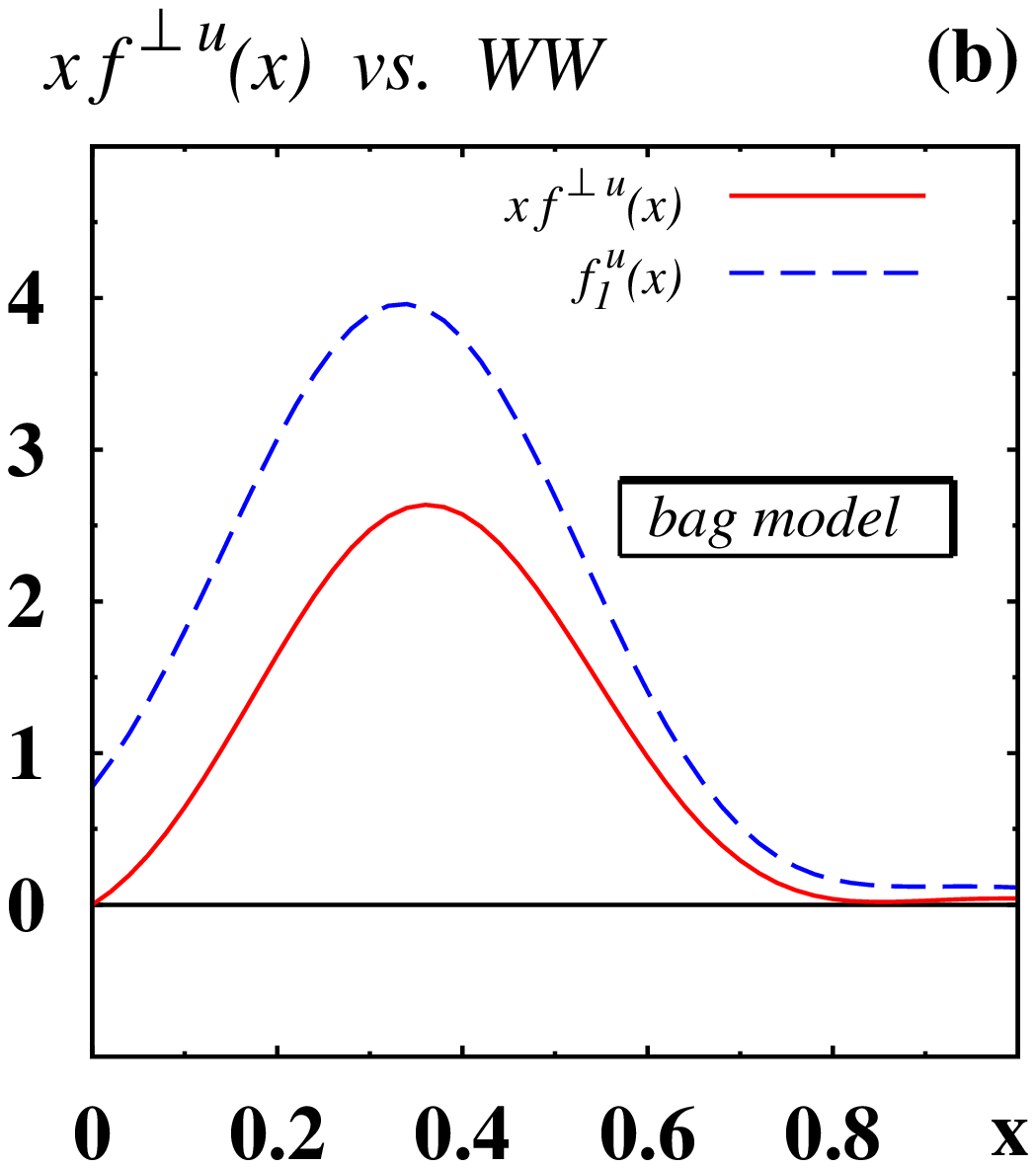}
    \includegraphics[width=5cm]{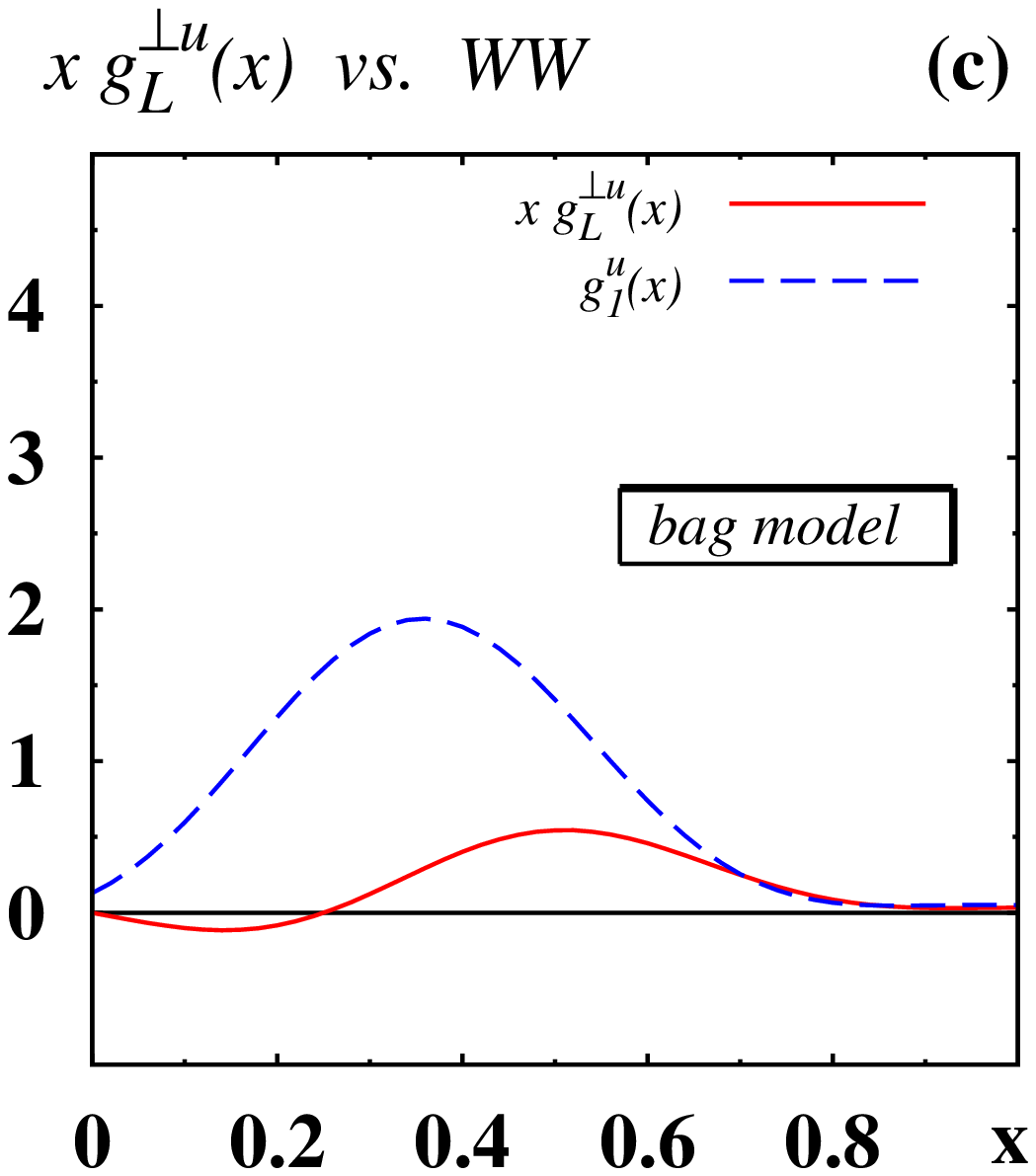}

    \includegraphics[width=5cm]{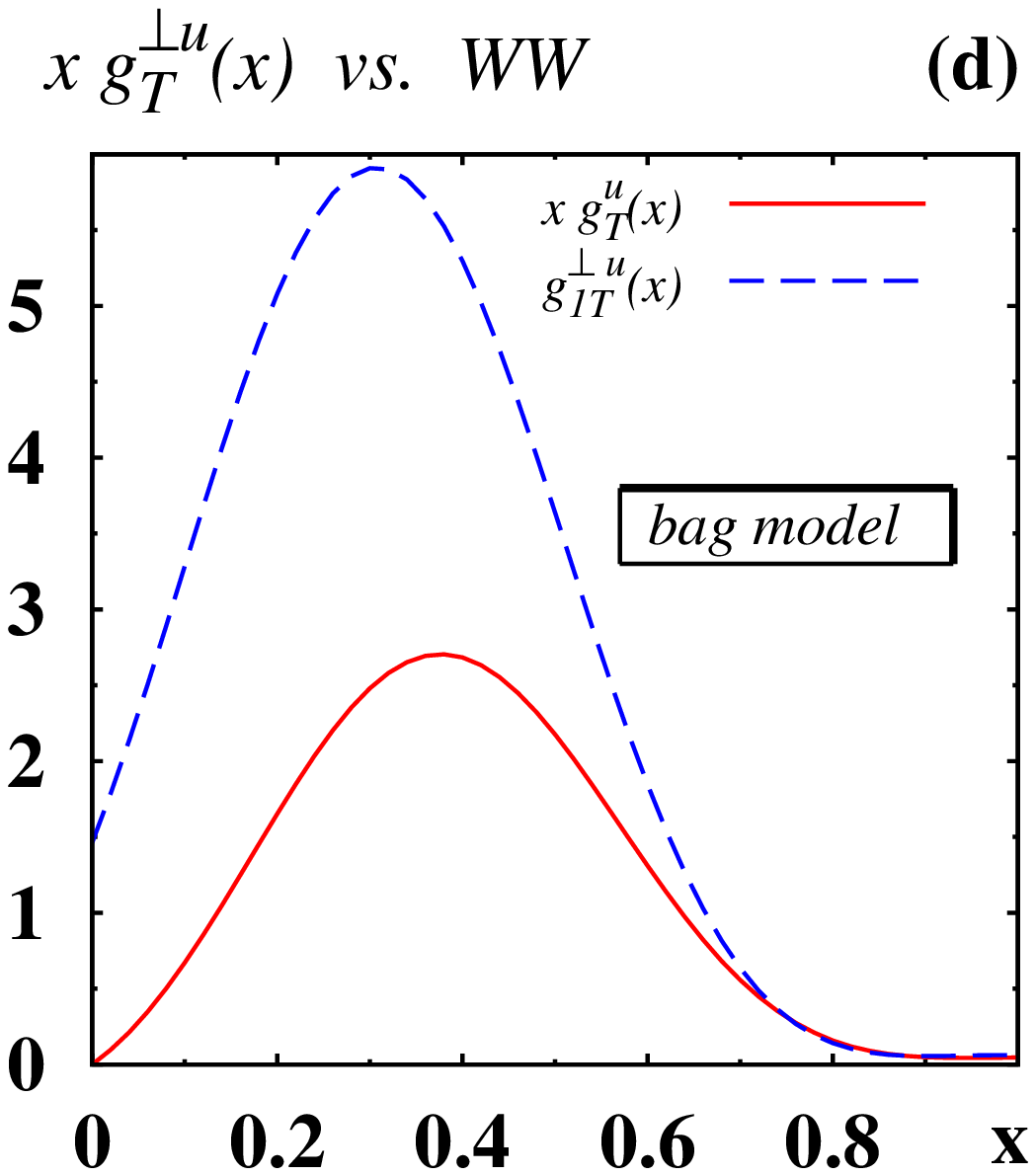}
    \includegraphics[width=5cm]{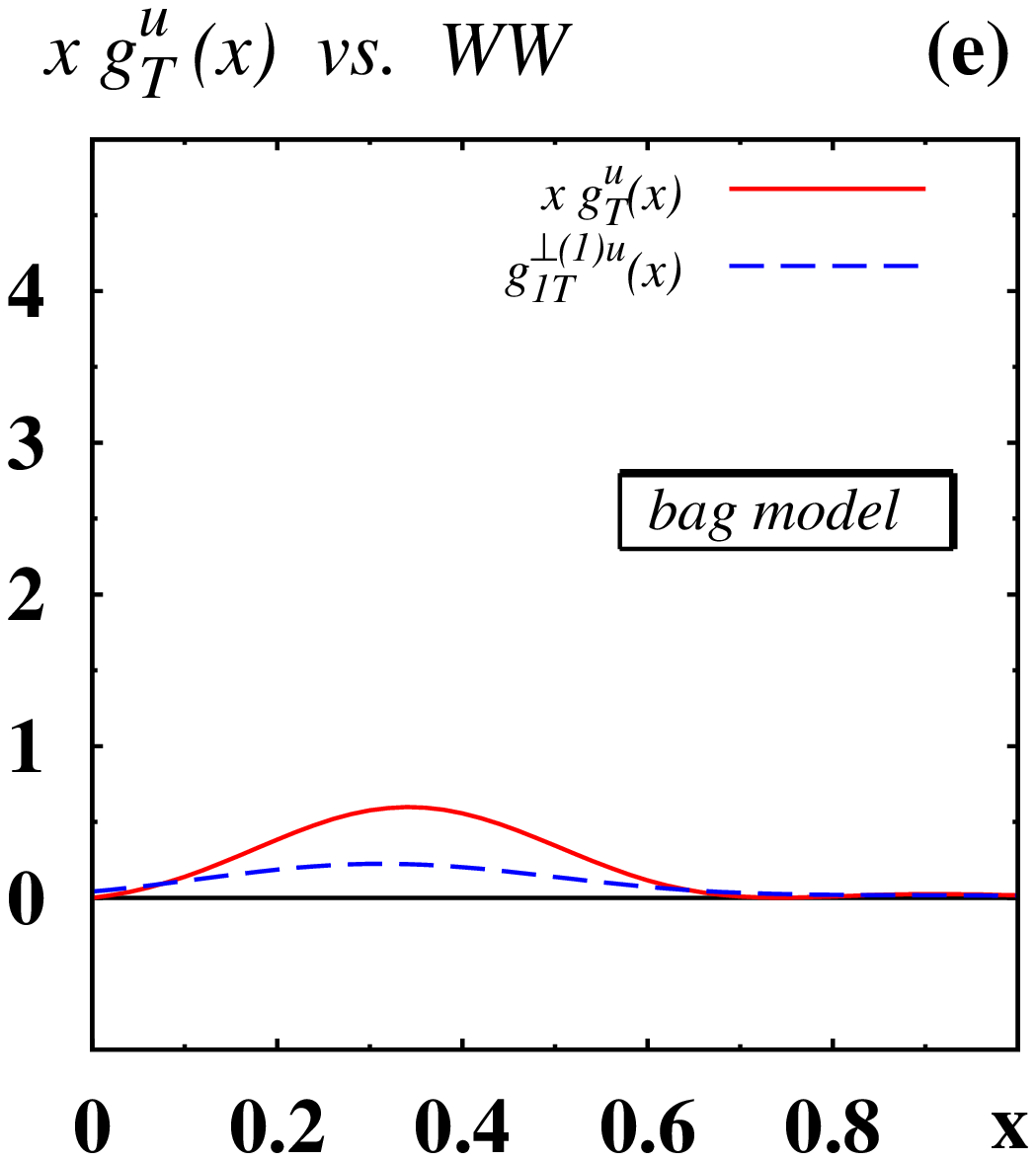}
    \includegraphics[width=5cm]{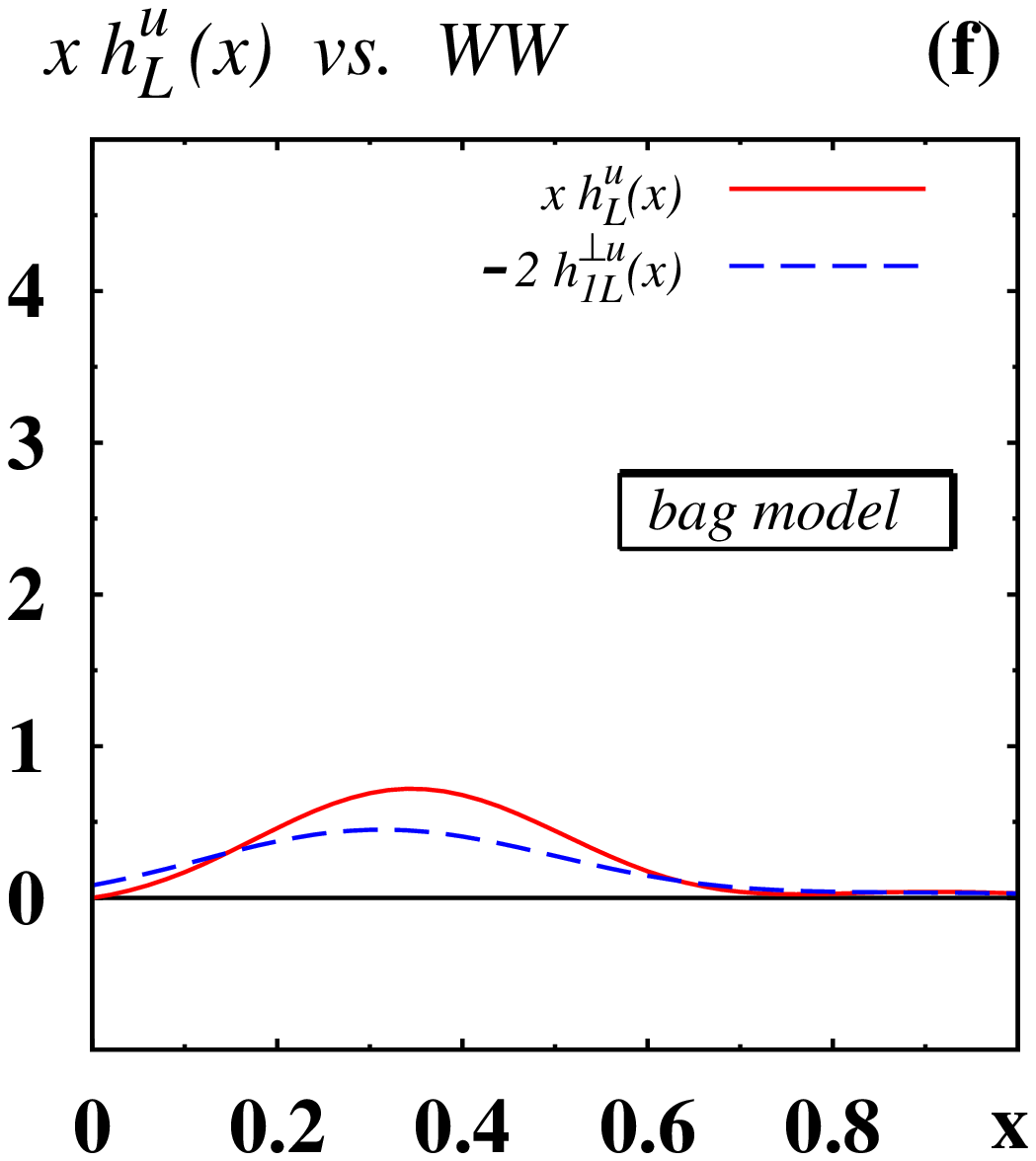}

    \includegraphics[width=5cm]{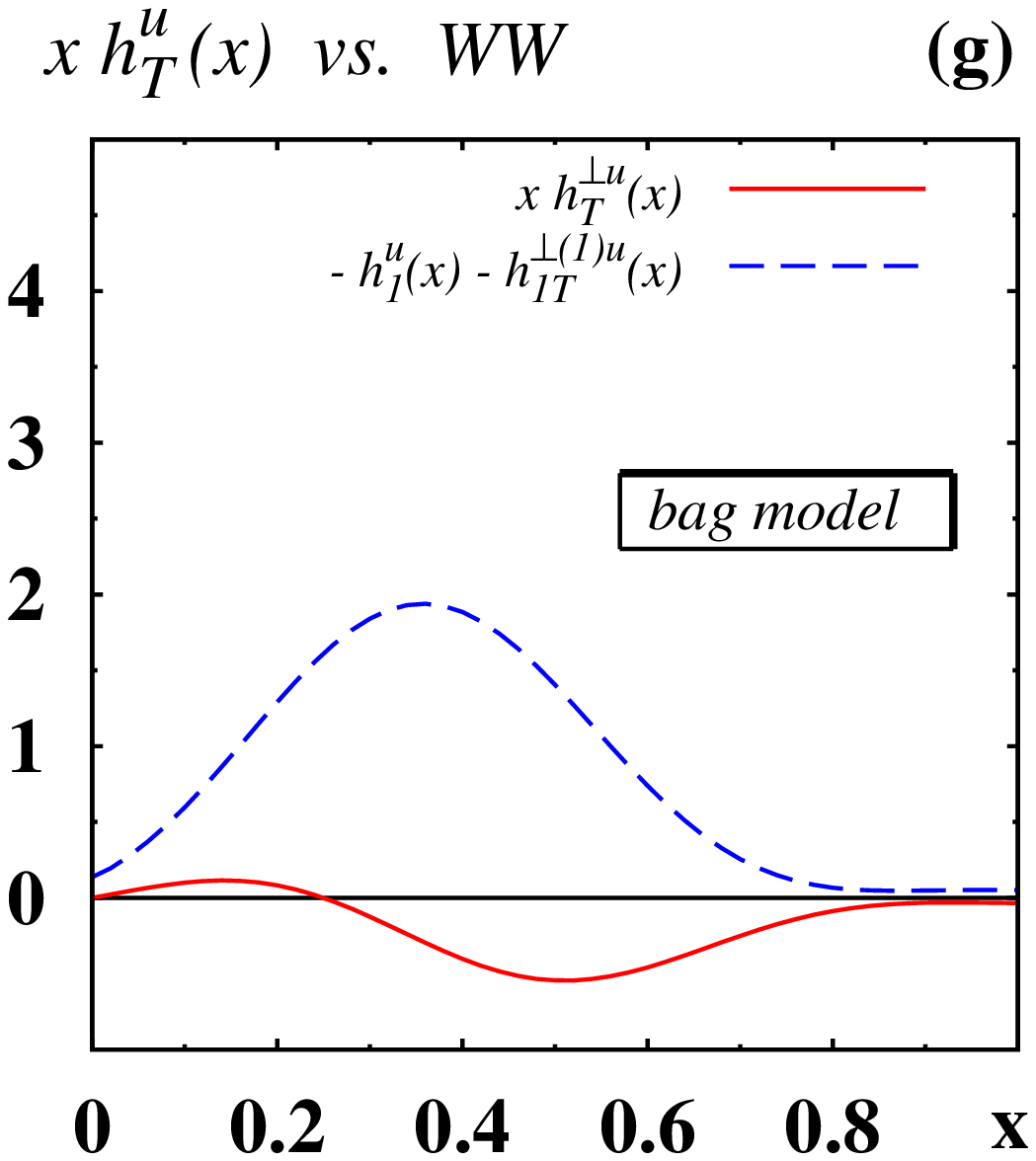}
    \includegraphics[width=5cm]{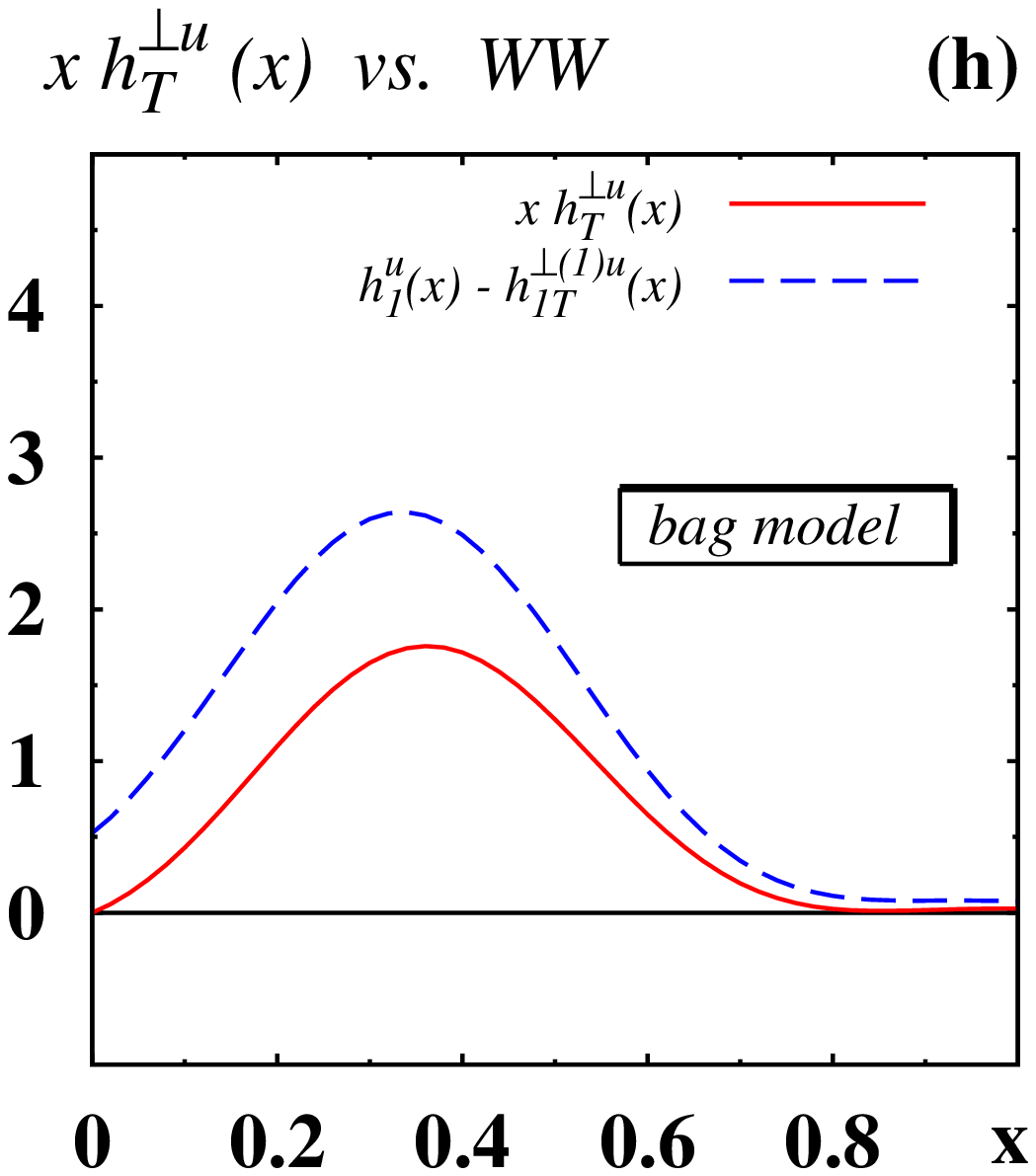}
    \includegraphics[width=5cm]{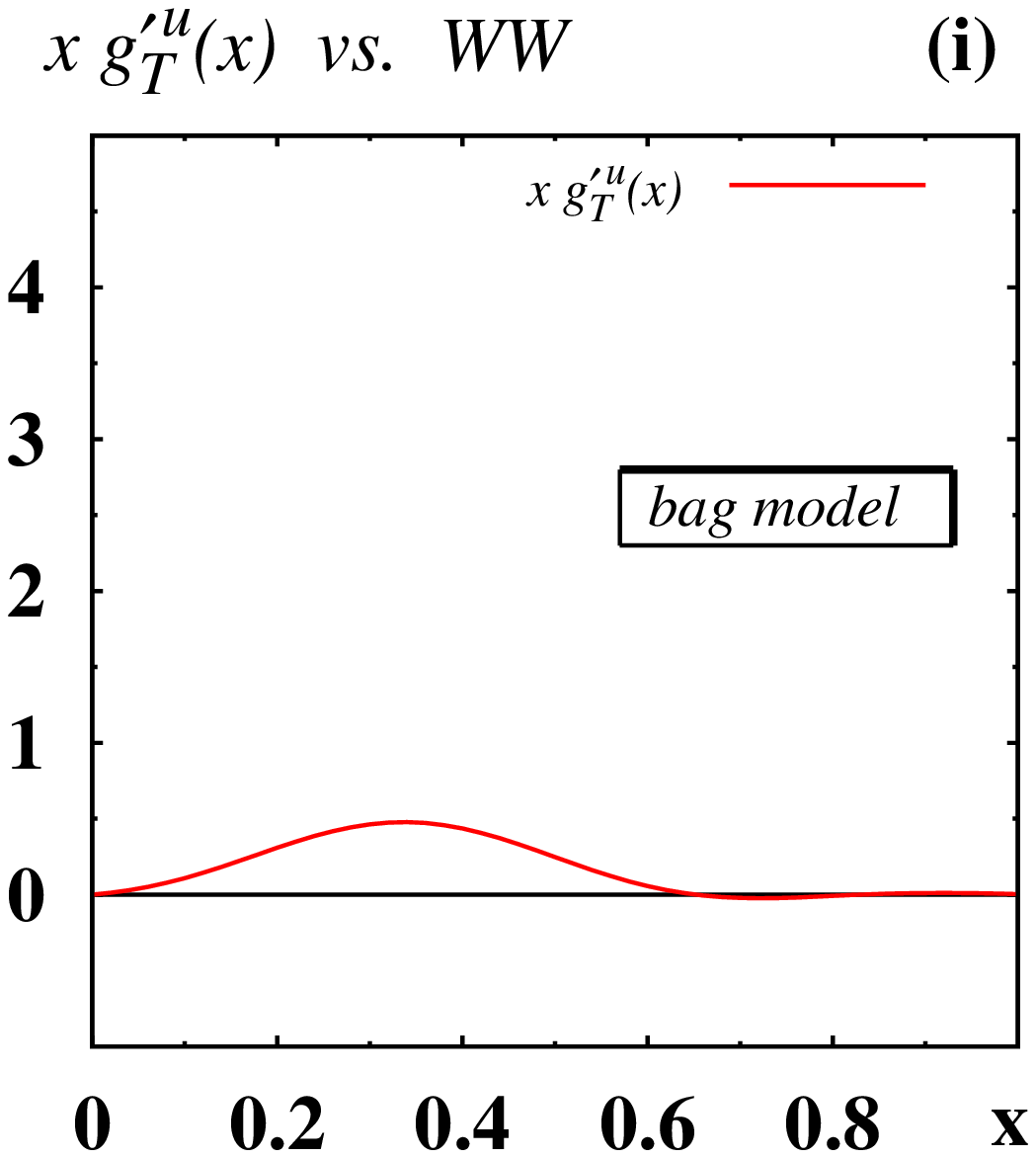}

\caption{\label{Fig02:WW-type-approx}
    Test of WW-type approximations in the bag model at the low scale.
    The solid lines show $x$ times twist-3 TMDs, as given on the
    left-hand-sides on the Eqs.~(\ref{Eq:WW-type-1}--\ref{Eq:WW-type-9}).
    The dashed lines are twist-2 TMDs (or their transverse moments or
    linear combinations thereof) as given on the right-hand-sides of 
    the Eqs.~(\ref{Eq:WW-type-1}--\ref{Eq:WW-type-9}).
    If the tilde- and mass-terms defined in terms of the 
    {\sl QCD equations of motion} vanished, the solid and dashed lines
    would coincide, see text.
    In Figs.~(a) and (i) no dashed lines appear because the 
    right-hand-sides of the Eqs.~(\ref{Eq:WW-type-1}) and 
    (\ref{Eq:WW-type-9}) are in the WW-type approximation.}
\end{figure}

Figs.~\ref{Fig02:WW-type-approx}a--\ref{Fig02:WW-type-approx}i 
compare respectively the left- (solid lines) and right-hand-sides 
(dashed lines) of the Eqs.~(\ref{Eq:WW-type-1}--\ref{Eq:WW-type-9}),
assuming that ${\cal O}(\varepsilon)$ is zero in each case.
In all Figures the same scale is used in order to better compare
the magnitudes of the various functions. 
First we observe that $xe^q(x)$ and $xg_T^{\prime q}(x)$
are not zero, as one would expect on the basis on the WW-type
approximations, though the functions are not large, see 
Figs.~\ref{Fig02:WW-type-approx}a and \ref{Fig02:WW-type-approx}i.

Next we remark that the WW-type approximations for 
$xf^{\perp q}(x)$, $xh_L^q(x)$, $xh_T^{\perp q}(x)$ can be considered 
as roughly supported by the bag model, see 
Figs.~\ref{Fig02:WW-type-approx}b, \ref{Fig02:WW-type-approx}f, 
\ref{Fig02:WW-type-approx}h.
In the remaining cases, however, the WW-type approximations
work only modestly, for example Fig.~\ref{Fig02:WW-type-approx}e,
or not at all, see Figs.~\ref{Fig02:WW-type-approx}c,
\ref{Fig02:WW-type-approx}d, \ref{Fig02:WW-type-approx}g.

At first glance one might be surprised that the WW-type approximations
are not exactly fulfilled in a no-gluon model with massless quarks, as 
apparently there are no contributions from quark-gluon and mass terms.
However, here one has to recall that the WW-type approximations
originate from applying {\sl QCD equations of motion}, and
separating leading and subleading twist terms.
In principle, one could repeat this game in the bag model, too.
As the quarks are {\sl not free} (but confined by the bag), 
one would consequently encounter certain ``{\sl interaction dependent}'' 
tilde-terms in the model, too.
These bag-model tilde-terms can be seen directly in the case of 
$e^q(x)$, $g_T^{\prime q}(x)$ in 
Figs.~\ref{Fig02:WW-type-approx}a,~\ref{Fig02:WW-type-approx}i. 
In the other plots in Fig.~\ref{Fig02:WW-type-approx} they are
apparent as the differences between the solid and dashed lines.
It has been argued that the bag, which is a model for confinement,
in some sense mimics the effects of gluons \cite{Jaffe:1991ra}.
However, to which extent the bag-model interaction--dependent terms 
are able to estimate reliably the QCD interaction--dependent terms,
remains to be seen.

In any case, it is interesting to observe that the bag model
roughly supports the WW-type approximation for $xf^{\perp u}(x)$,
see Fig.~\ref{Fig02:WW-type-approx}b, which played an important role 
in the interpretation \cite{Anselmino:2005nn} of the EMC data on the 
azimuthal asymmetry $A_{UU}^{\cos\phi}$ in unpolarized SIDIS 
\cite{Arneodo:1986cf} as being due to the Cahn effect \cite{Cahn:1978se}.

\begin{figure}[b!]
    \includegraphics[width=8cm]{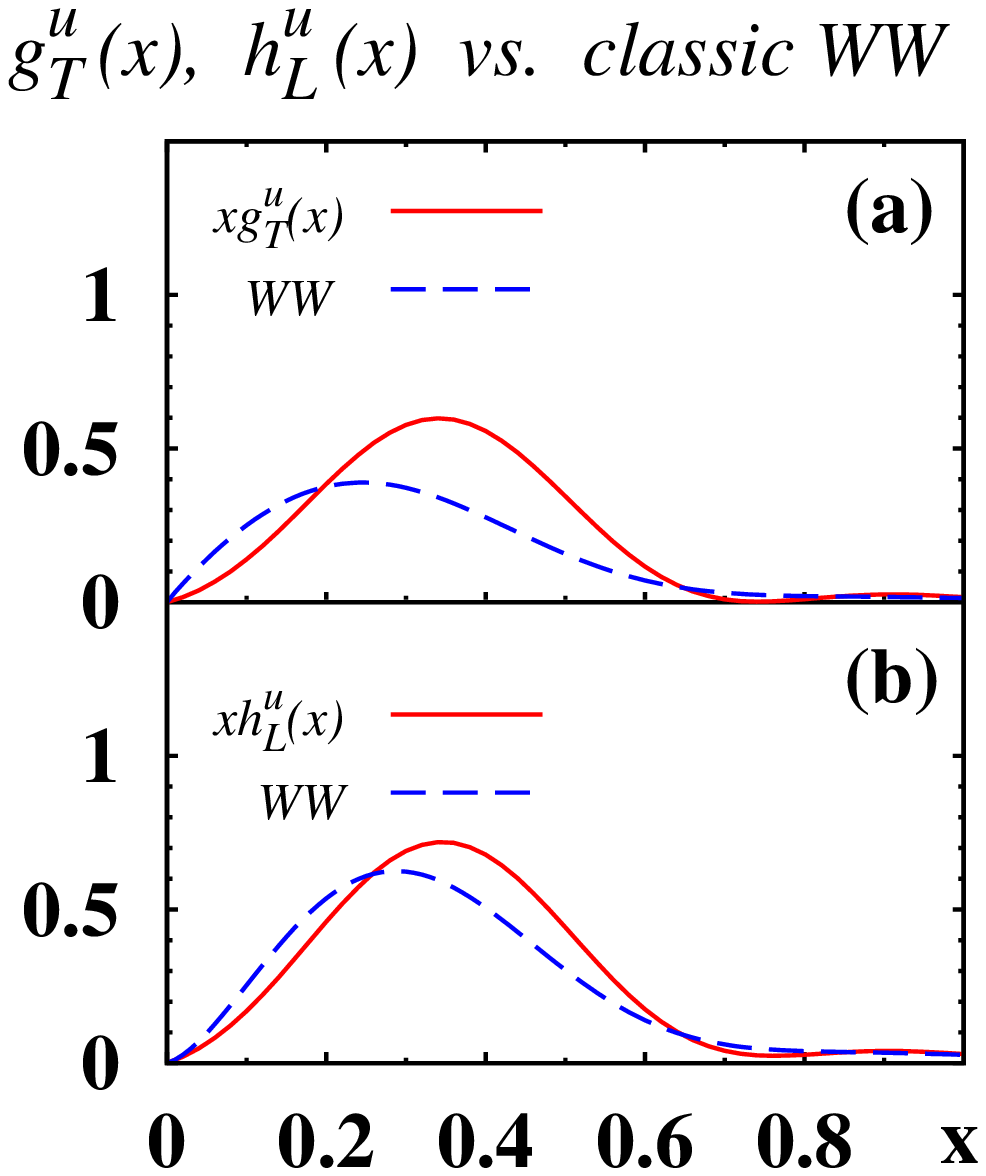}
    \includegraphics[width=8cm]{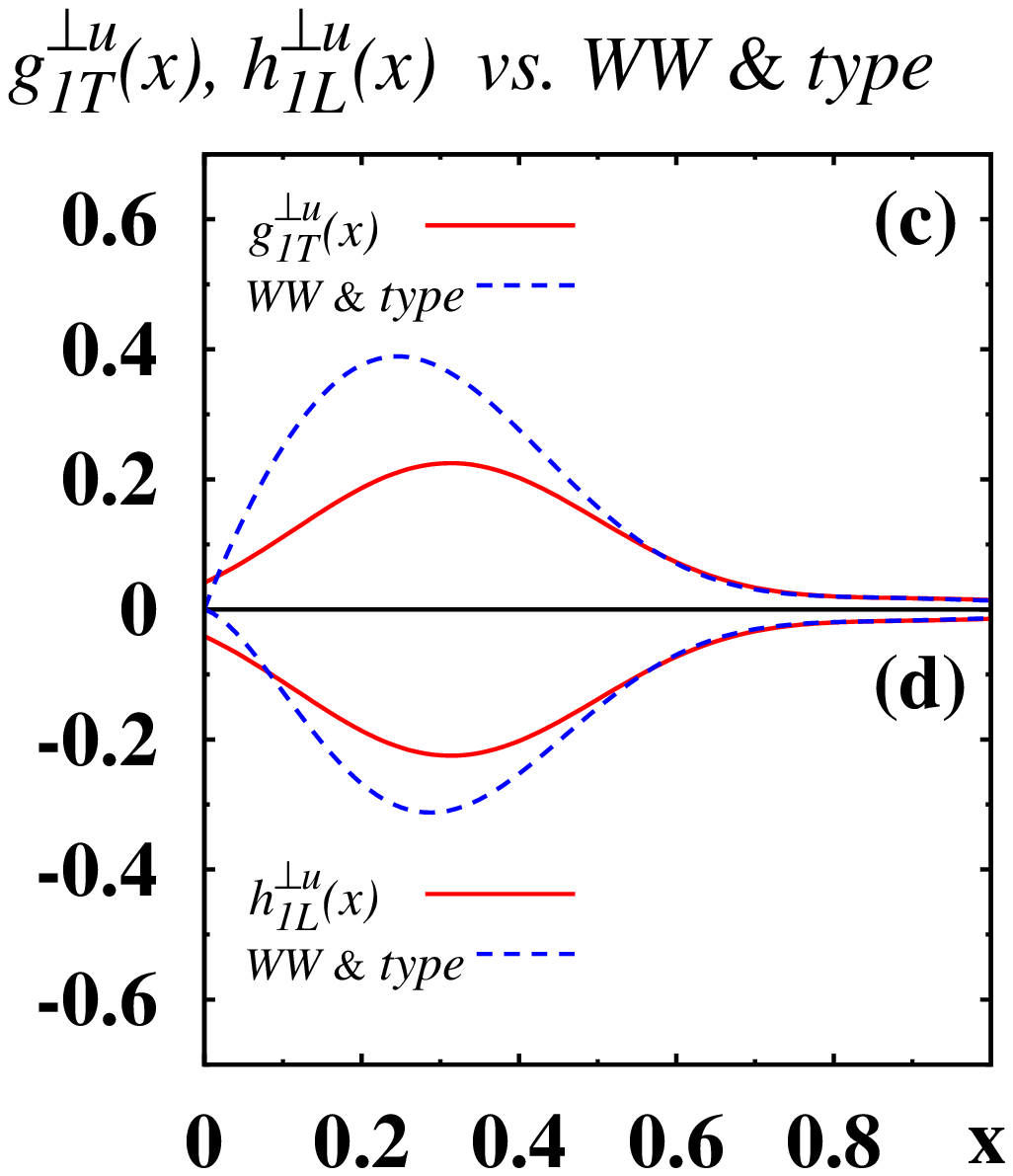}    

\caption{\label{Fig03:WW-classic+type}
    (a) and (b): 
    The test of the classic WW approximations for $g_T^q(x)$ 
    \cite{Wandzura:1977qf} and $h_L^q(x)$ \cite{Jaffe:1991ra},
    see Eqs.~(\ref{Eq:WW-approx-gT},~\ref{Eq:WW-approx-hL}), 
    in bag model. 
    (c) and (d): 
    Test of the approximations 
    for $g_{1T}^{\perp(1)q}(x)$ and $h_{1L}^{\perp(1)q}(x)$
    in Eqs.~(\ref{Eq:WW-approx-g1T},~\ref{Eq:WW-approx-h1L}),
    which result from combining 
    Eqs.~(\ref{Eq:WW-type-5},~\ref{Eq:WW-type-6})
    and (\ref{Eq:WW-approx-gT},~\ref{Eq:WW-approx-hL}).
    All results refer to the low bag model scale.}
\end{figure}

\newpage
Having discussed the WW-type approximations for TMDs, whose
usefulness remains to be tested \cite{Metz:2008ib}, it is
interesting to have a look back on the ``classic WW-approximations''
for $g_T^q(x)$ and $h_L^q(x)$ \cite{Wandzura:1977qf,Jaffe:1991ra},
which are distinguished from the WW-type approximations in that
in their derivations the notion and complications of transverse 
parton momenta does not need to be involved
\cite{Wandzura:1977qf,Jaffe:1991ra}, though can be considered
\cite{Belitsky:1997ay,Accardi:2009au}.
These are therefore in a certain sense ``collinear'' approximations.
They are given by
\ba
   	g_T^q(x)&\stackrel{\rm WW}{\approx}& \phantom{2x}\!
	\int_x^1\frac{\di y}{y\;}\,g_1^a(y) \;,
	\label{Eq:WW-approx-gT}\\
    	h_L^q(x)&\stackrel{\rm WW}{\approx}& 
	2x\!\int_x^1\frac{\di y}{y^2\;}\,h_1^a(y)\;.
	\label{Eq:WW-approx-hL}\ea
Figs.~\ref{Fig03:WW-classic+type}a and \ref{Fig03:WW-classic+type}b
show to which extent the WW approximations are supported by the
bag model: moderately in the case of $g_T^q(x)$, and somewhat 
better in the case of $h_L^q(x)$. (In 
Figs.~\ref{Fig03:WW-classic+type}a and \ref{Fig03:WW-classic+type}b
we compare $x$ times the functions and their WW-approximations,
because at small $x\lesssim 0.1$ the approximations 
(\ref{Eq:WW-approx-gT},~\ref{Eq:WW-approx-hL})
are poorly supported, which should not worry us as the bag model 
is expected to be more meaningful in the valence-$x$ region, see 
Sec.~\ref{Sec-2:TMDs-in-bag}.)

Finally, we remark that by combining the classic WW approximations 
for $g_T^q(x)$ in Eq.~(\ref{Eq:WW-approx-gT}) \cite{Wandzura:1977qf}, 
and $h_L^q(x)$ in Eq.~(\ref{Eq:WW-approx-hL}) \cite{Jaffe:1991ra},
with respectively the WW-type approximations in 
Eqs.~(\ref{Eq:WW-type-5}) and (\ref{Eq:WW-type-6}),
one obtains in principle a further class of approximations
\cite{Metz:2008ib}, which we shall denote by
``WW \& type'' (short for WW and WW-type) approximations.
These approximations relate leading twist TMDs to twist-2 
parton distributions as follows \cite{Avakian:2007mv,Metz:2008ib}
\ba
   	g_{1T}^{\perp(1)a}(x)
        &\stackrel{\rm WW\:\&\:type}{\approx}& 
        \phantom{-}x\,\int_x^1\frac{\di y}{y\;}\,g_1^a(y) \;,
	\label{Eq:WW-approx-g1T}\\
    	h_{1L}^{\perp(1)a}(x)
        &\stackrel{\rm WW\:\&\:type}{\approx}& 
        -x^2\!\int_x^1\frac{\di y}{y^2\;}\,h_1^a(y)\;.
	\label{Eq:WW-approx-h1L}\ea
These approximations were used in literature in order to make 
estimates for certain double \cite{Kotzinian:2006dw} and 
single \cite{Avakian:2007mv} spin asymmetries in SIDIS.
In Figs.~\ref{Fig03:WW-classic+type}c and \ref{Fig03:WW-classic+type}d
we test the quality of these approximations in the bag model at the
low scale. In both cases we observe that the approximations tend to
overestimate the magnitude of the true model results for 
$g_{1T}^{\perp(1)a}(x)$ and $h_{1L}^{\perp(1)a}(x)$, somewhat 
more in the former case and less in the latter case.

%
%

It is, however, difficult to suspect on the basis of these observations 
at the low scale of the bag model, whether the predictions from 
Refs.~\cite{Kotzinian:2006dw,Avakian:2007mv} will similarly
overestimate data. The evolution effects may play an important role.
Though the precise evolution pattern of polarized transverse
momentum dependent distribution functions is not yet fully
understood, it seems reasonable to expect the evolution 
from a low hadronic to experimentally relevant scales 
will ``shift'' the $x$-shape of the TMDs towards smaller $x$,
see also \cite{Cherednikov:2007tw}. In other words, the region
of valence- and large-$x$ in the bag model at its low scale, 
say $0.3 \lesssim x \lesssim 0.8$, could be what will be just
relevant at higher scales in the experiments at COMPASS, 
Jefferson Lab or HERMES. Remarkably, in the region of 
larger $x\gtrsim 0.3$ the approximations 
(\ref{Eq:WW-approx-g1T},~\ref{Eq:WW-approx-h1L}) work reasonably well, 
see Figs.~\ref{Fig03:WW-classic+type}c and \ref{Fig03:WW-classic+type}d.

\subsection{Orbital angular momentum}
\label{Sec-5d:OAM}

For completeness we include here also the bag model results for the
nucleon spin decomposition. Of course, it is well known that in the 
bag model $65\%$ of the nucleon spin is due to the intrinsic spin of 
the quarks, and the remaining $35\%$ are due to quark orbital motion
\cite{Chodos:1974je}, see also the study of this issue in the bag model in 
the context of generalized parton distribution functions \cite{Ji:1997gm}.
However, what is new here is that we obtain this information from 
the pretzelosity distribution, namely
\ba
     2L^3_u\equiv -2\int\di x\;h_{1T}^{\perp(1)u}(x)=  0.46\,,\hspace{4mm} &
     2L^3_d\equiv -2\int\di x\;h_{1T}^{\perp(1)d}(x)= -0.11\,,\hspace{5mm} &
     2L^3_Q   =    2L^3_u + 2L^3_d = 0.35\,,     \hspace{12mm} \\
     2S^3_u\equiv  \int\di x\;g_1^u(x)=  0.87\,, \hspace{15mm}  &
     2S^3_d\equiv  \int\di x\;g_1^d(x)= -0.22\,, \hspace{17mm}  &
     2S^3_Q   =   2S^3_u + 2S^3_d = 0.65\,,                     \\
     2J^3_u\, =   2L^3_u + 2S^3_u =  \frac43\,,  \hspace{20mm}  & \displaystyle
     2J^3_d\, =   2L^3_d + 2S^3_d = -\frac13\,,  \hspace{22mm}  &
     2J^3_Q\, =   2L^3_Q + 2S^3_Q = 1 \;.
\ea
This is a typical result for relativistic quark models at low
hadronic scales \cite{Schweitzer:2001sr,Thomas:2008ga}.
We remark that the MIT bag also reasonably well describes the 
axial coupling constant: the Bjorken sum rule yields
$g_A=\int\di x\;(g_1^u-g_1^d)(x)=1.09$ vs.\ 1.26 in experiment.
(At this point it is interesting to note that experimentally 
the Bjorken sum rule is verified with not much better accuracy
than that \cite{Alekseev:2010hc}, see also \cite{Pasechnik:2008th} 
for a recent comparison of QCD and data.)

However, in order to compare such numbers to phenomenology
or lattice QCD \cite{Hagler:2009ni} one needs to carefully take 
into account evolution effects 
\cite{Wakamatsu:2007ar} which, in the case of
bag model, is supposed to start at a very low hadronic scale 
\cite{Stratmann:1993aw}.
Evolution techniques possibly suitable for that were discussed 
in the context of the Bjorken sum rule in \cite{Pasechnik:2008th}.

\newpage
\section{Conclusions}
\label{Sec-6:conclusions}

We presented a study of leading- and subleading-twist TMDs in the MIT 
bag model. Since this model lacks explicit gluon degrees of freedom,
the Wilson-link needed in QCD to render the definitions of TMDs color 
gauge invariant is absent. As a consequence T-odd TMDs vanish.
Attempts to model T-odd TMDs in the bag were presented in
\cite{Yuan:2003wk,Courtoy:2008dn,Cherednikov:2006zn}.
In this work we have focused on the 14 T-even TMDs (6 leading- and 
8 subleading-twist). 

Another consequence of the absence of the Wilson-link 
(in any effective approach with global color symmetry only), 
is that certain relations hold among TMDs, the so-called LIRs 
\cite{Mulders:1995dh,Teckentrup:2009tk}. There are 5 such LIRs
among the 14 T-even leading- and subleading-twist TMDs, and we have
proven that they are all satisfied in the bag model.

Recently further relations among TMDs were found in models. 
One of the motivations of this work was therefore to shed some light 
on how such quark model relations arise.
We have shown that in the MIT bag model there are not more and not less
than 9 linear and 2 non-linear relations among the 14 T-even 
leading- and subleading-twist TMDs. We reviewed in detail that some of
these relations are supported also in other quark models
\cite{Jakob:1997wg,Pasquini:2008ax,Efremov:2009ze,Bacchetta:2008af} .

One of those linear relations, found  in our previous bag model study 
\cite{Avakian:2008dz}, connects the difference of $g_1^q$ and $h_1^q$
to the (1)-moment of pretzelosity. It was confirmed in several other 
\cite{Pasquini:2008ax,Efremov:2009ze,Bacchetta:2008af} though not all 
\cite{Bacchetta:2008af} quark models. 
What makes this relation particularly interesting, is the observation 
\cite{She:2009jq} that such a difference between helicity and transversity 
distributions is related in a light-cone SU(6) quark-diquark model 
\cite{Ma:1998ar} to quark orbital angular momentum (OAM).

Although intuitively the idea of quark orbital motion is associated 
with TMDs, this is to best of our knowledge the first 'rigorous' 
connection of a TMD to OAM --- in a model, of course.
While in gauge theories there is no gauge-invariant definition of an 
OAM field operator, in quark models the situation is simpler and
there are no ambiguities. Another important motivation for our study 
was therefore to investigate whether there exists any connection 
between pretzelosity and quark OAM in the bag model.
The answer is yes, agrees with the findings of \cite{She:2009jq}, and reads
\be\label{Eq:OAM-pretzel}
    L^3_q = (-1)\int\di x\;h_{1T}^{\perp(1)q}(x)\;.
\ee
Thus, by measuring the single spin asymmetry $A_{UT}^{\sin(3\phi-\phi_S)}$
due to $h_{1T}^{\perp q}$ \cite{Mulders:1995dh} one 
could access information on OAM. Of course, the
relation (\ref{Eq:OAM-pretzel}) is model-dependent. But it is supported
by two independent approaches, bag model (here)
and light-cone SU(6) quark-diquark model \cite{She:2009jq}.
Moreover, at least in the context of bag model, the information on OAM 
gained from pretzelosity, Eq.~(\ref{Eq:OAM-pretzel}), is equivalent to 
what one can learn from generalized parton distribution functions 
\cite{Ji:1997gm}. 
It will be exciting to see to what extent experimental information on 
TMDs and generalized parton distribution will, on the basis of a quark 
model inter\-pretation, converge to give a compatible picture of OAM.

The pretzelosity distribution $h_{1T}^{\perp q}$ seems to play in this
context a central role. It is also related to the non-sphericity of 
the spin-distribution in the nucleon \cite{Miller:2007ae}. 
It is interesting to ask, whether a quark model relation of 
the type (\ref{Eq:OAM-pretzel}) may inspire a way to establish a
rigorous connection between TMDs and OAM in QCD.
The task is demanding, as we observe  (\ref{Eq:OAM-pretzel}) 
on the level of matrix elements only. Further studies in 
effective approaches, and numerical results from
lattice QCD could provide valuable insights. For first attempts
to study TMDs on a lattice see \cite{lattice-TMD}.

The third main result of this work concerns practical aspects which are 
of interest for phenomenology. For example, in many phenomenological studies 
it is assumed that in TMDs the $x$- and $p_T$-dependencies factorize, and
the latter is ``Gaussian''. Many authors have stressed that 
in their models the $x$- and $p_T$-dependencies of TMDs are 
non-factorizing and non-Gaussian.
But in practical applications the Gaussian Ansatz works with a useful accuracy, 
e.g.\ \cite{Collins:2005ie,Anselmino:2005nn,D'Alesio:2004up}.

How to reconcile these observations? In order to address this question,
we introduced the notion of a (in general $x$-dependent) 'Gaussian width' 
which can be applied to any TMD. This effective Gaussian width is designed 
such that it reproduces the $p_T$-dependence of the TMD exactly in a vicinity 
of $p_T=0$ by definition. 
Although in the bag we also observe non-factorizing, non-Gaussian 
$x$- and $p_T$-dependencies, in this way we made two interesting observations. 
In the valence-$x$ region (and we speak here about a low hadronic scale),
this effective 'Gaussian width' turns out to be only weakly $x$-dependent. 
Moreover, such an effective Gaussian Ansatz approximates reasonably well
the exact model results not only in the vicinity of $p_T=0$, but up to 
$p_T\lesssim {\cal O}(M_N)$.

This is good news for phenomenological studies for two reasons:
azimuthal asymmetries in Drell-Yan or SIDIS are
dominated by intrinsic transverse parton momenta 
\cite{Cahn:1978se,Konig:1982uk,Chiappetta:1986yg}, and
one expects azimuthal and (single) spin phenomena to be sizable 
in the valence-$x$ region. And this is where we find
the Gaussian Ansatz to be a useful {\sl approximation}.
Surely, care is required when sea-quark effects start to play a role,
and for a precision treatment of transverse momenta one has to use a rigorous
approach such as the Collins-Soper-Sterman formalism \cite{Collins:1984kg}
as implemented in \cite{Landry:2002ix}.

Finally, we used the model results to test the Wandzura-Wilczek-type 
approximations 
\cite{Kotzinian:2006dw,Avakian:2007mv,Metz:2008ib,Teckentrup:2009tk} which 
were suggested as, at the presently early stage, useful tools for first 
interpretations of data. These approximations consist in neglect
'pure-interaction-dependent' terms (in QCD: quark-gluon correlations,
in the bag model: bag boundary conditions).
We observe that, for some TMDs these are fair 
approximations in the valence-$x$ region.

To conclude, though obtained in a simple model, our results bear many 
interesting insights, and we hope they will stimulate further studies
in quark models.

 \vspace{0.5cm}                                                       

\noindent{\bf Acknowledgements.}
A.~E.~is supported by the Grants RFBR 09-02-01149 and
07-02-91557, RF MSE RNP 2.1.1/2512 (MIREA) and by the
Heisenberg-Landau Program of JINR.
The work was supported in part by DOE contract DE-AC05-06OR23177, under
which Jefferson Science Associates, LLC,  operates the Jefferson Lab.
F.~Y.~is grateful to RIKEN, Brookhaven National Laboratory and the U.S.\
Department of Energy (contract number DE-AC02-98CH10886) for providing
the facilities essential for the completion of this work.

\appendix
\section{Proofs of LIRs}
\label{App-A:prove-LIRs}

In this Appendix we prove the LIRs (\ref{eq:LIR1}--\ref{eq:LIR5}).
Thanks to the model relations discussed in 
Sec.~\ref{Sec-3b:linear-relation-in-bag-model}, 
we do not need to prove every LIR.
For example, if we prove the LIR (\ref{eq:LIR1}) then 
also the LIR (\ref{eq:LIR2}) holds due to the relations 
(\ref{Eq:rel-IV},~\ref{Eq:measure-of-relativity},~\ref{Eq:rel-VIII}), 
which can be seen conveniently by combining 
(\ref{Eq:measure-of-relativity},~\ref{Eq:rel-VIII}) to form 
(\ref{Eq:new-interesting}). 
Similarly, if we prove the LIR (\ref{eq:LIR4}) then it is clear,
that also the LIR (\ref{eq:LIR3}) holds, due to the model
relations (\ref{Eq:rel-V},~\ref{Eq:rel-VI}).
Finally, (\ref{eq:LIR5}) is evidently true, c.f.\ the model relation 
(\ref{Eq:rel-IX}).
Thus, it is sufficient to demonstrate, for example, that the LIRs 
(\ref{eq:LIR1}) and (\ref{eq:LIR4}) hold. 

In order to prove (\ref{eq:LIR4}) we rewrite the expression for 
$h^{\perp(1)}_{1T}(x)$ in a convenient for our purposes way. 
Recalling that $k$ is a function of $k_\perp$ and 
$k_z=xM_N-\omega/R_0$ according to (\ref{Eq:notation}),
we may write
\ba\label{appEq:01}
     h_{1T}^{\perp(1)q}(x) \;            
     = P_q\,A \int\di^2k_\perp\;\frac{k_\perp^2}{2M_N^2}\biggl[-2\widehat{M}_N^2\,t_1^2(k)\biggr]
     = P_q\,A \int\di^3q\int\frac{\di\tau}{2\pi}\;e^{i(k_z-q\cos\theta)\tau} 
     \biggl[- \sin^2\theta\,t_1^2(q)\biggr]
\ea
where we replaced the transverse momentum integral by an integration
over the independent variables $\vec{q}$ with $q=|\vec{q}\,|$, using a 
delta-function $\int\di\tau/({2\pi})\,e^{i(k_z-q_z)\tau} = \delta(q_z-k_z)$, and
the spherical coordinates $q_z=q\cos\theta$ and  $q_\perp^2=q^2\sin^2\theta$.
Next we differentiate (\ref{appEq:01}) with respect to $x$, recalling that the
$x$-dependence is 'hidden' only in $k_z$ according to (\ref{Eq:notation})
with $\frac{\di\;}{\di x}\,k_z=M_N$. We obtain
\ba
     \frac{\di\;}{\di x}\;  h^{\perp(1)q}_{1T}(x)  
     &=& 
     P_q\,A \int\di^3q\int\frac{\di\tau}{2\pi}\;e^{i(k_z-q\cos\theta)\tau}\,(iM_N\tau) 
     \biggl[-\sin^2\theta\,t_1^2(q)\biggr]\nonumber\\
     &=& 
     P_q\,A \int\di^3q\int\frac{\di\tau}{2\pi}\;\biggl\{-\,\frac{M_N}{q}\,
     \frac{\di\;}{\di\cos\theta}\,e^{i(k_z-q\cos\theta)\tau}\biggr\}
     \biggl[-\sin^2\theta\,t_1^2(q)\biggr]\nonumber\\
     &=& 
     P_q\,A \int\di^3q\int\frac{\di\tau}{2\pi}\;\frac{M_N}{q}\,
     e^{i(k_z-q\cos\theta)\tau}
     \biggl[-2\cos\theta\,t_1^2(q)\biggr]\nonumber\\
     &=& 
     P_q\,A \int\di^3q\;\delta(q_z-k_z)\biggl[\,2M_Nq_z\,\frac{t_1^2(q)}{q^2}\biggr]
     \nonumber\\
     &=&
     P_q\,A \int\di^2k_\perp\biggl[\,2\widetilde{M}_N\widetilde{k}_z\,t_1^2(k)\biggr]
     \nonumber\\
     &\equiv&-h_T^q(x)
\label{appEq:02}\ea
where we interchanged the order of the integrations 
and differentiation, which is legitimate in our case, integrated by parts 
in the third step, and finally recovered the expression for $(-1)h_T^q(x)$, 
which completes the proof of the LIR~(\ref{eq:LIR4}).
The proof of the LIR~(\ref{eq:LIR1}) is straightforward, and consists of
repeating steps analog to (\ref{appEq:01},~\ref{appEq:02}).

It is interesting to remark that a different version of the LIR~(\ref{eq:LIR4})
is the following
\be\label{eq:LIR4-version} 
     h^{(1)q}_T(x) \; \stackrel{\rm LIR}{=} \;-\,\frac12\; 
     \frac{\di }{\di  x} h^{\perp(2)q}_{1T}(x) \, , \\
\ee
with the subtlety that the (2)-moment of pretzelosity is divergent,
but its derivative with respect to $x$ is finite, i.e.\ the same careful
treatment is required as in the case of the (1)-moment of $f_1^q$ discussed in 
detail in Sec.~\ref{Sec-5b:pT-dependence}.
The numerical bag model results for TMDs satisfy all LIRs including the version
(\ref{eq:LIR4-version}).

\newpage
\section{Proofs of inequalities}
\label{App-B:prove-ineq}

This Appendix contains the explicit demonstrations 
that the quark TMDs from the bag model satisfy the inequalities 
(\ref{Ineq:f1-g1-h1}--\ref{Ineq:h1Lperp}). 
We recall that, in the present version of the model, the 
T-odd TMDs $f_{1T}^{\perp q}$ and $h_1^{\perp q}$ are absent
and that the inequalities for antiquarks are violated, 
see Sec.~\ref{Sec-2:TMDs-in-bag} for a detailed discussion.

In order to check the inequalities in (\ref{Ineq:f1-g1-h1})
we work directly with the model expressions. 
With $\widehat{k}_z=k_z/\sqrt{k_\perp^2+k_z^2}$ we have 
$-1<\widehat{k}_z<1$ which we use below in (\ref{appEq:f1-positivity}), 
$-1\le (2\widehat{k}_z^2-1) \le 1$ we use in (\ref{appEq:g1-positivity}), 
and $\widehat{k}_z^2<1$ used in (\ref{appEq:h1-positivity}),
\ba
&& \label{appEq:f1-positivity}
f_1^q(x,k_\perp) 
    =   N_q  A\biggl[t_0^2+2\widehat{k}_z\,t_0t_1 +t_1^2\biggr] 
    \ge N_q  A\,(t_0-t_1)^2 \ge 0\;,\\
&&\label{appEq:g1-positivity}
g_1^q(x,k_\perp) 
    =   P_q  A\biggl[t_0^2+2\widehat{k}_z\,t_0t_1 +(2\widehat{k}_z^2-1)t_1^2\biggr]
    \le P_q  A\biggl[t_0^2+2\widehat{k}_z\,t_0t_1 + t_1^2\biggr]
    =  \frac{P_q}{N_q}\; f_1^q(x,k_\perp) \;,\\
&&\label{appEq:h1-positivity}   
   h_1^q(x,k_\perp)  
   =   P_q\,A\biggl[t_0^2+2\widehat{k}_z\,t_0t_1 +\widehat{k}_z^2\,t_1^2\biggr]
   \le P_q\,A\biggl[t_0^2+2\widehat{k}_z\,t_0t_1 + t_1^2\biggr]
   = \frac{P_q}{N_q}\; f_1^q(x,k_\perp) \;,
\ea
The inequalities (\ref{appEq:g1-positivity},~\ref{appEq:h1-positivity}) imply 
$|g_1^q(x,k_\perp)| < f_1^q(x,k_\perp)$ and 
$|h_1^q(x,k_\perp)| < f_1^q(x,k_\perp)$, because for the nucleon in SU(6)
\be\label{AppIneq:SU6}
   \biggl|\frac{P_q}{N_q}\biggr|<1 \;,\;\;q=u,\; d.
\ee

In order to check the inequalities (\ref{Ineq:Soffer},~\ref{Ineq:pretzel}) it 
is convenient to explore the model relations (\ref{Eq:rel-I}--\ref{Eq:rel-IX}).
For example, because of 
(\ref{AppIneq:SU6}) the relation (\ref{Eq:rel-I}) immediately implies 
that for the nucleon the Soffer inequality (\ref{Ineq:Soffer}) is a true 
inequality $|h_1^q(x,k_\perp)|<\frac12(f_1^q(x,k_\perp)+g_1^q(x,k_\perp))$,
see also \cite{Barone:2001sp}. Similarly, by using the relations
(\ref{Eq:rel-I},~\ref{Eq:measure-of-relativity}) to eliminate 
transversity, we conclude with (\ref{AppIneq:SU6}) that 
(\ref{Ineq:pretzel}) is a true inequality, i.e.\ 
$|h_{1T}^{\perp q}(x,k_\perp)|<\frac12(f_1^q(x,k_\perp)-g_1^q(x,k_\perp))$.

In order to verify the inequalities 
(\ref{Ineq:g1Tperp},~\ref{Ineq:h1Lperp}) we explore the 
linear (\ref{Eq:rel-I}--\ref{Eq:rel-IX}) and non-linear 
(\ref{Eq:non-lin-2},~\ref{Eq:non-lin-1}) model relations. 
It is sufficient, thanks to the absence of the T-odd TMDs 
$f_{1T}^{\perp q}$ and $h_1^{\perp q}$ and the relation (\ref{Eq:rel-IV}),
to prove one of these inequalities, let us say (\ref{Ineq:h1Lperp}).
For that we multiply (\ref{Eq:non-lin-2}) by $(k_\perp^2/2M_N^2)^2$,
and eliminate transversity by means of the relation (\ref{Eq:rel-I}) 
and $h_{1T}^{\perp(1)q}(x,k_\perp)$ by means of the relations 
(\ref{Eq:rel-I},~\ref{Eq:measure-of-relativity}). This yields
\be\label{AppEq:pretzel-f1-g1}
    h_{1L}^{\perp(1)q}(x,k_\perp)^2 = \frac{k_\perp^2}{4M_N^2} \biggl(
    \frac{P_q^2}{N_q^2}\;f_1^q(x,k_\perp)^2-g_1^q(x,k_\perp)^2 \biggr)
    < \frac{k_\perp^2}{4M_N^2} 
    \biggl(f_1^q(x,k_\perp)^2-g_1^q(x,k_\perp)^2 \biggr)\,,
\ee 
where in the last step we explored (\ref{AppIneq:SU6}). 

It is interesting to remark that, except for $f_1^q(x,k_\perp)\ge 0$, 
all the other inequalities in (\ref{Ineq:f1-g1-h1}--\ref{Ineq:h1Lperp})
are true inequalities for the proton in SU(6), i.e.\ they are never
saturated. For other baryons in SU(6) saturations may occur, for example,
for strange quarks in the $\Lambda$ hyperon, where $N_s=P_s=1$,
see \cite{Barone:2001sp} for a related discussion.


\end{document}